\documentclass[%
 reprint,
superscriptaddress,
 amsmath,amssymb,
 aps,
prd,
]{revtex4-2}

\usepackage{graphicx}
\usepackage{dcolumn}
\usepackage{bm}
\usepackage{slashed}
\usepackage{orcidlink}
\usepackage{ragged2e}  
\usepackage{subcaption}
\usepackage[capitalise]{cleveref}
\usepackage{amsmath}
\usepackage{amsfonts}
\usepackage{lipsum}
\usepackage{dsfont}
\usepackage{amssymb}
\hypersetup{colorlinks=true, linkcolor=blue, urlcolor=blue, citecolor=blue} 
\usepackage{hyperref} 
\usepackage[normalem]{ulem}

\newcommand{\orcidauthorA}{\orcidlink{0000-0001-7569-7197}}  
\newcommand{\orcidauthorB}{\orcidlink{0000-0003-1728-0304}}  
\newcommand{\orcidauthorC}{\orcidlink{0000-0002-1255-7338}}  
\newcommand{\orcidauthorD}{\orcidlink{0009-0003-5075-3107}}  
\newcommand{\orcidauthorE}{\orcidlink{0000-0002-9976-2405}}  

\begin{document}

\preprint{APS/123-QED}

\title{Particlelike solutions of the Einstein-Dirac-Higgs equations: \\
ground, excited, and many-fermion states
}

\author{Reinosuke Kusano~\orcidauthorA}
\affiliation{SUPA Physics and Astronomy, University of St Andrews, North Haugh, St Andrews, Fife KY16~9SS, UK} 
\author{Keith Horne~\orcidauthorB}
\affiliation{SUPA Physics and Astronomy, University of St Andrews, North Haugh, St Andrews, Fife KY16~9SS, UK}
\author{Peter E. D. Leith~\orcidauthorC}
\affiliation{SUPA Physics and Astronomy, University of St Andrews, North Haugh, St Andrews, Fife KY16~9SS, UK} 
\author{Miguel Yulo Asuncion~\orcidauthorD} 
\affiliation{Nottingham Centre of Gravity, University of Nottingham, Nottingham NG7~2RD, UK}
\affiliation{School of Mathematical Sciences, University of Nottingham, Nottingham NG7~2RD, UK}
\author{Chris A. Hooley~\orcidauthorE} 
\affiliation{Centre for Fluid and Complex Systems, Coventry University, Coventry CV1~2TT, UK}

\date{\today}

\begin{abstract}
We present an extended study of the gravitationally localized soliton-like solutions to the minimally-coupled Einstein-Dirac-Higgs equations, embedded in asymptotically Minkowski spacetimes. The equations of motion describing these Yukawa-coupled Dirac stars are generalized to any even number of constituent fermions. We first expand the discussion of two-fermion ground-state solutions initially analyzed in Leith~\textit{et al.}~[Phys.~Rev.~D. \textbf{107}, 106020 (2023)]. Upon extending our analysis to excited states and many-fermion states, we find that both exhibit different behaviors to their two-fermion ground-state counterparts. In these excited states and many-fermion states, the Higgs field may exhibit stepwise increases and at times decrease within the soliton, which has not been seen for the two-fermion ground states. We also propose the potential cause of a significant degree of ADM-to-fermion mass disparity, which is present in states with strong Yukawa-coupling. 
\end{abstract} 

\maketitle 

\section{\label{sec:intro}Introduction}

As a consistent theory of quantum gravity is yet to be fully realized, various approximative approaches are applied to describe the coupling of gravity to fundamental fields. Examples of localized configurations resulting from such approaches include geons~\cite{wheeler1955}, boson stars~\cite{kaup1968}, Proca stars~\cite{brito2016}, and Bartnik-McKinnon solutions~\cite{bartnikmckinnon1988}, which describe classical fields trapped in their own gravitational well. 

Within this Einstein-wave matter framework~\cite{dafermos2003}, similar compact objects to the above can be constructed with fermions as well. One such approach is detailed in a work by Felix Finster, Joel Smoller, and Shing-Tung Yau~(FSY hereafter), in which they present soliton-like solutions to the minimally-coupled Einstein-Dirac equations~\cite{fsy1999}. In this formalism, the classical Dirac action sources the matter sector of the Einstein field equations. The resulting configurations describe the localization of a self-gravitating system comprised of an even number of Dirac fermions. While not being a semiclassical framework as the matter fields are not second-quantized~\cite{herdeiro2017,herdeiro2020,kain2023}, this toy model of ``Einstein-Dirac solitons'' or ``Dirac stars'' may still provide insights into how spin-$1/2$ fermions bind themselves to each other gravitationally. 

The above FSY Einstein-Dirac~(ED) system has been explored extensively. The formalism has been extended to charged~\cite{fsy1999_EDM}, spinning~\cite{herdeiro2019,herdeiro2022},  
many-fermion~\cite{leith2020, leith2021, sun2024}, and multi-shell~\cite{iwasawa2025}
configurations. A recent development has seen a semiclassical extension to the problem, wherein the matter sector is sourced by a canonically quantized Dirac field~\cite{kain2023}. Some analytical solutions exist~\cite{bakuczcanario2020, blazquezsalcedo2020, leith2026}, though not all are normalizable. A series of proofs also finds that such objects do not exist in static spherically symmetric black hole spacetimes, with results generalized to include couplings to electroweak gauge fields~\cite{fsy1999_noBHEDM, fsy2000_noRNBH, fsy2000_noKerrBH, finster2002, bernard2006}. 

One may also consider the inclusion of additional Standard Model mechanisms, such as to Yukawa-couple the Dirac fermions to a real Klein-Gordon scalar field. This is motivated by the aim of understanding the dynamical mass generation of fermions in curved spacetime, and may also be linked to an eventual goal of classically modeling the reheating process in cosmology~\cite{freese2018,datta2024,litsa2021,bhusal2026}. These ``Einstein-Dirac-Higgs''~(EDH) solitons, first formulated in Ref.~\cite{leith2023}, are the focus of the present work. Unlike other Dirac-scalar field systems~\cite{liang2023, berti2024}, we refer to our scalar field as a ``Higgs'' field, as the chosen shape of the scalar field potential in the Lagrangian is a double well.

The present work explores even parity particlelike solutions to this EDH system, wherein fermion masses depend on the radial profile of the Higgs field as a result of Yukawa-coupling. We detail the exotic phenomenologies of ground- and excited-state solutions in this approach, with extensions to many-fermion states. We also propose a qualitative explanation for the cause of a behavior seen in ground-state solitons, where the sum of the asymptotic fermion masses far exceed the gravitational mass of the soliton~\cite{leith2023}. We attribute this to a large discrepancy between the interior and exterior fermion masses. 

The present work is structured as follows. In Sec.~\ref{sec:eom}, we derive the five equations of motion for an arbitrary even number of fermions in the EDH system. We then outline the expected structure of numerical solutions in Sec.~\ref{sec:num_sol}. In Sec.~\ref{sec:mc_dynamics}, we present new two-fermion ground-state solutions, and clarify the reasons behind the deviations of these families from the FSY ED case. Furthermore, we introduce excited states~(Sec.~\ref{sec:mc_exc}) and many-fermion states~(Sec.~\ref{sec:many_fermion}), which both exhibit different behaviors to the two-fermion ground-state solutions. In Sec.~\ref{sec:MSS1} and Sec.~\ref{sec:MSS2}, we propose a probable cause of a ``mass-scale separation'' found in ground-state solutions, the effects of which we show to be validated by various phenomenologies of our solutions. We conclude in Sec.~\ref{sec:conc}. 

Data for individual solutions are given in Appendix~\ref{app:indivsol}.

\section{\label{sec:eom}Equations of motion}

In this section, we 
derive the five coupled differential equations that describe the EDH system. A two-fermion derivation is provided in Ref.~\cite{leith2023}, but we generalize to many-fermion states here. We follow the procedures and ansätze of Ref.~\cite{fsy1999_noBHEDM}, and adopt the mostly-positive metric signature~($-,\,+,\,+,\,+$).

The total EDH action takes the form: 
\begin{equation}\label{eq:S_EDH}
    S_\mathrm{EDH}=\int\left(\mathcal{L}_\mathrm{g} +\mathcal{L}_\mathrm{m}\right) \sqrt{-g} \, \mathrm{d}^4 x, 
\end{equation}
with the gravitational Lagrangian $\mathcal{L}_\mathrm{g}$ taking the standard Einstein-Hilbert form: 
\begin{equation}\label{eq:L_grav}
    \mathcal{L}_\mathrm{g}=\dfrac{R}{16\pi G}.
\end{equation}
$R$ is the Ricci scalar and $G$ is Newton's gravitational constant. $\hbar=c=1$ here and henceforth, while we keep $G$ in our equations explicitly. Thus, all quantities are given in Planck units: masses and energies are given in units of the Planck mass~$m_\mathrm{P}=\sqrt{\hbar c/G}$ and lengths are in units of the Planck length~$\ell_\mathrm{P}=\sqrt{\hbar G/c^3}$~\cite{leith2022_phd}.

For the present minimally-coupled EDH system, the classical matter Lagrangian $\mathcal{L}_\mathrm{m}$ is the sum of the Dirac and Higgs actions~\cite{leith2022_phd}:
\begin{equation}\label{eq:L_mat}
\begin{aligned}
    \mathcal{L}_\mathrm{m}=\overline{\Psi} \left(\slashed{D}-\mu h\right)\Psi - \dfrac{1}{2}\left(\nabla^\nu h\right)\left(\nabla_\nu h\right) - V(h).
    \end{aligned}
\end{equation}
$\Psi$ is the Dirac spinor field, while $h$ is a real-valued Klein-Gordon scalar field. $\slashed{D}$ denotes the Dirac operator, and $\mu$ is the dimensionless constant determining the strength Yukawa-coupling between the fermions and the scalar field. The Higgs potential $V(h)$ takes the double well form 
\begin{equation}\label{eq:V(h)}
    V(h)=\lambda(h^2-v^2)^2,
    \quad \frac{{\rm d}V}{{\rm d}h} = 4\lambda h\left(h^2 - v^2 \right),
\end{equation}
with extrema at $h=0$ (local maximum) and $h=\pm v$ (local minima). 
A stable vacuum with $V(h)$ bounded from below requires the dimensionless quartic Higgs self-coupling parameter $\lambda>0$. To recover asymptotically flat spacetime with no vacuum energy density from the Higgs field at large $r$, one requires $h\rightarrow\pm v$ so that $V(h)\rightarrow0$ outside the localized fermion distribution. We choose to impose $h\rightarrow +v$ at large $r$ such that for $\mu>0$ our fermion masses are asymptotically positive, and assign $+v$ as our vacuum expectation value~(VEV) for the Higgs field~\cite{leith2023}. 

From the curvature at the minimum of $V(h)$, we identify the Higgs mass $m_\mathrm{H}$: 
\begin{equation}\label{eq:Higgs_mass}
    m_\mathrm{H}^2\equiv\left.\frac{\mathrm{d}^2V}{\mathrm{d}h^2}\right|_{h=v}= 8 \lambda v^2.
\end{equation}

In line with FSY~\cite{fsy1999}, the metric ansatz for our static and spherically symmetric spacetime is: 
\begin{equation}\label{eq:metric_ansatz}
    g_{\mu\nu}=\text{diag}\left(-\dfrac{1}{T^2(r)}, \dfrac{1}{A(r)},r^2,r^2\sin^2\theta\right),
\end{equation}
Each individual solution in an ED(H) family can be characterized by its ``central (gravitational) redshift''~$z$~\cite{bakuczcanario2020} due to staticity:
\begin{equation}\label{eq:redshift_def}
    z=\frac{1}{\sqrt{-g_{tt}(r=0)}}-1 =T(r=0)-1.
\end{equation}
We also choose a static scalar field $h=h(r)$. 

In the following subsections, we derive the Dirac, Einstein, and Higgs equations for the present system for an arbitrary even number of fermions. 

\subsection{Dirac equations}\label{subsec:Dirac_eq}

Taking the variation of the action $S_{\rm EDH}$~\eqref{eq:S_EDH} with respect to $\overline{\Psi}$, we arrive at the Dirac equation in curved spacetime:
\begin{equation}\label{eq:Dirac}
    (\slashed{D}-\mu h)\Psi = 0.
\end{equation}
We can determine the Dirac operator $\slashed{D}$ with
\begin{equation}
\begin{aligned}
    \slashed{D}=i\gamma^\mu\partial_\mu &+ \dfrac{i}{2}\nabla_\mu\gamma^\mu,
\end{aligned}
\end{equation}
wherein the gamma matrices $\gamma^\mu$ are described using the spherical Pauli matrices $\sigma^r$, $\sigma^\theta$, and $\sigma^\phi$~\cite{leith2022_phd}:
\begin{equation}
        \begin{aligned}
        \gamma^t =T\begin{pmatrix}
            \mathds{1} & 0\\
            0 & -\mathds{1}
            \end{pmatrix},\quad
        \gamma^r 
        =\sqrt{A}\begin{pmatrix}
            0 & \sigma^r\\
            -\sigma^r & o
            \end{pmatrix},\\
        \gamma^\theta 
        =\dfrac{1}{r}\begin{pmatrix}
            0 & \sigma^\theta\\
            -\sigma^\theta & 0
            \end{pmatrix},\quad
        \gamma^\phi 
        =\dfrac{1}{r}\begin{pmatrix}
            0 & \sigma^\phi\\
            -\sigma^\phi & 0
            \end{pmatrix},
            \end{aligned}
\end{equation}
such that 
\begin{equation}\label{eq:Dirac_operator}
\begin{aligned}
\slashed{D}
    =\,& i\gamma^t\partial_t + i\gamma^r\left(\partial_r + \dfrac{1}{r}\left(1-\dfrac{1}{\sqrt{A}}\right) - \dfrac{T'}{2T}\right) 
    \\&\hspace{3cm}
    +i\gamma^\theta\partial_\theta + i\gamma^\phi\partial_\phi.
\end{aligned}
\end{equation}

The eigenvector basis is $\Psi^\pm_{jk\omega}$, with a fermion (kinetic) energy $\omega\in\mathbb{R}$. The quantum numbers $j$ and $k$ span $j=1/2,3/2,5/2,...$ and $k=-j,-j+1,...,j$~\cite{fsy1999_noBHEDM}. We consider a singlet state with $N=|\kappa|=2j+1$ identical fermions in a filled shell, which due to $N\in2\mathbb{N}$ maintains the spherical symmetry of the system. This can be seen by describing the system using the Hartree-Fock formalism~\cite{fsy1999_noBHEDM}: 
\begin{equation}
    \Psi^\text{HF}=\Psi^c_{j,k=-j,\omega}\land\Psi^c_{j,k=-j+1,\omega}\land...\land\Psi^c_{j,k=-j,\omega}.
\end{equation}

The spinor ansatz is
\begin{equation}\label{eq:spinor_ansatz}
    \Psi^c_{jk\omega}=\dfrac{\sqrt{T}}{r}\begin{pmatrix}
        \chi^k_{j\mp\frac{1}{2}}\alpha\\
        i\chi^k_{j\pm\frac{1}{2}}\beta
    \end{pmatrix}e^{-i\omega t},
\end{equation}
with the real-valued functions $\alpha$ and $\beta$ respectively corresponding to the matter and antimatter components of the Dirac spinor. The 2-spinors $\chi^k_{j}$ are:
\begin{equation}\label{eq:2spinors}
    \begin{aligned}
        \chi^k_{j-\frac{1}{2}}&=\sqrt{4\pi}\Bigg(\sqrt{\dfrac{j+k}{2j}}Y^{k-\frac{1}{2}}_{j-\frac{1}{2}}\begin{pmatrix}
            1\\0
        \end{pmatrix}\\
        &\hspace{1.9cm}+\sqrt{\dfrac{j-k}{2j}}Y^{k+\frac{1}{2}}_{j-\frac{1}{2}}\begin{pmatrix}
            0\\1
        \end{pmatrix}\Bigg),\\
        \chi^k_{j+\frac{1}{2}}&=\sqrt{4\pi}\Bigg(\sqrt{\dfrac{j+1-k}{2j+2}}Y^{k-\frac{1}{2}}_{j+\frac{1}{2}}\begin{pmatrix}
            1\\0
        \end{pmatrix}\\
        &\hspace{1.75cm}
        -\sqrt{\dfrac{j+1+k}{2j+2}}Y^{k+\frac{1}{2}}_{j+\frac{1}{2}}\begin{pmatrix}
            0\\1
        \end{pmatrix}\Bigg),
    \end{aligned}
\end{equation}
with standard spherical harmonics $Y^l_m (\theta,\phi)$. This ansatz reduces the $N$-particle spinor $\Psi^\pm_{jk\omega}$ so that it can be described by the two real functions $\alpha$ and $\beta$. 

From this, our two Dirac equations~\eqref{eq:Dirac} are now 
\begin{equation}\label{eq:dirac_matrix}
\begin{aligned}
    \Bigg[i\gamma^t&\partial_t + i\gamma^r\left(\partial_r + \dfrac{1}{r}\left(1-\dfrac{1}{\sqrt{A}}\right)-\dfrac{T'}{2T}\right)\\&+i\gamma^\theta\partial_\theta+i\gamma^\phi\partial_\phi - \mu h\Bigg]\dfrac{\sqrt{T}}{r}\begin{pmatrix}
        \chi^k_{j\mp\frac{1}{2}}\alpha\\
        i\chi^k_{j\pm\frac{1}{2}}\beta
    \end{pmatrix}e^{-i\omega t}=0.
\end{aligned}
\end{equation}
The Dirac equations from each row of~\eqref{eq:dirac_matrix} are then
\begin{equation}\label{eq:Dirac_EDH}
\begin{aligned}
    \sqrt{A}\alpha'&=+\dfrac{\kappa\alpha}{2r}-\left(\omega T + \mu h\right)\beta,\\[1ex]
    \sqrt{A}\beta'&=-\dfrac{\kappa\beta}{2r} + \left(\omega T - \mu h\right)\alpha.
\end{aligned}
\end{equation}
These Dirac equations appear identical to those of the non-Yukawa-coupled ED system~\cite{fsy1999, fsy1999_noBHEDM, bakuczcanario2020,leith2020}, though here the fermion mass $\mu h(r)$ is a function of $r$, which takes an asymptotic value of $m_\mathrm{F}=\mu v$.

\subsection{Einstein equations}\label{subsec:Einstein_eq}

To obtain the equations for $T(r)$ and $A(r)$, we vary the action~\eqref{eq:S_EDH} with respect to the metric~\eqref{eq:metric_ansatz}, which yields the Einstein equations:
\begin{equation}\label{eq:Einstein}
    G_{\mu\nu}=8\pi G \, T_{\mu\nu}.
\end{equation}
$G_{\mu\nu}$ is the Einstein tensor encoding spacetime curvature, while $T_{\mu\nu}$ is the energy-momentum tensor containing the contributions from matter and energy sources: 
\begin{equation}\label{eq:T_munu}
\begin{aligned}
    T_{\mu\nu}=-\sum^{N}_{a=1}\Re&\{\overline{\Psi}_a(i\gamma_\mu \partial_\nu)\Psi_a\} + (\partial_\mu h)(\partial_\nu h)\\
    &-g_{\mu\nu}\left[\dfrac{1}{2}(\partial^\sigma h)(\partial_\sigma h) + V(h)\right],
\end{aligned}
\end{equation}
which is equivalent to that given in Ref.~\cite{leith2023}, though here we have generalized our system to $N=|\kappa|$ fermions. 

The energy-momentum tensor is diagonal, and its mixed nonzero components are given by:
\begin{equation}\label{eq:T_munu_comps}
\begin{aligned}
    {T^t}_t&=-\dfrac{|\kappa|\omega}{r^2}T^2(\alpha^2+\beta^2)-\dfrac{A}{2}\,{h'\,}^2-V(h),\\
    {T^r}_r&=\dfrac{|\kappa|}{r^2}T\sqrt{A}(\alpha\beta'-\alpha'\beta)+\dfrac{A}{2}\,{h'\,}^2-V(h),\\
    {T^\theta}_\theta={T^\phi}_\phi&=\dfrac{\kappa|\kappa|}{2r^3}T\alpha\beta-\dfrac{A}{2}\, {h'\,}^2-V(h),
\end{aligned}
\end{equation}
with equal angular components owing to spherical symmetry. Primes denote radial derivatives $\mathrm{d}/{\rm d}r$. 

Mixed components of the Einstein tensor $G_{\mu\nu}$ are 
\begin{equation}\label{eq:G_munu_comps}
\begin{aligned}
    {G^t}_t&=\dfrac{1}{r^2}(-1+A+rA'),\\
    {G^r}_r&=\dfrac{1}{r^2}\left(-1+A-\dfrac{2rAT'}{T}\right),\\
    {G^\theta}_\theta={G^\phi}_\phi&=\dfrac{A'}{2r}-\dfrac{A'T'}{2T}+\dfrac{2A\, {T'\,}^2}{T^2}-\dfrac{AT'}{rT}-\dfrac{AT''}{T}.
\end{aligned}
\end{equation}
As the Einstein equations are divergenceless, in spherical symmetry two of the equations of motion can always be expressed in terms of the other two, so we choose the $tt$ and $rr$ components. The Einstein equations for our system then read: 
\begin{equation}\label{eq:Einstein_EDH}
    \begin{aligned}
        -1+A+rA'&=8 \pi G r^2\Big(-\dfrac{|\kappa|\omega}{r^2}T^2(\alpha^2+\beta^2)\\
        &\hspace{2.25cm}-\dfrac{A}{2}\,{h'}^2-V(h)\Big),\\
        -1+A-\dfrac{2rAT'}{T}&=8\pi G r^2\Big(\dfrac{|\kappa|}{r^2}T\sqrt{A}(\alpha\beta'-\alpha'\beta)\\
        &\hspace{2.25cm}+\dfrac{A}{2}\,{h'\,}^2-V(h)\Big).
    \end{aligned}
\end{equation}
where both sides of~\eqref{eq:Einstein} are multipled by $r^2$ for convenience. 

\subsection{Higgs equation}\label{subsec:Higgs_eq}

We now present the Higgs equation, which is obtained through the variation of the action~\eqref{eq:S_EDH} with respect to $h$. This gives
\begin{equation}\label{eq:Higgs} 
    \nabla_\nu \nabla^\nu h = \mu\overline{\Psi}\Psi + \dfrac{\mathrm{d}V}{\mathrm{d}h}.
\end{equation}
Using~\eqref{eq:spinor_ansatz}, $\overline{\Psi}\Psi$ is
\begin{equation}\label{eq:psipsi}
        \overline{\Psi}\Psi=\sum^j_{k=-j}\overline{\Psi}^c_{jk\omega}{\Psi}^c_{jk\omega}
        =\dfrac{\lvert\kappa\rvert T}{r^2}(\alpha^2-\beta^2),
\end{equation}
whereas $\nabla_\nu \nabla^\nu h$ is:
\begin{equation}
\begin{aligned}
        \nabla_\nu \nabla^\nu h 
        &= Ah''-A\left(\dfrac{T'}{T}-\dfrac{A'}{2A}-\dfrac{2}{r}\right)h'.
\end{aligned}
\end{equation}

Our second-order Higgs equation is therefore:
\begin{equation}\label{eq:Higgs_EDH}
    Ah''-A\left(\dfrac{T'}{T}-\dfrac{A'}{2A}-\dfrac{2}{r}\right)h'=\dfrac{\lvert\kappa\rvert\mu T}{r^2}(\alpha^2-\beta^2)+\dfrac{{\rm d} V}{{\rm d} h}.
\end{equation}

\section{Solution structure\label{sec:num_sol}}

\eqref{eq:Dirac_EDH},~\eqref{eq:Einstein_EDH}, and~\eqref{eq:Higgs_EDH} describe $N=\lvert\kappa\vert\in2\mathbb{N}$ identical Dirac fermions minimally coupled to a Higgs field in Einstein gravity. The ED system of FSY~\cite{fsy1999} is recovered in the limit $\lambda\rightarrow\infty$, which fixes the scalar field at $h=v$ with a constant fermion mass throughout spacetime. 

In Sec.~\ref{subsec:char_const}, we present the characteristic constants associated with each individual solution. 
We then detail the expected radial structure of these numerical solutions~(Sec.~\ref{subsec:zonal_structure}) by streamlining the terminology of Ref.~\cite{bakuczcanario2020} to be applicable to the present EDH system, and also present a new singular solution corresponding to a maximally compressed soliton. We also highlight the expected behavior of the Higgs field with reference to the zonal structure~(Sec.~\ref{subsec:higgs_mech}). 

A few points of discussion in the present section are already established. However, we revisit and expand on some technicalities here, such that the present work is self-contained and the framework can easily be understood without reference to previous works. 

\subsection{Characteristic constants}\label{subsec:char_const}

Nonsingular localized gravitational solitons require imposing both interior and exterior boundary conditions. The interior boundary conditions are
\begin{equation}\label{eq:bounds_int}
        \alpha(r=0)=0,\hspace{0.5cm}
        \beta(r=0)=0.
\end{equation}
Meanwhile, the exterior boundary conditions are
\begin{equation}\label{eq:bounds_ext}
    \begin{aligned}
        A(r\rightarrow\infty)=1,&\hspace{0.5cm}
        T(r\rightarrow\infty)=1,\\
        h(r\rightarrow\infty)=v,&\hspace{0.5cm}
        h'(r\rightarrow\infty)=0.
        \end{aligned}
\end{equation}
At infinity, the fermion fields decay to zero, $T$ and $A$ go to their values in Minkowski space, and the Higgs field asymptotes to its VEV $v$. We also impose a normalization condition on the fermion fields $\alpha$ and $\beta$ such that 
\begin{equation}\label{eq:norm_rescaled}
    4\pi\int^\infty_0\dfrac{T(r)}{\sqrt{A(r)}}(\alpha^2(r)+\beta^2(r))\,\mathrm{d}r=1.
\end{equation}
The numerical strategy used to solve this system is described in detail in Appendix~\ref{app:num}. The procedure can be used to solve for any excited state, which is denoted by $n$ referring to the $n^\mathrm{th}$ excited state. $n$ is even for the even parity states we are considering~\cite{leith2022_phd}, and for an arbitrarily-excited solution~($n>0$), $\alpha$ and $\beta$ each have $n/2$ nodes.

We parametrize solution families for a given $\kappa$ using the Higgs field VEV $v>0$ and a constant $\xi>0$. $\xi$ is defined as the ratio between the Higgs mass~\eqref{eq:Higgs_mass} and the asymptotic fermion mass $m_\mathrm{F}=\mu v$:
\begin{equation}\label{eq:xi}
    \xi \equiv\frac{m_\mathrm{H}}{m_\mathrm{F}}= \dfrac{2\sqrt{2\lambda}}{\mu}.
\end{equation}

Having generated a normalized solution to the EDH equations using the procedure in Appendix~\ref{app:num}, we can obtain different constants characterizing each solution. 

One such quantity is the gravitational mass of the soliton. Equating $A(r)$ to the $-1/g_{rr}$ component of the Schwarzschild metric~($-g_{tt}=1/g_{rr}=1-2GM/r$) and rearranging, one can obtain the quasilocal Misner-Sharp mass function $M(r)$ measuring the enclosed gravitational mass at a given radius $r$: 
\begin{equation}\label{eq:Misner_Sharp}
    M(r)=\dfrac{r}{2G}(1-A(r)).
\end{equation}
The gravitational mass measured from infinity, the Arnowitt-Deser-Misner~(ADM) mass $M$, coincides with the Misner-Sharp mass~\eqref{eq:Misner_Sharp} evaluated at radial infinity: 
\begin{equation}\label{eq:ADM_mass}
M=\lim_{r\rightarrow\infty}\left(\dfrac{r}{2G}(1-A(r))\right).
\end{equation}

Other evaluable quantities are the fermion binding energy, 
\begin{equation}\label{eq:Eb_F}
    E^{\rm F}_\mathrm{b}\equiv 
   m_\mathrm{F} 
    - \omega,
\end{equation}
the gravitational binding energy, 
\begin{equation}\label{eq:Eb_grav}
    E_\mathrm{b}^\mathrm{g}\equiv M-\lvert\kappa\rvert \omega ,
\end{equation}
and the total binding energy:
\begin{equation}\label{eq:Eb_tot}
    E_\mathrm{b}^\mathrm{tot}\equiv E_\mathrm{b}^\mathrm{g} - \lvert\kappa \rvert E^{\rm F}_\mathrm{b}=M-\lvert\kappa\rvert m_\mathrm{F}.
\end{equation}
The root-mean-square (RMS) radius of the soliton $\overline{R}$~\cite{leith2022_phd} is obtained through the following formula:
\begin{equation}\label{eq:R_soliton}
    \overline{R}\equiv\left(4\pi\int^\infty_0{\dfrac{r^2 T}{\sqrt{A}}\left(\alpha^2+\beta^2\right)\mathrm{d}r}\right)^{1/2},
\end{equation}
which we use as a measure of the approximate radial extent of the localized solution. 

One last feature that we consider is the null geodesic structure. Evaluating the radii of circular null geodesics, known as photon spheres, can illustrate how relativistic our soliton is. This can be done by observing whether the fermion fields localize around photon spheres, because in highly relativistic~($z\gg1$) solutions fermion trajectories become approximable by null geodesics~\cite{leith2020}. 

From the null geodesic equation, we arrive at: 
\begin{equation}\label{eq:geodesics}
    \left(\dfrac{\mathrm{d}r}{\mathrm{d}\phi}\right)^2={r^4T^2A}\left(\dfrac{1}{b^2}-U(r)\right),
\end{equation}
where $b$ is the impact parameter of the massless particle, and the effective null potential $U(r)$ is 
\begin{equation}\label{eq:Ueff}
    U(r)=\dfrac{1}{r^2T(r)^2}.
\end{equation}
Photon spheres occur at radii $r=r_\mathrm{ph}$, which we define as extrema of $U(r)$ where 
\begin{equation}\label{eq:dUeff}
    U'= \left. \frac{-2\left( T + r \,T'\right)}{\left(r\,T\right)^3}\right|_{r=r_\mathrm{ph}} = 0,
\end{equation}
which is equivalent to finding where $(r\,T(r))'=0$. 

Each family can be characterized by the parameters $\{\kappa,\,n,\,\xi,\,v\}$. Each solution within such a family is identified by its central gravitational redshift $z$~\eqref{eq:redshift_def}.

\subsection{Zonal structure}\label{subsec:zonal_structure} 

\begin{figure}
    \centering
    \includegraphics[width=0.999\linewidth]{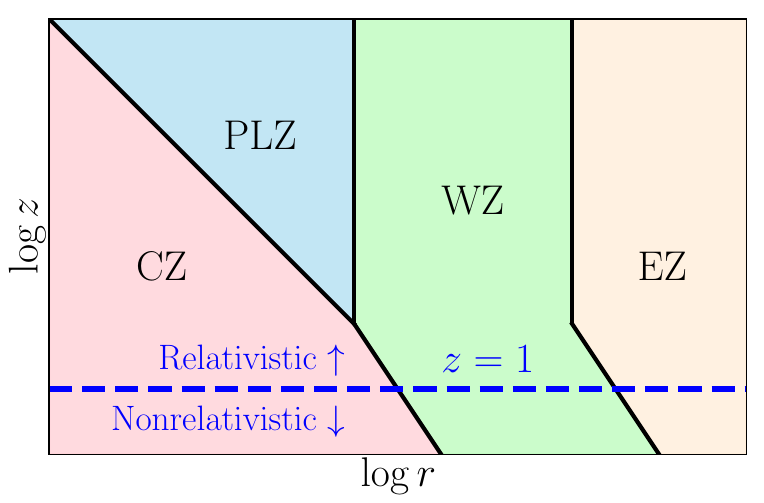}
    \caption{\justifying Schematic diagram of the zonal structure of excited-state Einstein-Dirac(-Higgs) solitons, against logarithmically scaled radius $r$ and redshift $z$~\eqref{eq:redshift_def}. Core zone: CZ~(pink), power-law zone: PLZ~(light blue), wave zone: WZ~(light green), evanescent zone: EZ~(peach). For the ground state~($n=0$), the wave zone is omitted such that the power-law zone is between the core zone and evanescent zone. At $z\rightarrow\infty$, the power-law zone starts at $r=0$. The crossover between the relativistic and nonrelativistic regimes at $z=1$ is marked by the blue dashed line.} 
    \label{fig:zonal_structure}
\end{figure}

A schematic diagram of the radial structures of ED(H) solitons is given in Fig.~\ref{fig:zonal_structure}. 
Like in non-Yukawa-coupled ED systems~\cite{bakuczcanario2020}, there are up to four distinct ``zones'' in the radial structure of nonsingular EDH solitons. 
In order of increasing $r$, these are named~\cite{bakuczcanario2020}:
\\ \indent 1. Core zone,
\\ \indent 2. Power-law zone, absent at low redshift,
\\ \indent 3. Wave zone, absent in the ground state,
\\ \indent 4. Evanescent zone.
\\
We briefly discuss these in the subsections below.

\subsubsection{Core zone}

The soliton is initialized in the core, which includes the origin. 
A small-$r$ Taylor expansion of the fields in the equations of motion gives the following approximations: 
\begin{equation}\label{eq:rel_CZ}
\begin{aligned}
    \alpha &= r^{|\kappa|/2} \, \left( \alpha_1 + \mathcal{O}(r^2) \right), 
     & \alpha_1>0 .
\\  \beta &= r^{|\kappa|/2}
    \left( \beta_2\, r + \mathcal{O}(r^3) \right) , & \beta_2>0.
\\  T &= T_0 + T_2 \, r^2 + ...,
    & T_0 > 1.
\\ A & = A_0 + A_2 \, r^2 + ...,
    & A_0=1.
\\ h &= h_0 + h_2\, r^2 + ... ,
    & h_0 < v.
    & \quad
\end{aligned}
\end{equation}
The detailed relations between the power-law coefficients can be found in Appendix~\ref{app:num}.

\subsubsection{Power-law zone}
Our fermion soliton stars may be referred to as ``nonrelativistic''~($z\lesssim1$) or ``relativistic''~($z\gtrsim1$), with the transition near $z=1$ marked by the dashed blue line on Fig.~\ref{fig:zonal_structure}. The power-law zone emerges above this transition at some redshift $z>1$. In this zone, $\alpha,\,\beta\propto r^1$, $T\propto r^{-1}$, $A(r)$ dips to $A\approx1/3$~\cite{bakuczcanario2020}, and we find that $h(r)\propto r^1$. As $z$ increases, the power-law zone extends further inwards in radius. 

At the infinite-$z$ limit, the power-law zone replaces the core as the innermost region of the soliton.
This limit is defined by the following analytic power-law solution to the EDH equations, found by assuming that $\mu h\ll\omega$~\cite{leith2022_phd}: 
\begin{equation}\label{eq:infz}
    \begin{gathered}
        \alpha(r) = \sqrt{\dfrac{\omega}{12\pi G \kappa^2\kappa_-}}\,r, \qquad \beta(r) = \sqrt{\dfrac{\omega}{12\pi G \kappa^2\kappa_+}}\,r,\\[2.5ex]
        T(r) 
        = \frac{ \sqrt{\kappa_+\,\kappa_-}}{\omega}r^{-1}
    , \qquad A(r) = \dfrac{1}{3},
        \\[2.5ex]
        h(r)=h_0 + \dfrac{\mu}{6\sqrt{3}\pi G}\dfrac{1}{\kappa\sqrt{\kappa_+\kappa_-}}\,r,
    \end{gathered}
\end{equation}
where the auxiliary parameters $\kappa_\pm$ are given by 
\begin{equation}\label{eq:infz_params}
    \kappa_\pm \equiv \dfrac{\kappa}{2} \pm \dfrac{1}{\sqrt{3}}.
\end{equation}
The above, valid for even-parity states, is presented in Ref.~\cite{leith2023} for the two-fermion system, but it is generalized to $\lvert\kappa\rvert$ fermions here following Ref.~\cite{leith2022_phd}.

A numerical solution generated using the power-law solution of~\eqref{eq:infz} as a small-$r$ boundary condition for the EDH system corresponds to a singular soliton with $z\rightarrow+\infty$. The corresponding solution for the FSY system is named the ``infinite redshift'' or the ``lightlike singularity'' solution~\cite{blazquezsalcedo2020,bakuczcanario2020}. In the spiral spectra given in later sections, the pole of the spiral corresponds to this state~\cite{bakuczcanario2020}. It is found that for finite-redshift solutions with a power-law zone, the fields initially miss their analytic power-law profiles in the transition from the core, and the fields exhibit damped oscillations near and outside this matching region. 

\subsubsection{Wave zone}
Moving further outwards in radius $r$, the wave zone occurs exclusively for excited state~$(n > 0)$ solutions, which have one or more nodes in $\alpha$ and $\beta$. The spinor fields in this region exhibit oscillations in their radial structure, resembling trapped spherical standing waves~\cite{bakuczcanario2020}. 

\subsubsection{Evanescent zone}

The transition to the evanescent zone from the core/power-law/wave zone approximately starts at a radius $r_\mathrm{EZ}$ where~\cite{bakuczcanario2020}
\begin{equation}
    \omega\,T(r_\mathrm{EZ})=\mu h(r_\mathrm{EZ}).
\end{equation}
Upon entering the evanescent zone, the spinor fields $\alpha$ and $\beta$ decay exponentially, and the metric fields $T$ and $A$ reduce to the Schwarzschild metric expected outside a localized solution~\cite{bakuczcanario2020}. The Higgs field $h$ asymptotes exponentially to the VEV~\cite{leith2022_phd}. 

The asymptotic fermion mass $\mu v$ defines the decay rate of the fermion fields. This is to be expected; in the evanescent zone, we know that $h\approx v$. One may formulate the asymptotic relations between the fermion and metric fields through the following large-$r$ Taylor expansion~\cite{leith2022_phd}: 
\begin{equation}
    \begin{aligned}
        \alpha(r)&=\alpha_\infty r^\gamma(1 + \mathcal{O}(r^{-1})+...)\exp(- r/d),\\
        \beta(r)&=\beta_\infty r^\gamma(1 + \mathcal{O}(r^{-1})+...)\exp(- r/d),\\
        T(r)&=1+GM r^{-1}+\mathcal{O}(r^{-2})...,\\
        A(r)&=1-2GM r^{-1}+\mathcal{O}(r^{-2})...,\\
        h(r)&=v - r^\sigma (\mathcal{O}(1)+...)\exp^{-\epsilon r},
\end{aligned}
\end{equation}
where the metric is practically Schwarzschild. We find that the decay rates of $h(r)$ and its next-to-highest order expansions are irrelevant for this particular problem. It has been shown that the Higgs field can decay faster than the fermion fields in some regions of the phase space~\cite{leith2022_phd}.

Attempting a similar analysis to that undertaken in Ref.~\cite{bakuczcanario2020}, Ref.~\cite{leith2022_phd} finds that the fermion decay length scale $d$ is
\begin{equation}\label{eq:decayexp}
    d=\dfrac{1}{\sqrt{\mu^2v^2-\omega^2}}.
\end{equation}
The ratio between the leading power-law coefficients in the $\alpha$ and $\beta$ decays is 
\begin{equation}\label{eq:E0D0}
    \dfrac{\beta_\infty}{\alpha_\infty}=\sqrt{\dfrac{\mu v - \omega}{\mu v + \omega}},
\end{equation}
and the fermions' power-law decay scale $\gamma$ is 
\begin{equation}\label{eq:gamma}
    \gamma=GM\dfrac{2\omega^2-\mu^2v^2}{\sqrt{\mu^2v^2-\omega^2}}.
\end{equation}.

\subsection{Higgs field dynamics}\label{subsec:higgs_mech}

\begin{figure}
    \centering 
    \begin{subfigure}{0.235\textwidth}
        \includegraphics[width=\textwidth]{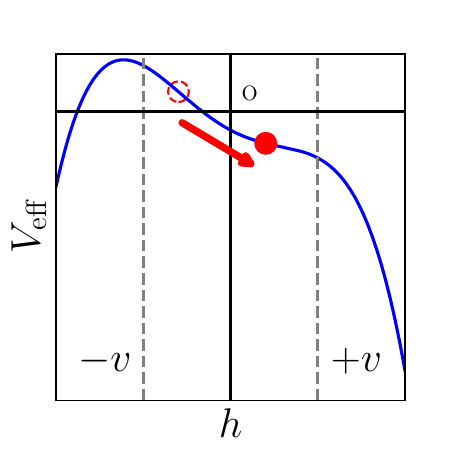}
        \caption[]%
        {Core/power-law zone.}
        \label{fig:higgs_mech_tilt}
    \end{subfigure}
    \hfill
    \begin{subfigure}{0.235\textwidth}  
        \includegraphics[width=\textwidth]{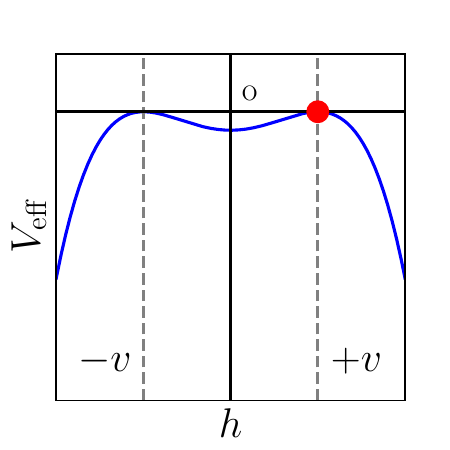}
        \caption[]%
        {Evanescent zone.} 
        \label{fig:higgs_mech_notilt}
    \end{subfigure}
    \caption[]
    {\justifying A schematic diagram illustrating the behavior of the Higgs field and its corresponding effective potential $V_\mathrm{eff}$, with reference to the zonal structure~(Fig.~\ref{fig:zonal_structure}) of an EDH soliton. Red circle: Higgs field $h$, solid blue line: Effective Higgs potential $V_\mathrm{eff}$.} 
    \label{fig:higgs_mech} 
\end{figure} 

We now clarify the expected behavior of the Higgs field and the resulting effective fermion mass~\cite{leith2022_phd,leith2023} in each individual zone of a ground-state soliton. This is illustrated schematically in Fig.~\ref{fig:higgs_mech}.

The Higgs equation~\eqref{eq:Higgs} may be reformulated as
\begin{equation}
    \nabla_\nu\nabla^\nu h=-\partial_h V_\mathrm{eff}(h),
\end{equation}
where $V_\mathrm{eff}$ is an effective Higgs potential 
\begin{equation}\label{eq:Veff_Higgs}
    V_\mathrm{eff}=-V(h)-\dfrac{\lvert\kappa\rvert\mu T}{r^2}(\alpha^2-\beta^2)h.
\end{equation}
The Higgs equation~\eqref{eq:Higgs_EDH} may thus be interpreted as one describing the effective acceleration $h''$ experienced by the Higgs field~$h$ under a potential $V_\mathrm{eff}$~\eqref{eq:Veff_Higgs}, that makes it accelerate towards $+v$. This requires the initial starting value of $h(r)$ to be less than $+v$.

As the Higgs potential~$V(h)$~\eqref{eq:V(h)} is inverted in $V_\mathrm{eff}$~\eqref{eq:Veff_Higgs}, $V_\mathrm{eff}$ has maxima, not minima, at $h=\pm v$. Despite the fact that $h=\pm v$ are stable minima of $V(h)$, they are unstable spatially in $r$~\cite{leith2023}. 

The effective potential $V_\mathrm{eff}$ also contains a term that is linear in $h$, dubbed the ``fermion tilt''~\cite{leith2023}. This linear term tilts the effective potential in a direction that depends on the sign of $\alpha^2 - \beta^2$. For particle-dominated states such as the ones explored in the present work, $\alpha^2>\beta^2$ for most of the radial extent of the soliton, so the fermion tilt term contributes an overall negative slope to $V_\mathrm{eff}$,
This raises the peak near $-v$ and lowers the peak near $+v$, as illustrated in Fig.~\ref{fig:higgs_mech_tilt}. 

Having discussed both contributions to the effective potential $V_\mathrm{eff}$, we now discuss the behavior of the Higgs field with explicit reference to the zonal structure of the soliton~(Sec.~\ref{subsec:zonal_structure} and Fig.~\ref{fig:zonal_structure}). 

Starting in the core, with the overall tilt contribution being negative with $\alpha^2-\beta^2>0$, $h(r)$ is accelerated towards $+v$. As can be inferred from~\eqref{eq:infz}, in the power-law zone $h''=0$ and $h'>0$. Therefore, for higher-redshift solutions in which such a zone is present, the Higgs field $h$ increases towards the VEV $v$ at an effective terminal velocity induced by the fermion tilt. The resulting behavior of the Higgs field ``rolling down'' the tilted effective potential is visualized in Fig.~\ref{fig:higgs_mech_tilt}, where the Higgs field~(solid red circle) departs from its starting point~(hollow red circle) and increases towards the VEV~$v$. 

In the evanescent zone, the fermion fields decay exponentially, so the fermion tilt effectively disappears. As a result, the Higgs field decelerates, and comes to rest exactly at $h(r)\rightarrow +v$ as $r\rightarrow\infty$. This is illustrated in Fig.~\ref{fig:higgs_mech_notilt}, which locates the Higgs field at its required asymptotic value in the evanescent zone. 

The present discussion is a simplified case mainly applicable to the two-fermion ground-state solutions presented in the subsequent section. While the general phenomenon is seen in the excited state~($n>0$: Sec.~\ref{sec:mc_exc}) and many-fermion~($\kappa>2$: Sec.~\ref{sec:many_fermion}) solutions, the Higgs field and its effective potential exhibit behaviors not discussed in the present subsection. This is, for example, due to the presence of the wave zone in excited state solutions, and the absence of fermion tilt near $r=0$ and multi-shell structures~\cite{leith2020} in the many-fermion solutions. 

\section{Ground-state two-fermion configurations\label{sec:mc_dynamics}}

In this section, we generate some examples of ground-state~($n=0$) two-fermion~($\kappa=2$) EDH solitons embedded in asymptotically flat spacetimes. We provide hitherto unpresented explanations of the specific effects of $\{\xi,\,v\}$ on solution families, as well as their influence on deviations away from the FSY ED system. 

As a consequence of two of our parameters being set, each individual soliton in the present section can be parametrized by values of $\{\xi\,,v,\,z\}$, with families of solutions characterized by $\{\xi\,,v\}$.

\subsection{Individual states}\label{subsec:2-ferm_indiv_sol}

\begin{figure*}
    \centering 
    \begin{subfigure}{0.999\textwidth}
        \includegraphics[width=\textwidth]{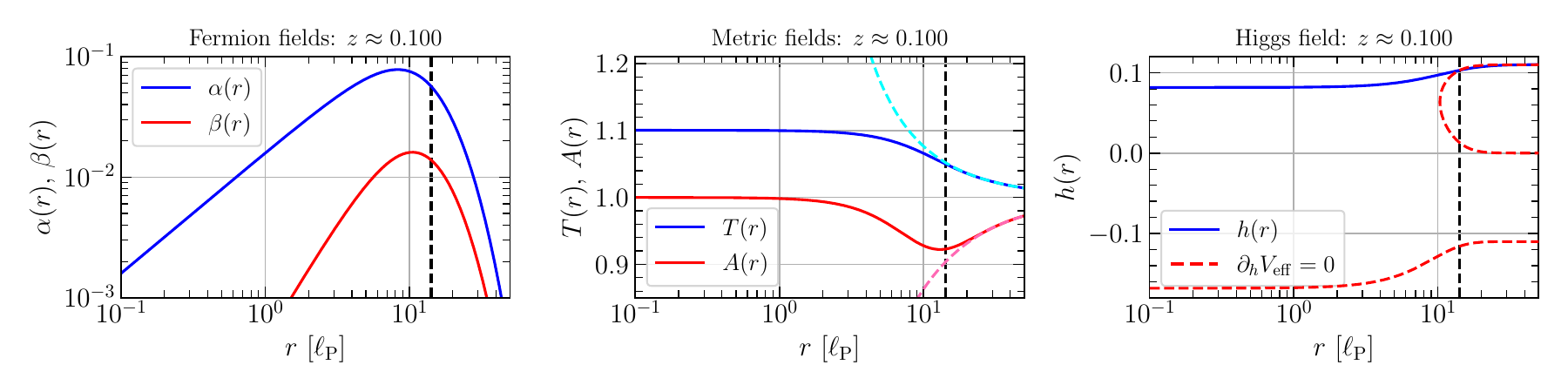}
        \caption[]%
        {A low-redshift/nonrelativistic solution.}
        \label{fig:edh_lowz}
    \end{subfigure}
    \hfill
    \begin{subfigure}{0.999\textwidth} 
        \includegraphics[width=\textwidth]{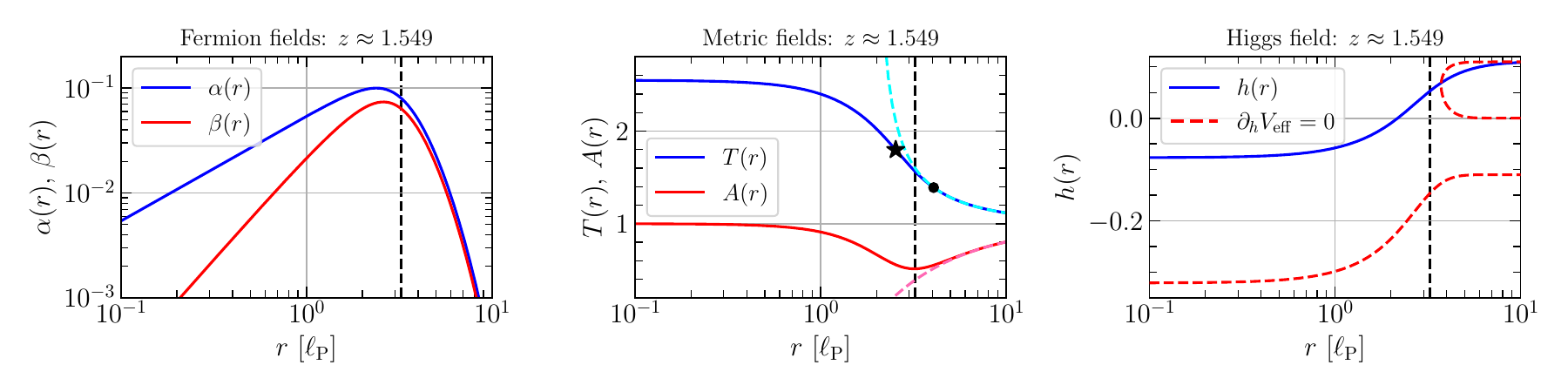}
        \caption[]%
        {Maximally bound state.} 
        \label{fig:edh_maxb}
    \end{subfigure}
    \hfill
    \begin{subfigure}{0.999\textwidth}  
        \includegraphics[width=\textwidth]{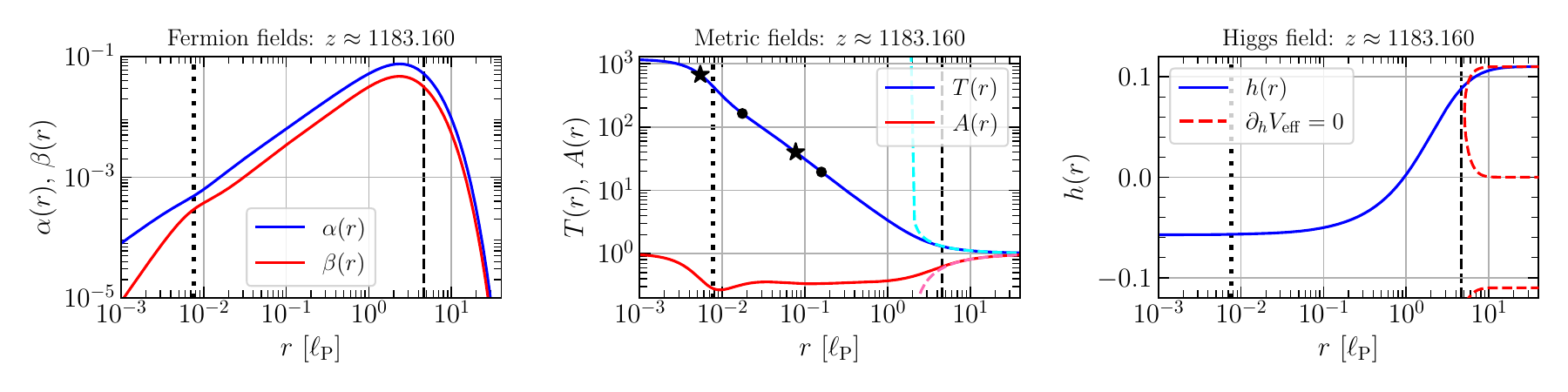}
        \caption[]%
        {A high-redshift/relativistic solution.} 
        \label{fig:edh_highz}
    \end{subfigure}
    \caption[]
    {\justifying Three individual solutions for EDH solitons from the $\{\xi,v\}=\{0.36,0.11\}$ family of solutions. 
    Left panel: fermion fields $\alpha(r)$ (blue) and $\beta(r)$ (red) as functions of the radial coordinate $r$, middle panel: metric fields $T(r)$ (blue) and $A(r)$ (red) over $r$, right panel: Higgs field $h(r)$ over $r$. 
    The vertical dotted and dashed black lines separate the core from the power-law zone and the power-law zone from the evanescent zone, respectively. 
    Dashed cyan/pink curves denote the Schwarzschild metric fields to which $T(r)$ and $A(r)$ match onto in the evanescent zone. 
    Dashed red lines on the Higgs field plots locate where $\partial_hV_\mathrm{eff}=0$, where $V_\mathrm{eff}$ is the effective Higgs potential as given in~\eqref{eq:Veff_Higgs}. 
    Black stars and circles on $T(r)$ respectively correspond to the locations of stable and unstable photon spheres~\eqref{eq:dUeff}. 
    \label{fig:indiv_kappa=2} 
    }
\end{figure*} 

We present some examples of particlelike solutions of the EDH system of equations. Just as in Ref.~\cite{leith2023}, we consider $\kappa=2$ and $n=0$. The characteristic constants for each individual solution are in Appendix~\ref{app:indivsol}. 

Three examples of solutions selected from the $\{\xi,v\}=\{0.36,0.11\}$~family are shown in Fig.~\ref{fig:indiv_kappa=2}: Fig.~\ref{fig:edh_lowz} corresponds to a low-redshift solution~($z\approx 0.100$). We further evaluate the maximally bound state, which corresponds to the solution with the most negative value of $E_\mathrm{b}^\mathrm{tot}$~\eqref{eq:Eb_tot}. 
This state is shown in Fig.~\ref{fig:edh_maxb}, and has a redshift of $z\approx1.549$. A high redshift solution~($z\approx1183$) is shown in Fig.~\ref{fig:edh_highz}. As expected from Ref.~\cite{leith2023}, the presented soliton solutions resemble those of FSY~\cite{fsy1999}. 

The low-redshift solution (Fig.~\ref{fig:edh_lowz}) is radially expansive, with the RMS soliton radius~\eqref{eq:R_soliton} taking a value of $\overline{R}\approx 11.569$~[$\ell_\mathrm{P}$]. As can be seen on the leftmost subpanel showing the fermion fields on a log-log plot, the core directly matches onto the evanescent zone at the dashed black line located at a radius where $\omega T(r)=\mu h(r)$, and the maxima of the fermion fields are seen to be quite disparate in their magnitude. In the rightmost subpanel, the Higgs field is initialized at $h(r=0)=0.082<v$, rolls upwards towards larger $h$, and comes to rest at $v=0.11$; thus, $h$ does not start very far from the VEV. Only one root of $\partial_h V_\mathrm{eff}=0$~(red dashed line) is obtainable at smaller $r$ up to $r\approx10~[\ell_\mathrm{P}]$. When the evanescent zone starts, the fermion tilt subsides as the $\alpha$ and $\beta$ fields decay exponentially. As can be inferred from~\eqref{eq:V(h)} and~\eqref{eq:Veff_Higgs}, $V_{\rm eff}$ is a quartic in $h$ and hence $\partial_h V_{\mathrm{eff}}$ is a cubic in $h$, with the tilt corresponding to a vertical offset in $\partial_h V_\mathrm{eff}$. 
The reduction of the magnitude of the fermion tilt means that the cubic function loses this vertical offset. This corresponds to the cubic $\partial_h V_\mathrm{eff}$, which previously only possessed one root when $V_\mathrm{eff}$ was tilted, gaining two additional roots. After a sufficient exponential decay in the fermion fields, the extrema of $V_\mathrm{eff}$ are located at $h=0$ and $\pm v$ as expected from~\eqref{eq:V(h)}. This corresponds to an effectively untilted $V_\mathrm{eff}$.

The $z\approx1.549$ solution~(Fig.~\ref{fig:edh_maxb}) is the maximally bound member of the family, or the solution with the most negative value of $E_\mathrm{b}^\mathrm{tot}$. Much like the previous solution, there is no power-law zone where both $\alpha$ and $\beta$ are proportional to $r^1$, as can be seen from the different slopes of the fermion fields. Hence, the core once again transitions directly into the evanescent zone, as can be inferred from Fig.~\ref{fig:zonal_structure}. 
However, at this slightly higher redshift we find $h(r=0)<0$, such that the fermions have negative mass for a range of radii in the core. We also notice the emergence of photon spheres, despite the 
soliton being only mildly relativistic. One of these photon spheres is located inside the core-to-evanescent-zone transition, while one is located outside. 

In contrast to the two lower-redshift solutions, the high-redshift solution~(Fig.~\ref{fig:edh_highz}) exhibits some noticeable deviations from the low-redshift case. For one, the soliton itself is more compact ($\overline{R}\approx3.080$~[$\ell_\mathrm{P}$]) than the previous two cases. As a result of the large redshift, the power-law zone is evident on the log-log fermion and metric field plots. On the metric field plot, $A(r)$ drops to around $1/3$ and exhibits damped oscillations around this value, while $T(r)$ oscillates similarly around a $1/r$ power law. Neither behavior is seen in either low-redshift solution. There are now two pairs of photon spheres in the metric fields. For high-redshift solutions, the core zone transitions into the power-law zone at an approximate radius~(dotted black line) on the order of:
\begin{equation}
    r\approx\dfrac{\sqrt{\kappa_+ \kappa_-}}{\omega (1+z)},
\end{equation}
which is found through matching the core zone relations~\eqref{eq:rel_CZ} and the power-law zone relations~\eqref{eq:infz} at a finite and nonzero radius~\cite{bakuczcanario2020}. The power-law zone continues until $\omega T(r)=\mu h(r)$, at which point $T$ and $A$ match onto their exterior Schwarzschild profiles upon entering the evanescent zone. 

Regarding the radial profile of $h(r)$ and the extrema of $V_\mathrm{eff}$~(Fig.~\ref{fig:edh_highz}, rightmost panel), this is similar to that of the low-redshift solution, with $h(r)<0$ for much of the core. This, aside from meaning that the fermions have locally negative masses in the core of the soliton, shows that the fermion tilt is much steeper than in the low-redshift case. This may be explained through referring to the effective Higgs potential~\eqref{eq:Veff_Higgs}. While $\alpha^2-\beta^2$ indeed determines the direction of the fermion tilt, the fermion tilt is large owing to the large value of $T(r)/r^2$ at small $r$ for high-redshift solutions. There is therefore a correlation between the central gravitational redshift $z$ and the difference between $h(r=0)$ and the VEV~$v$.

\subsection{Families of states}\label{subsec:2-ferm_fam_sol}

\begin{figure*}
    \centering 
    \begin{subfigure}{0.999\textwidth}
        \includegraphics[width=\textwidth]{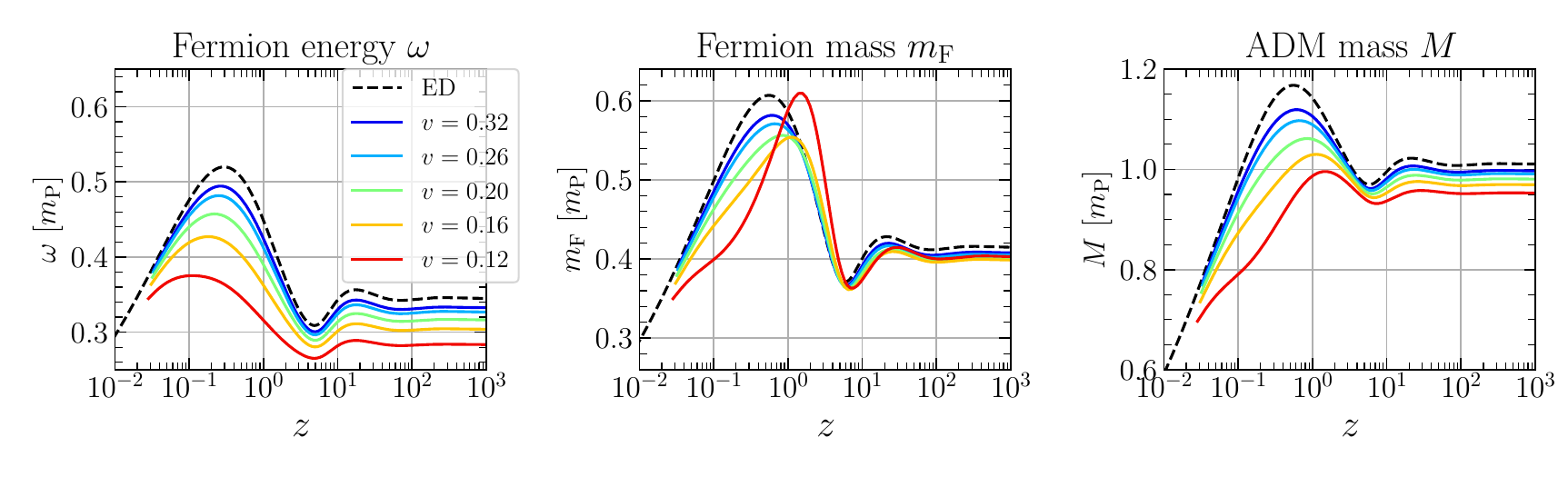}
        \caption[]%
        {Fermion energy ($\omega$), asymptotic fermion mass ($m_\mathrm{F}$), and ADM mass ($M$) over a range of redshifts $z$.}
        \label{fig:log_plots_mc_v}
    \end{subfigure}
    \hfill
    \begin{subfigure}{0.999\textwidth}  
        \includegraphics[width=\textwidth]{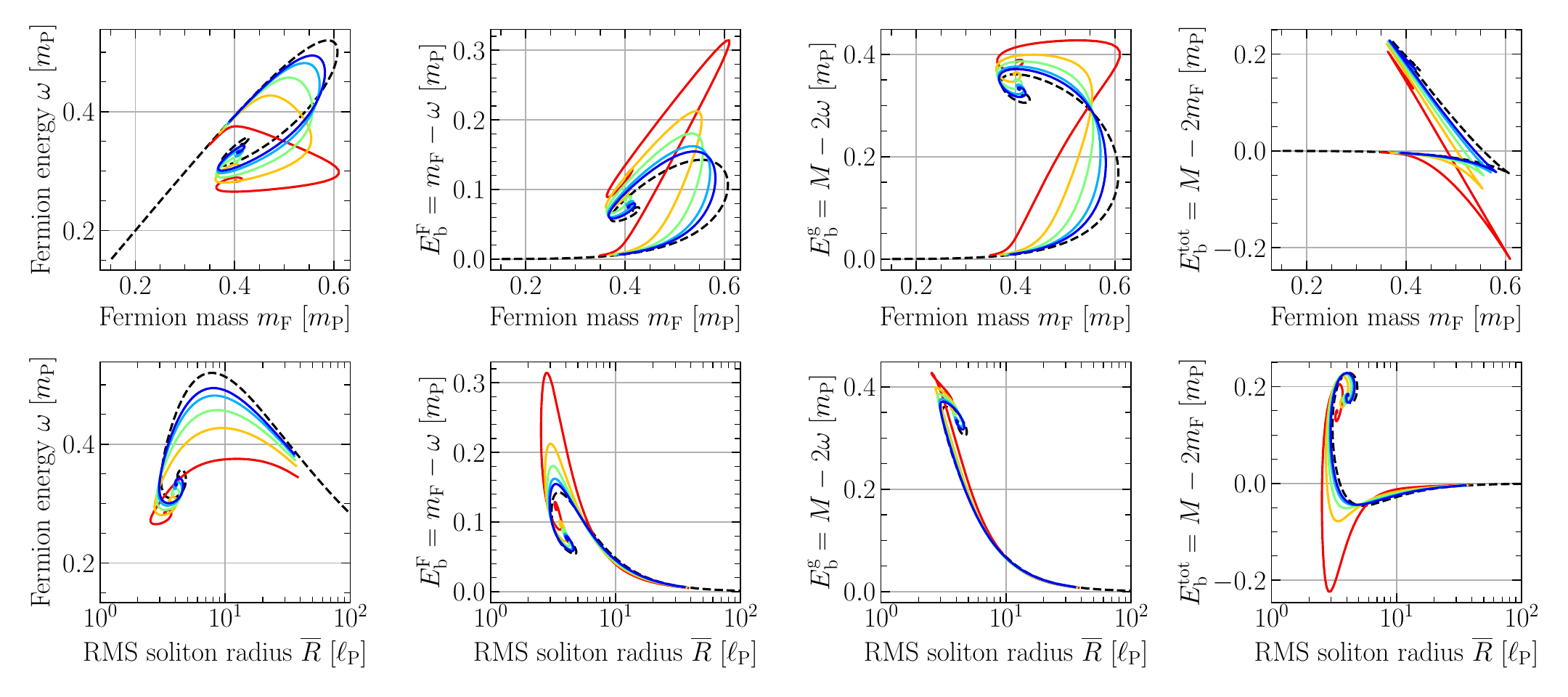}
        \caption[]%
        {Spiral spectra of characteristic energies against asymptotic fermion mass and RMS soliton radius.} 
        \label{fig:mc_spirals_v}
    \end{subfigure}
    \caption[]
    {\justifying Families of the particlelike solutions of the EDH equations for various $v$ values, with $\xi=0.6$ and $\kappa=2$. The horizontal axes are asymptotic fermion mass $m_\mathrm{F}$ and RMS soliton radius $\overline{R}$~\eqref{eq:R_soliton}, whereas the columns are the fermion (kinetic) energy $\omega$, fermion binding energy $E_\mathrm{b}^\mathrm{F}$~\eqref{eq:Eb_F}, gravitational binding energy~\eqref{eq:Eb_grav}, and total binding energy $E_\mathrm{b}^\mathrm{tot}$~\eqref{eq:Eb_tot}, from left to right. The dashed black line corresponds to the non-Yukawa-coupled ED case with $\kappa=2$.} 
    \label{fig:families_v} 
\end{figure*}

\begin{figure*}
    \centering 
    \begin{subfigure}{0.999\textwidth}
        \includegraphics[width=\textwidth]{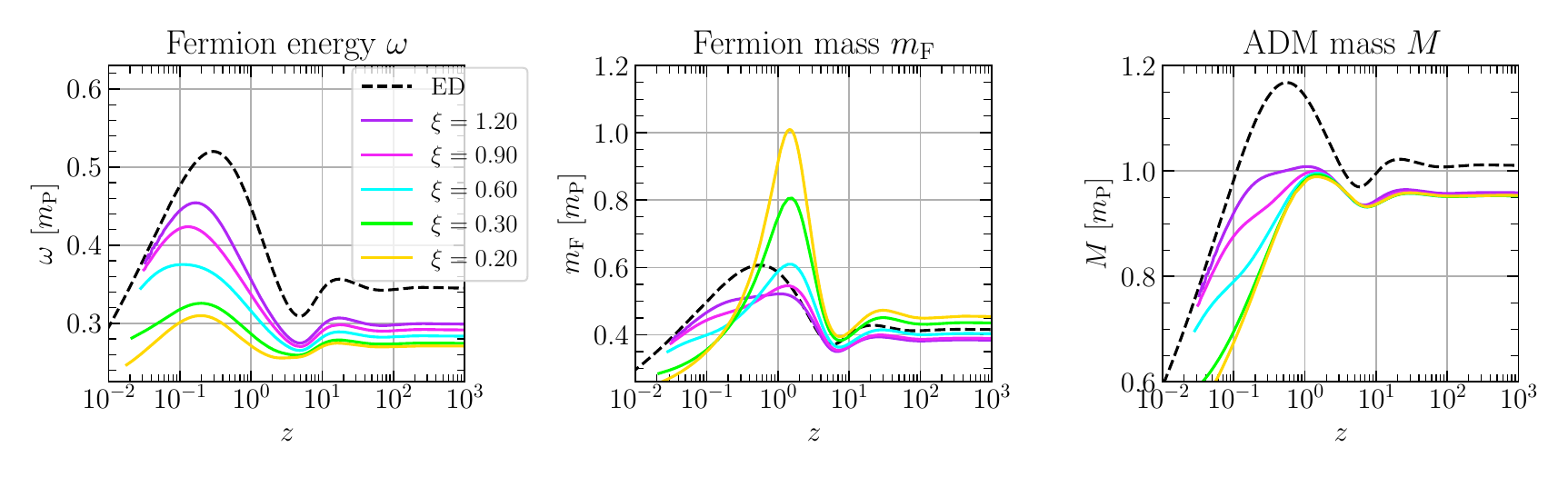}
        \caption[]%
        {Fermion energy ($\omega$), asymptotic fermion mass ($m_\mathrm{F}$), and ADM mass ($M$) over a range of redshifts $z$.}
        \label{fig:log_plots_mc_xi}
    \end{subfigure}
    \hfill
    \begin{subfigure}{0.999\textwidth}  
        \includegraphics[width=\textwidth]{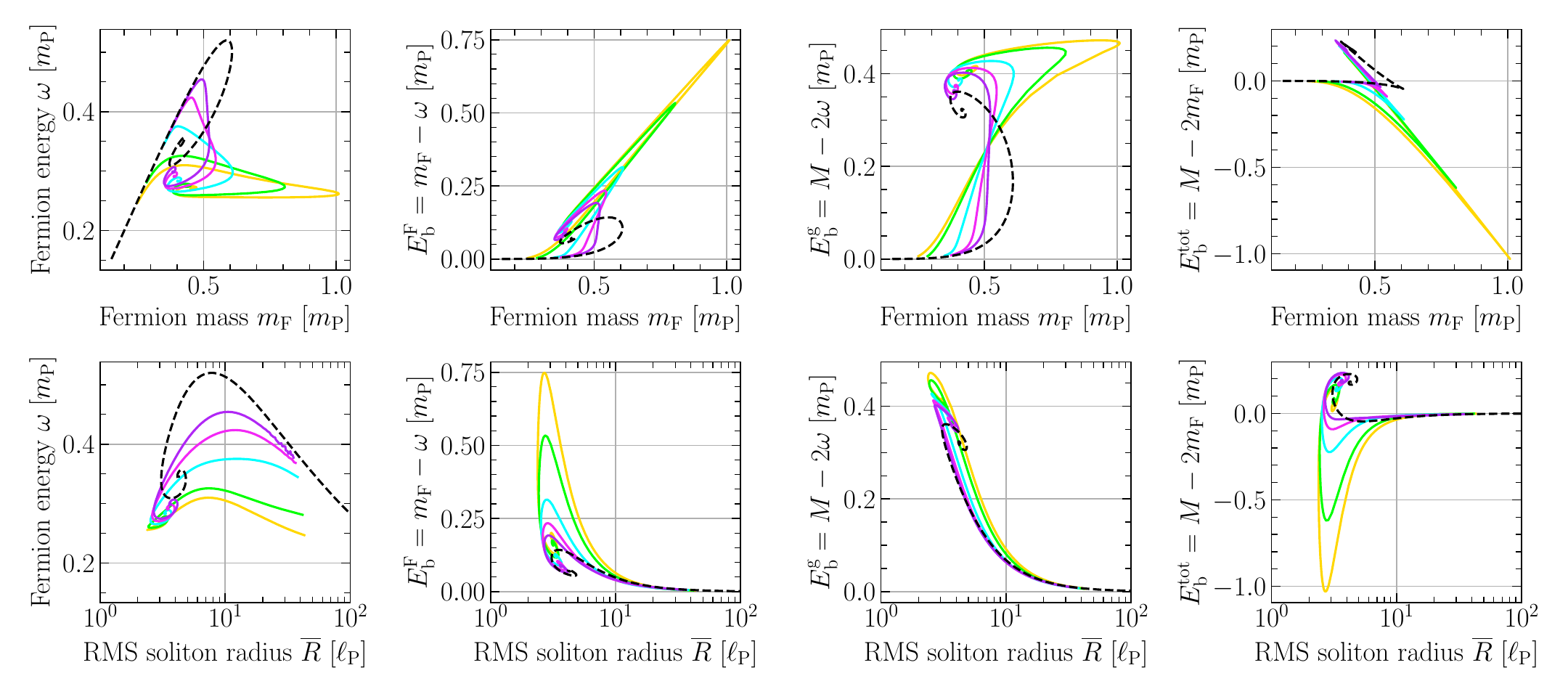}
        \caption[]%
        {Mass- and radius-spectra of EDH families.} 
        \label{fig:mc_spirals_xi}
    \end{subfigure}
    \caption[]
    {\justifying Families of the particlelike solutions of the EDH equations for various $\xi$ values, with $v=0.12$ and $\kappa=2$. The horizontal axes are asymptotic fermion mass $m_\mathrm{F}$ and RMS soliton radius $\overline{R}$~\eqref{eq:R_soliton}. The vertical axes are the fermion (kinetic) energy $\omega$, fermion binding energy $E_\mathrm{b}^\mathrm{F}$~\eqref{eq:Eb_F}, gravitational binding energy~\eqref{eq:Eb_grav}, and total binding energy $E_\mathrm{b}^\mathrm{tot}$~\eqref{eq:Eb_tot} from left to right. The dashed black line corresponds to the FSY ED case with $\kappa=2$.} 
    \label{fig:families_xi} 
\end{figure*}

We present examples of different families of states in the EDH system. While several spectra of states are presented in Ref.~\cite{leith2023}, it is for the first time here that we explain in detail each deviation that our solution families exhibit from the standard FSY ED system. 

In Fig.~\ref{fig:families_v}, we vary the VEV $v$ for a constant $\xi$, and vice versa for Fig.~\ref{fig:families_xi}. Figs.~\ref{fig:log_plots_mc_v} and~\ref{fig:log_plots_mc_xi} show the variations in fermion energy $\omega$, asymptotic fermion mass $m_\mathrm{F}$, and ADM mass $M$~\eqref{eq:ADM_mass} of the two-fermion soliton systems over decades in redshift $z$, while Figs.~\ref{fig:mc_spirals_v} and~\ref{fig:mc_spirals_xi} show the spectra of solitons for different asymptotic fermion masses and RMS soliton radii. For each solution, the relevant characteristic constants are extracted and plotted as shown. 

For each state, the characteristic constants $\omega$, $m_\mathrm{F}$, $M$, $\overline{R}$ etc. exhibit damped oscillations in redshift~(Fig.~\ref{fig:log_plots_mc_v}, Fig.~\ref{fig:log_plots_mc_xi}) that are slightly out of phase with one another. These oscillations in turn create the spiral structures presented in Fig.~\ref{fig:mc_spirals_v} and Fig.~\ref{fig:mc_spirals_xi}. Increasing redshift to high values corresponds to the family of solutions converging at the pole of these spirals, where the infinite-redshift solution~\eqref{eq:infz} is located.

While the ED system has only one family of states for the $\kappa=2$ ground-state system~(black dashed line in Fig.~\ref{fig:families_v} and Fig.~\ref{fig:families_xi}), this is not true for the EDH cases. In these Yukawa-coupled cases, there is a two-dimensional manifold of families parametrized by $\{\xi,v\}$. 

Overall, although there are significant departures in these EDH families from their FSY ED cousin, the RMS soliton radii~$\overline{R}$ are mostly the same from one family to another~(Fig.~\ref{fig:mc_spirals_v}, Fig.~\ref{fig:mc_spirals_xi}). The deviations from the ED family are then observed in the fermion energy $\omega$, fermion mass $m_\mathrm{F}$, ADM mass $M$~\eqref{eq:ADM_mass}, and as a direct consequence, the fermion binding energy $E_\mathrm{b}^\mathrm{F}$~\eqref{eq:Eb_F}, the gravitational binding energy $E_\mathrm{b}^\mathrm{g}$~\eqref{eq:Eb_grav}, and the total binding energy $E_\mathrm{b}^\mathrm{tot}$~\eqref{eq:Eb_tot}. In all cases, these families start with $\mu v\approx\omega$ and $M\approx2\mu v$ in the nonrelativistic regime. In the relativistic regime, the solution families depart from these low-redshift relations and exhibit spiraling behavior, before settling into the infinite-redshift pole of the spiral. 

Varying $v$ while keeping $\xi$ constant~(Fig.~\ref{fig:families_v}), we see that families of solutions with larger $v$ align more closely with the ED family~(dashed black line) compared to those with small $v$. This is because smaller $v$ values lead to the quartic $V(h)$ becoming increasingly more sensitive to the fermion tilt, while large $v$ values are associated with more confining potentials. 

If $v$ is large, the restoring force at the minimum of $V(h)$ is strong even for small displacements away from $v$. Also, by referring to~\eqref{eq:V(h)}, it is clear that the unstable maximum of $V_\mathrm{eff}$ at $h=v$ is more unstable for larger $v$ values, requiring that the fermion tilt must be tuned very finely and must never be too steep. Subsequently, the Higgs field does not vary too much from the VEV $v$ inside the soliton. In such cases, in the limit of an infinitely-confined potential~($\lambda\rightarrow\infty$) one recovers the non-Yukawa-coupled ED family for which the fermion mass does not change as a function of $r$. It is then natural that solution families with larger $v$ approach the ED family, while those with small $v$ deviate. 

On the other hand, if we keep $v$ constant and instead change $\xi$~(Fig.~\ref{fig:families_xi}), we see a phenomenon where cases of strong Higgs-to-fermion coupling (small $\xi$ and large $\mu$) depart significantly from the ED family at moderate redshifts. Larger values of $\xi$ correspond to families of solutions that align more closely to the non-Yukawa-coupled ED cases. This again is explainable through referring to the Higgs potential $V(h)$~\eqref{eq:V(h)}. With $\xi\propto\sqrt{\lambda}/\mu$~\eqref{eq:xi}, larger $\xi$ correspond to larger $\lambda$ and smaller $\mu$. As $\lambda$ increases, the extrema of $V(h)$ are increasingly more confined like when $v$ is large. On the other hand, a decrease in $\mu$ indicates a weakening in the Yukawa-coupling between the fermions and Higgs field. At $\xi\rightarrow\infty$, we can straightforwardly intuit that $\lambda\rightarrow\infty$ and/or $\mu\rightarrow0$. In this limit, the Higgs field is necessarily confined to $h(r)=v$ for all $r$ and there is no Yukawa-coupling between the Dirac and Higgs fields, which just corresponds to the standard FSY ED system. The opposite limit where $\xi=0$ corresponds simply to the Higgs field $h$ being a free field in the absence of a self-interacting potential. 

From Fig.~\ref{fig:mc_spirals_v} and Fig.~\ref{fig:mc_spirals_xi}, it is seen that small values of $v$ and $\xi$ exhibit the largest deviations from the FSY ED family. Specifically, note in Fig.~\ref{fig:mc_spirals_xi} the increasingly prominent deviations from the FSY family in the values of $E_\mathrm{b}^{\rm tot}=M-2m_\mathrm{F}$~\eqref{eq:Eb_tot} and $E_\mathrm{b}^{\rm F}=m_\mathrm{F}-\omega$~\eqref{eq:Eb_F} as $\xi$ is decreased; one may interpret that departures from the standard ED family are more clearly seen with changes in $\xi$ rather than in $v$. The $E_\mathrm{b}^\mathrm{tot}$ plots in particular indicate increasingly negative total binding energies for decreasing $\xi$, meaning that solitons develop a larger degree of stability. This is linked directly to the peaking behavior in $m_\mathrm{F}$ at $z\approx1$ (middle panel of Fig.~\ref{fig:log_plots_mc_xi}), while no similar phenomenon is seen in the variation of ADM mass over redshift. It is furthermore interesting that despite the values taken by the fermion mass $m_\mathrm{F}$ departing drastically from their ED counterpart, a proportional response is absent for the fermion energy $\omega$~\cite{leith2023}, which is observed in the leftmost panel of Fig.~\ref{fig:log_plots_mc_xi} and the $E_\mathrm{b}^\mathrm{g}$~\eqref{eq:Eb_grav} column of Fig.~\ref{fig:mc_spirals_xi}. In fact, it seems that $\omega$ and $M$ both decrease as $\xi$ is decreased~\cite{leith2023}.

This peculiarity, known as the ``mass-scale separation'', is symptomatic of minimally coupled EDH solitons in the (two-fermion) ground-state configuration~\cite{leith2023}. This separation indicates the gravitational~(ADM) mass of a soliton being quantitatively much smaller than the sum of masses of the constituent fermions. This puzzling phenomenon may be seen most clearly in Fig.~\ref{fig:log_plots_mc_xi} for the $\xi=0.20$ case, where the maximum ADM mass obtained is $M<1.00$, while the mass for each constituent fermion is around $m_\mathrm{F}\approx1$ for the corresponding solution. Thus, the total of the constituent fermion masses is approximately double the gravitational mass. The trend suggests that reducing $\xi$ below 0.2 will produce an even more extreme mass-scale separation.

ADM-to-fermion mass disparities are expected for solitonic solutions, bound or unbound~\cite{herdeiro2022}. However, as we clarify in Sec.~\ref{sec:MSS1} and Sec.~\ref{sec:MSS2}, solutions of the ED family, even if bound, all have $|\kappa| m_\mathrm{F}\approx M$. With some small-$\xi$ EDH families, we find that these clearly deviate from and branch off of the non-Yukawa-coupled FSY family, such that the $\lvert\kappa\rvert m_\mathrm{F}\approx M$ relation no longer holds. It is this specific deviation that we classify as the mass-scale separation.

\section{Excited-state two-fermion configurations}\label{sec:mc_exc}

Moving on from the two-particle ($\kappa=2$) ground-state~($n=0$) system, we present a first exploration and analysis of even parity excited state ($n\in2\mathbb{N}$) particlelike solutions to the EDH equations, while retaining the particle number at $\kappa=2$ for now. These excited states have $n/2$ nodes in both $\alpha$ and $\beta$.

\subsection{Individual states}

\begin{figure*}
    \centering 
    \begin{subfigure}{0.999\textwidth}
        \includegraphics[width=\textwidth]{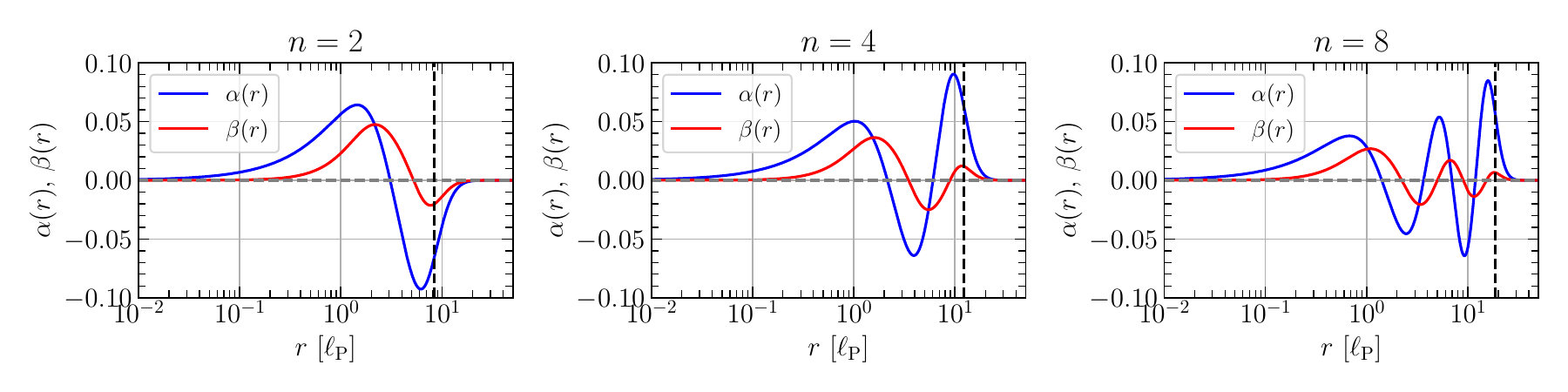}
        \caption[]%
        {Fermion fields.}
        \label{fig:edh_exc_fermion}
    \end{subfigure}
    \hfill
    \begin{subfigure}{0.999\textwidth}  
        \includegraphics[width=\textwidth]{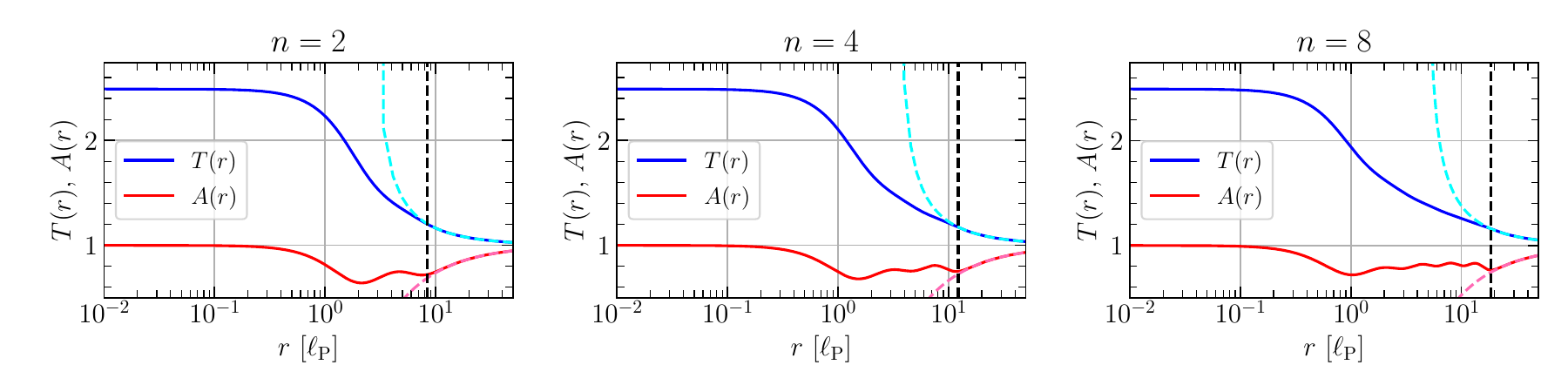}
        \caption[]%
        {Metric fields.} 
        \label{fig:edh_exc_metric}
    \end{subfigure}
    \hfill
    \begin{subfigure}{0.999\textwidth}  
        \includegraphics[width=\textwidth]{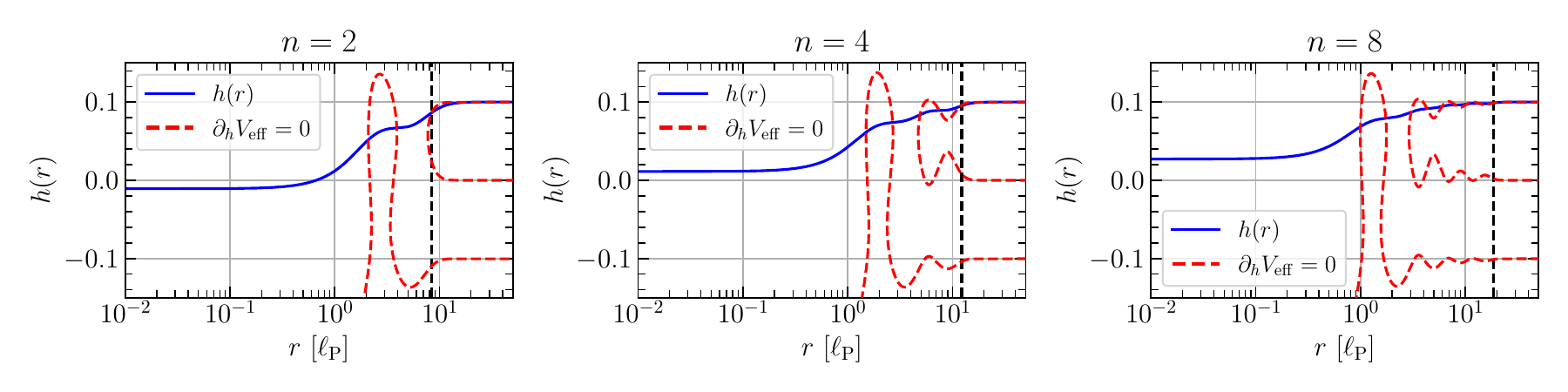}
        \caption[]%
        {Higgs fields.} 
        \label{fig:edh_exc_higgs}
    \end{subfigure}
    \caption[]
    {\justifying Three individual excited-state EDH solitons for $\{\xi,v\}=\{0.42,0.10\}$, all with redshifts $z\approx1.5$ near the maximally bound state. Just as in Fig.~\ref{fig:indiv_kappa=2}, for the EDH solitons: $\alpha$ and $\beta$ are denoted by blue and red on Fig.~\ref{fig:edh_exc_fermion}, $T$ and $A$ are blue and red on Fig.~\ref{fig:edh_exc_metric}, and $h$ and $\partial_h V_\mathrm{eff}=0$ contours are blue and dashed red on Fig.~\ref{fig:edh_exc_higgs}. The vertical dashed black line indicates the inner edge of the evanescent zone.}
    \label{fig:indiv_kappa=2_n>2} 
\end{figure*} 

In Fig.~\ref{fig:indiv_kappa=2_n>2} we present three examples of excited states, all at redshift $z \approx 1.5$. From left to right we have $n=2$, $4$, and $8$ for $\{\xi=0.42,\,v=0.10\}$. Each subfigure/row illustrates the radial profiles of the fermion~(Fig.~\ref{fig:edh_exc_fermion}), metric~(Fig.~\ref{fig:edh_exc_metric}), and Higgs~(Fig.~\ref{fig:edh_exc_higgs}) fields.

In essence, these excited-state solutions exhibit a high degree of similarity to the standard ED cases~\cite{leith2021}. However, the radial profiles of the Higgs field and the equilibria of its potential are both highly distinct from those of the ground-state solutions. Each time that the fermion field $\alpha$ changes sign, there is a pocket-like formation in the $\partial_h V_\mathrm{eff}=0$ contour~(red dashed line, Fig.~\ref{fig:edh_exc_higgs}).

The reasoning behind this phenomenon is straightforward when considering the nodal structures of the fermion fields. 
The nodes in $\alpha$ and $\beta$ alternate in the wave zone of excited states. Each $\alpha$ node naturally occurs in a region where $\beta^2>\alpha^2$, corresponding to a positive fermion tilt in the effective potential~\eqref{eq:Veff_Higgs}. As can be inferred from the earlier discussion in Sec.~\ref{subsec:higgs_mech}, a positive fermion tilt would decelerate the Higgs field's effective velocity in the $+v$ direction, such that $h''<0$.
However, as $\alpha$ and $\beta$ nodes occur in pairs, the deceleration experienced by $h$ around each $\alpha$ node is followed by a region of negative fermion tilt ($\alpha^2>\beta^2$). This produces $h''>0$ near and at the subsequent $\beta$ node, and $h$ is subsequently accelerated towards $+v$. The effective Higgs potential $V_\mathrm{eff}$ tilting back and forth between positive and negative gradients in the wave zone produces a staircase-like radial profile in $h(r)$. We refer to this as the ``Higgs staircase''. 

This back and forth tilting of $V_\mathrm{eff}$ also explains the loop-like structures in the contours of $\partial_h V_\mathrm{eff}=0$ (red dashed curves in Fig.~\ref{fig:edh_exc_higgs}).
Note that $h''$ reverses sign when $h(r)$ crosses a contour. These extrema of $V_\mathrm{eff}$ appear in pairs, split away from one another, and then converge and annihilate. The formation mechanism of the extrema is in principle the same as in the ground-state solutions shown in Fig.~\ref{fig:indiv_kappa=2}, except that in the $\kappa=2$ ground-state solutions, no extrema-annihilations occur. 

Notably, these loop-like structures do not occur for all of the nodes in $\alpha$ and $\beta$. The biggest loop in $\partial_h V_{\rm eff}=0$ occurs
on the first step of the Higgs staircase at the first $\alpha$ node. This is linked to how large $\beta$ is when $\alpha$ hits its first node, and it seems that the amplitudes of the peaks in $\beta$ decrease with increasing $r$, while the opposite applies to $\alpha$. Therefore, the loops denoting $\partial_h V_{\rm eff}=0$ are most prominent at the innermost node in $\alpha$, and gradually weaken after every subsequent node in the fermion fields. This is most clearly seen in Fig.~\ref{fig:edh_exc_higgs}. In the $n=2$ solution, the lower branch contour of $\partial_h V_\mathrm{eff}=0$~(dashed red line) starts well below $-v$, due to the strong fermion tilt in the soliton core. It then rises above $+v$ as the fermion tilt weakens and reverses sign, and then dips below $-v$ before rising to finish at $-v$. In contrast, for $n=4$ and $n=8$ the contours exhibit additional ripples beyond the first dip, owing to the extra nodes in the fermion fields that emerge as the effective potential tilts back and forth. 

Lastly, regardless of the aforementioned Higgs staircase, for the ground and excited states that we have generated we find that the Higgs field approaches $v$ from $h<v$. We find no states where the opposite occurs. 

\subsection{Families of states}

\begin{figure*}
    \centering
    \begin{subfigure}[b]{0.99\textwidth} 
        \includegraphics[width=\textwidth]{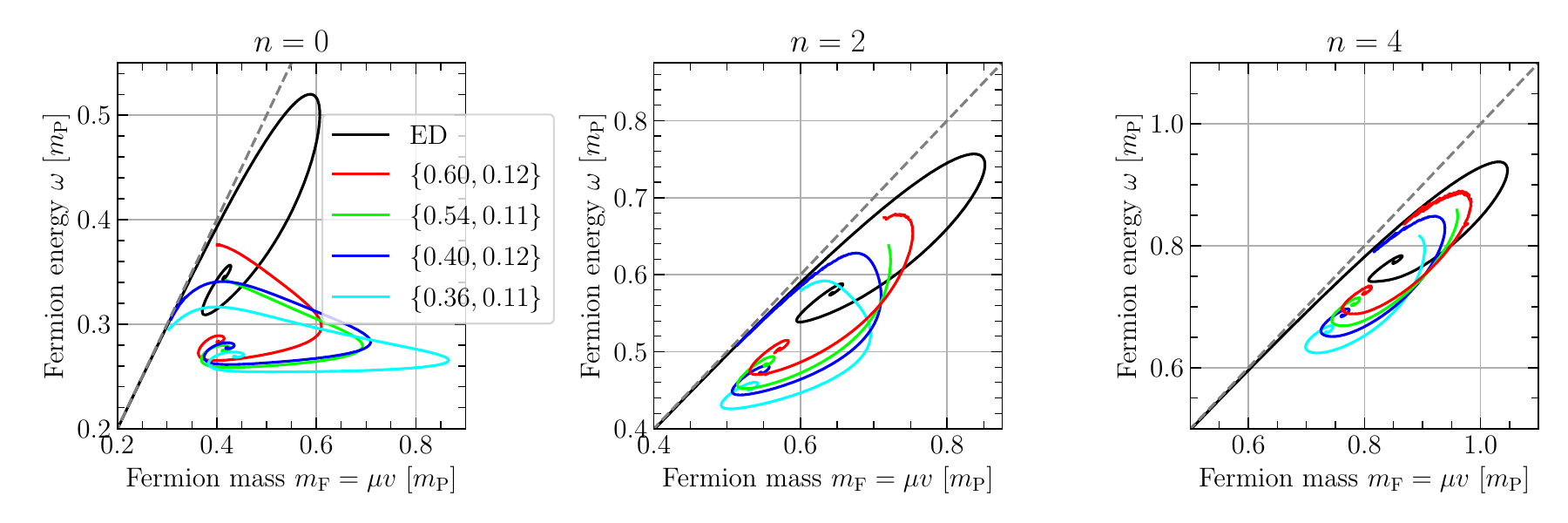} 
        \caption[]%
        {Fermion energy $\omega$.}
        \label{fig:mvw_n>0}
    \end{subfigure}
    \hfill
    \begin{subfigure}[b]{0.99\textwidth} 
        \includegraphics[width=\textwidth]{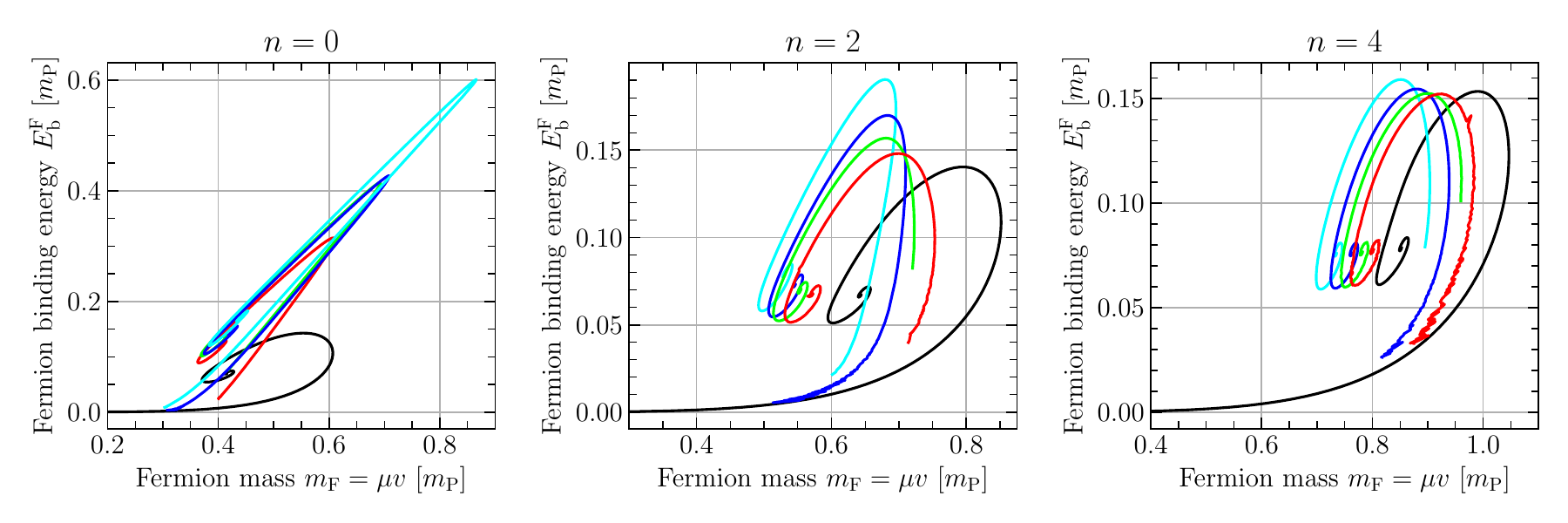} 
        \caption[]%
        {Fermion binding energy $E_{\rm b}^{\rm F}$.}
        \label{fig:mvEf_n>0}
    \end{subfigure}
    \hfill
    \begin{subfigure}[b]{0.99\textwidth}
        \includegraphics[width=\textwidth]{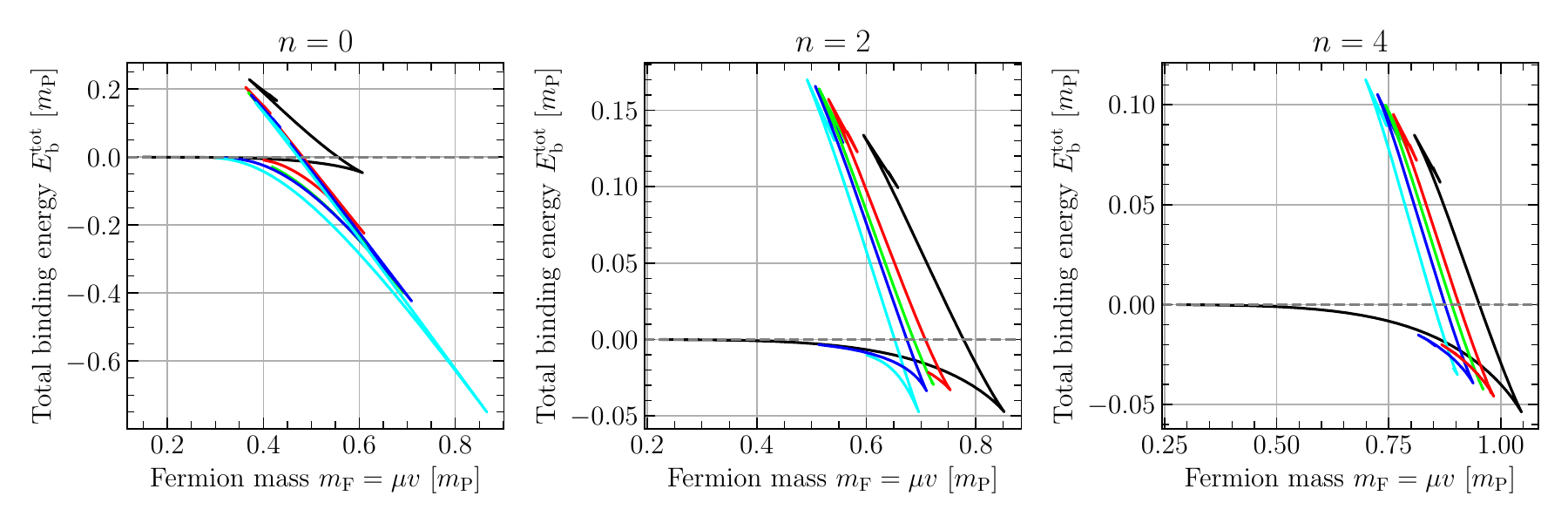} 
        \caption[]%
        {Total binding energy $E_\mathrm{b}^\mathrm{tot}$.}
        \label{fig:mvEb_n>0}
    \end{subfigure}
    \caption[] 
    {\justifying EDH soliton spectra for fermion energy $\omega$, fermion binding energy $E_\mathrm{b}^\mathrm{F}$~\eqref{eq:Eb_F}, and total binding energy $E_\mathrm{b}^\mathrm{tot}$~\eqref{eq:Eb_tot} versus asymptotic fermion mass $m_\mathrm{F}$, for ground state ($n=0$), second excited state ($n=2$), and fourth excited state ($n=4$) solitons. Each family is denoted by the $\{\xi,\,v\}$ label in the legend, and the FSY ED families are plotted for reference (black). The low-redshift solutions for the excited states are omitted because of high levels of numerical noise, but it is clear that each family approaches the ED family at the corresponding $z\rightarrow0$ limit.} 
    \label{fig:spirals_n>0} 
\end{figure*}

As the excited states are sensitive to the precision of our solver, we are unable to generate families of states with as much flexibility as in Sec.~\ref{sec:mc_dynamics} for the ground-state, especially at low redshifts $z\ll1$. Nevertheless, we are able to generate some families, the spectra for which we present in Fig.~\ref{fig:spirals_n>0}. These spectra are for fermion energy $\omega$~(Fig.~\ref{fig:mvw_n>0}), fermion binding energy $E_\mathrm{b}^\mathrm{F}$~(Fig.~\ref{fig:mvEf_n>0}), and total binding energy $E_\mathrm{b}^\mathrm{tot}$~(Fig.~\ref{fig:mvEb_n>0}). From left to right, we have $n=0$, $n=2$, and $n=4$. 

We see that the strong deviations from the FSY ED family (black) are clearly visible only in the ground-state families~($n=0$, left panels). These exhibit, among other other things, comparatively large $m_\mathrm{F}$ values relative to $\omega$, leading to the spectra stretching to the right on the plots. The fermion energy spectra~(Fig.~\ref{fig:mvEf_n>0}) also increase: in particular, the maximum $E_\mathrm{b}^\mathrm{F}$ value of the cyan $\{\xi=0.36,\,v=0.11\}$ family is nearly fivefold above that of the FSY ED family. As discussed in the previous section, these phenomena indicate a strong mass-scale separation discussed earlier.

While deviations from the FSY ED family can be large for the $n=0$ ground state, they diminish with increasing $n$, implying the subsidence of mass-scale separation as oscillations are added to build up the excited-state soliton's wave zone. This trend then strongly suggests that as the excited state number $n$ increases, each family approaches the FSY ED family. 

Momentarily taking ED solitons as a reference, it is a previously established result that higher excited states generally have larger magnitudes of negative total binding energies $E_\mathrm{b}^\mathrm{tot}$~\eqref{eq:Eb_tot}, such that for example the entire family of solutions is stable if e.g. $n=60$~\cite{leith2022_phd}. This corresponds to the energy density in the wave zone deepening the gravitational well. 

However, we find that the opposite is true for EDH solitons: ground-state solutions seem to have a larger magnitude of negative $E_\mathrm{b}^\mathrm{tot}$ compared to their excited-state counterparts. This thus implies that the wave zone diminishes the energetic stabilty of the soliton. Accordingly, we see here that excited-state EDH solitons for small $\xi$ exhibit \textit{less} mass-scale separation than the ground states. This can be seen for example in Fig.~\ref{fig:mvEb_n>0}, where the minimum $E_\mathrm{b}^\mathrm{tot}$ value taken by the $\{\xi=0.36,\,v=0.11\}$ family~(cyan) is $E_\mathrm{b}^\mathrm{tot}\approx-0.75$ if $n=0$~(leftmost panel), whereas for $n=2$ it recedes to $E_\mathrm{b}^\mathrm{tot}\approx-0.05$, and $E_\mathrm{b}^\mathrm{tot}\approx-0.03$ for $n=4$. 

This result is suggestive of the wave zone emergence acting as a direct suppressant for the mass-scale separation in an EDH soliton. This result will be justified in Sec.~\ref{sec:MSS1}.

\section{Many-fermion configurations}\label{sec:many_fermion}

We now extend our analysis to the many-fermion high-Rydberg states~($\kappa>2$) of EDH solitons.

\begin{figure*}[t!]
    \centering 
    \begin{subfigure}{0.999\textwidth}
        \includegraphics[width=\textwidth]{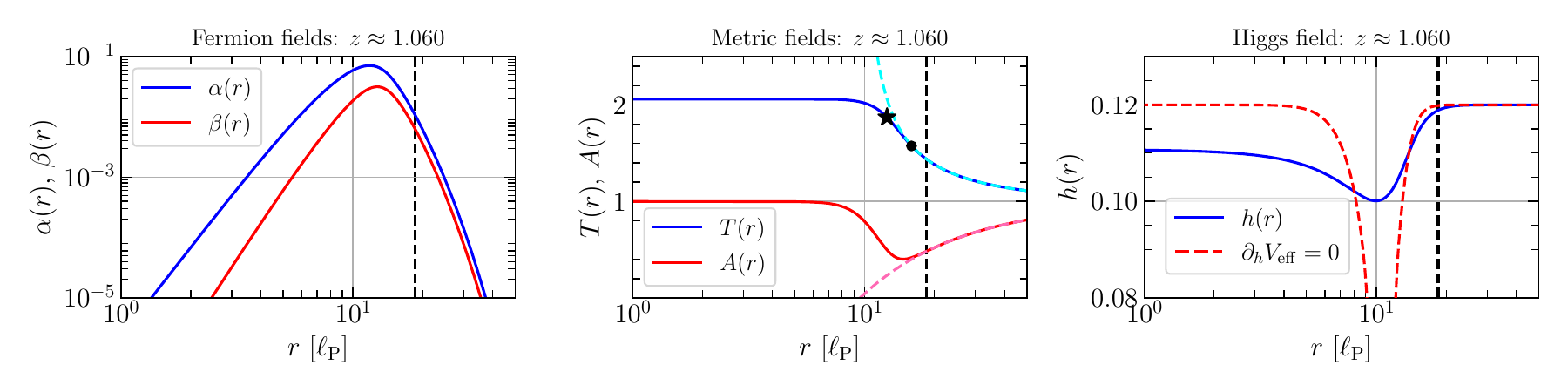}
        \caption[]%
        {A low-redshift solution.}
        \label{fig:edh_k>2_a}
    \end{subfigure}
    \hfill
    \begin{subfigure}{0.999\textwidth}  
        \includegraphics[width=\textwidth]{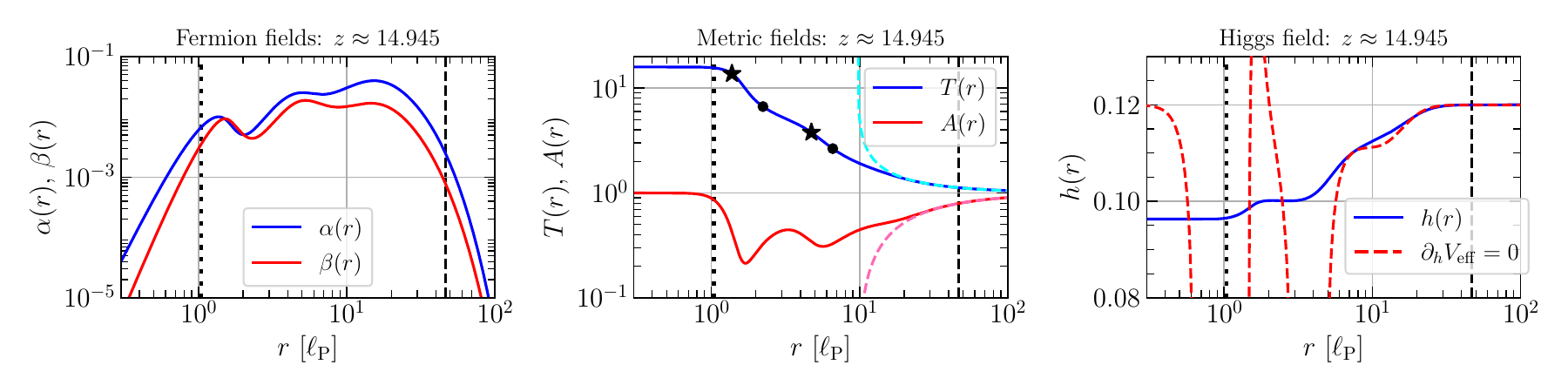}
        \caption[]%
        {A higher-redshift solution.} 
        \label{fig:edh_k>2_b}
    \end{subfigure}
    \hfill
    \begin{subfigure}{0.999\textwidth}  
        \includegraphics[width=\textwidth]{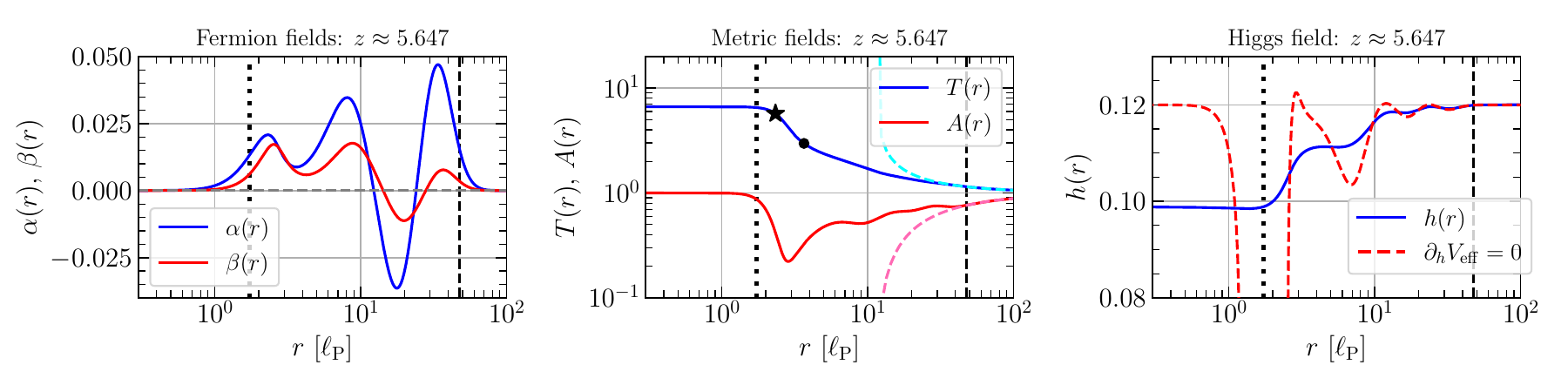}
        \caption[]%
        {An excited-state solution.} 
        \label{fig:edh_k>2_c}
    \end{subfigure}
    \caption[]
    {\justifying Three $\kappa=10$ solutions with different redshift $z$ for $\{\xi,\,v\}=\{0.65,0.12\}$, two of which~(Figs.~\ref{fig:edh_k>2_a} and Fig.~\ref{fig:edh_k>2_b}) are in the ground state ($n=0$) and one in the fourth excited state~(Fig.~\ref{fig:edh_k>2_c}, $n=4$). As in the previous figures for individual $\kappa=2$ solutions, $\alpha$~(left panel), $T$~(middle panel), and $h$~(right panel) are plotted in blue, while $\beta$~(left), $A$~(middle), and dashed contours of $\partial_h V_{\rm eff}=0$~(right) are plotted in red. The dotted and dashed vertical black lines respectively mark the approximate transition from the core to the power-law zone and from the power-law zone to the evanescent zone. Black circles and stars mark the radii of unstable and stable photon spheres~\eqref{eq:dUeff}.} 
    \label{fig:indiv_kappa>2_n} 
\end{figure*}

\begin{figure*}
    \centering
    \includegraphics[width=0.99\linewidth]{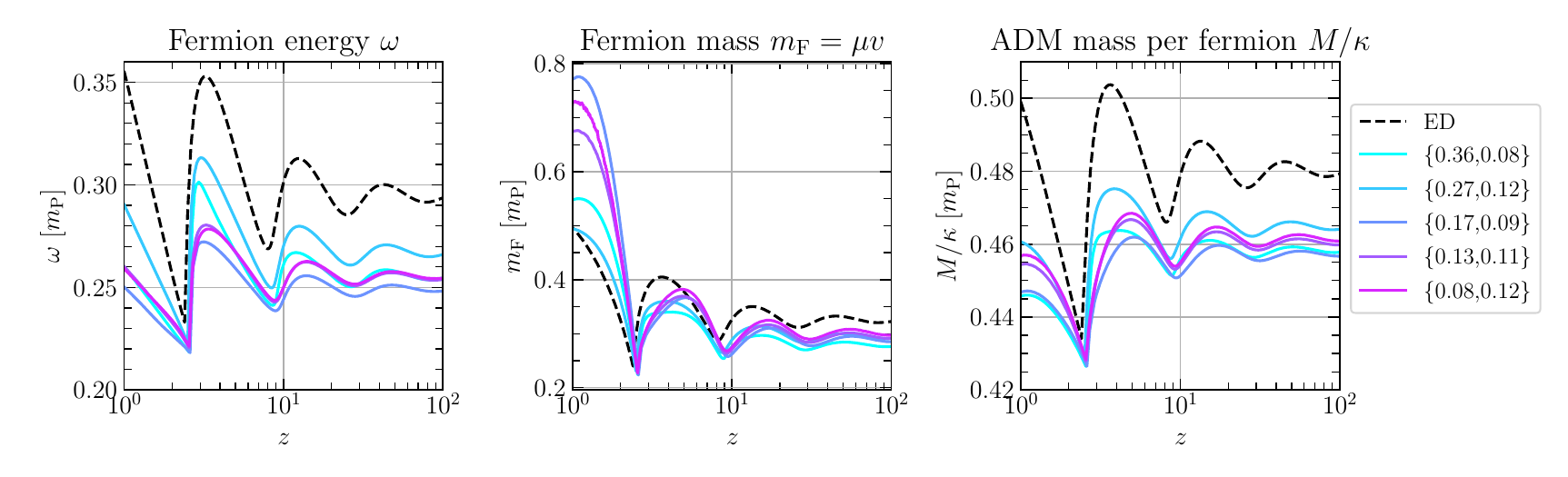}
    \caption{\justifying The variation of characteristic constants $\omega$, $m_\mathrm{F}$, and ADM mass-per-fermion $M/\kappa$ of ED~(dashed black) and EDH soliton solutions over two decades in redshift $z$, for various families of ground-state $\kappa=10$ solitons given in the legend by $\{\xi,v\}$.}
    \label{fig:logz_manyfermion}
\end{figure*}

\subsection{Individual states}

Plotting many-fermion/high angular momentum states for both ground- and excited-state EDH solitons in Fig.~\ref{fig:indiv_kappa>2_n}~(\{$\kappa=10,\,\xi=0.65,\,v=0.12\}$), we immediately notice major departures from the two-fermion system~(Sec.~\ref{sec:mc_dynamics}). 

Fig.~\ref{fig:edh_k>2_a}, showing a soliton solution of $z\approx1.060$, demonstrates one example of such a departure in the Higgs field dynamics. We see that, having started at a value of $h(r=0)\approx0.11$, $h(r)$ dips to a minimum at $r\approx10~[\ell_\mathrm{P}]$ before rising to the VEV. This unintuitive radial profile of $h$ can be explained as follows. It has been seen for the ED system that as $\kappa$ is increased, the soliton becomes more radially expansive~\cite{leith2020,leith2021}. This is because the fermions, being at higher angular momentum compared to $\kappa=2$ states, become localized and ``orbit'' at larger radii. 

By comparing Fig.~\ref{fig:edh_k>2_a} with Fig.~\ref{fig:edh_lowz} and recalling that low redshifts are associated with more radially disperse solitons, it is recognized that the present $\kappa=10$ solution has an RMS soliton radius $\overline{R}\approx12.314~[\ell_\mathrm{P}]$ that is larger than the low-redshift $\kappa=2$ soliton with $\overline{R}\approx11.569~[\ell_\mathrm{P}]$ despite being higher in redshift $z$ by more than a decade. It follows that, because $\alpha$ and $\beta$ respectively have $r^{\lvert\kappa\rvert/2}$ and $r^{\lvert\kappa\rvert/2+1}$ dependencies at small $r$~(see Appendix~\ref{app:num}), the fermion number density $n_\mathrm{F}(r) = \lvert\kappa\rvert T \left( \alpha^2+\beta^2\right)/r^2$ scales as $r^{\lvert\kappa\rvert-2}$ and is thus close to negligible in the core zone for $\lvert\kappa\rvert>2$. This leads to the fermion tilt effectively vanishing in the effective potential $V_\mathrm{eff}$ for a large portion of the core, and $V_\mathrm{eff}$ is just an inverted and untilted quartic of the form $V_\mathrm{eff}(h)\approx-\lambda (h^2-v^2)^2$~\eqref{eq:Veff_Higgs}. Recalling Fig.~\ref{fig:higgs_mech}, one can see that as $h(r)$ is not at an extremum of $V_\mathrm{eff}$ and is not being nudged towards larger values by the fermion tilt, it starts to roll down towards $h=0$ away from the VEV $v$. Then, at $r\approx10~[\ell_\mathrm{P}]$, the rising contributions from $\alpha$ and $\beta$ make the fermion tilt large enough to generate $h''>0$. The initial drop in $h$ hence arises not as a result of a positive fermion tilt, but rather from the lack of a tilt in the first place.

The high-redshift ($z\approx15)$ solution in Fig.~\ref{fig:edh_k>2_b} displays another phenomenon, which has previously been seen in the ED system but is now also shown in the EDH solutions. This is most clearly observed in $A(r)$ and in the fermion fields, which exhibit several large-amplitude damped oscillations around their power-law profiles in the power-law zone. This is associated with a phenomenon known as ``fermion self-trapping'' that emerges for higher-redshift multi-fermion states in the ED system~\cite{leith2020}. Here, the filled shell of $N=|\kappa|$ fermions splits into multiple layers, each localized around a stable photon sphere that forms as a result of a larger $z$, which also causes $T(r)$ to exhibit damped oscillations of larger amplitudes in the power-law zone~\cite{leith2020}. Photon spheres are circular null geodesics allowing either stable or unstable orbits of null particles~\cite{kusano2026_2}. As high-redshift solitons correspond to relativistic states, the fermions themselves in such states are highly relativistic and therefore have trajectories similar to those of massless particles~\cite{leith2020}. The fermion wavefunctions subsequently respond to these circular null geodesics similarly to null particles, and become trapped in the optical geometry of the metric~\cite{leith2020}. In a sense, this is a fermionic analogue to the accumulation and self-gravitation of null particles at stable photon spheres of ultracompact objects and black holes~\cite{difilippo2025,cunha2025, kusano2026_2}. 

Fig.~\ref{fig:edh_k>2_b} exhibits three maxima of the fermion fields, of which the inner two are trapped near the stable photon spheres. These shells are also visible in the metric fields, where both $T$ and $A$ perform damped oscillations in phase with those of the fermion fields. 

Phenomenologically, these damped oscillations in the power-law zone are a result of the fields attempting to match onto their power-law profiles~\eqref{eq:infz} and missing, repeatedly~\cite{bakuczcanario2020}. Though such a trapping phenomenon has previously been hypothesized to be possible for compact~\cite{karlovini2001} and ultracompact objects~\cite{karlovini2002}, Dirac stars are the first system for which this mechanism has been explicitly demonstrated~\cite{leith2020}. 

In the EDH system, departures from the two-fermion systems resulting from fermion self-trapping are not limited to the fermion and metric fields. The Higgs field and the extrema of its effective potential $V_\mathrm{eff}$~\eqref{eq:Veff_Higgs} also exhibit previously unseen behaviors. For example, we notice in Fig.~\ref{fig:edh_k>2_b} that a Higgs staircase is again visible. However, with the ground state fermion fields having no nodes, its origin is distinct to that from the excited-state scenario associated with the nodal structure of the fermion fields. The formation mechanism of the Higgs staircase in these many-fermion cases is a direct consequence of the aforementioned multi-shell fermion self-trapping. Immediately outside the first peak of the fermion fields, there is a region in which $\beta^2>\alpha^2$. This brief dominance of $\beta$ over $\alpha$ reverses the direction of the fermion tilt. The shift appears to be quite sudden, as evidenced by the almost-vertical profile of the $\partial_h V_\mathrm{eff}=0$ contour~(dashed red) followed by a flattening in $h$. This is canceled out by the subsequent dominance of $\alpha$, and the Higgs field is once again nudged towards the VEV $v$. 
Interestingly, the increase in $h$ does not seem uniform, but it is clear this arises as a result of $\alpha$, $\beta$, and $T$ exhibiting damped oscillations, leading to an irregular tilt. The appearance of additional fermion shells does not signify additional positive tilts, so it seems that the Higgs staircase arising from the self-trapping mechanism is predominantly at the first photon sphere. 

As can be seen from comparing Fig.~\ref{fig:edh_k>2_a} and Fig.~\ref{fig:edh_k>2_b}, this multi-shell trapping clearly does not manifest itself for all redshifts. This is also expected from the findings of Ref.~\cite{leith2020} for the ED system. 

For high-enough redshifts this trapping also applies to excited states as well, as can be seen in Fig.~\ref{fig:edh_k>2_c}. The Higgs staircase is again visible, but this time arises from both the multi-shell splitting of the fermion fields and the wave zone structure. This confirms that, just as in the many-fermion ED system, the multi-shell structure and the wave zone can coincide, as seen in the first two fermion field peaks. As previously investigated for the ED system, the coincidence of the fermion self-trapping and wave zone can lead to multivaluedness and nonlinearities in the characteristic constants over redshift~$z$~\cite{leith2021}, especially for high-$\kappa$ families. Therefore, we may quite straightforwardly infer that such nonlinear behaviors would also be visible in EDH systems as well.

\begin{figure}
    \centering
    \includegraphics[width=1\linewidth]{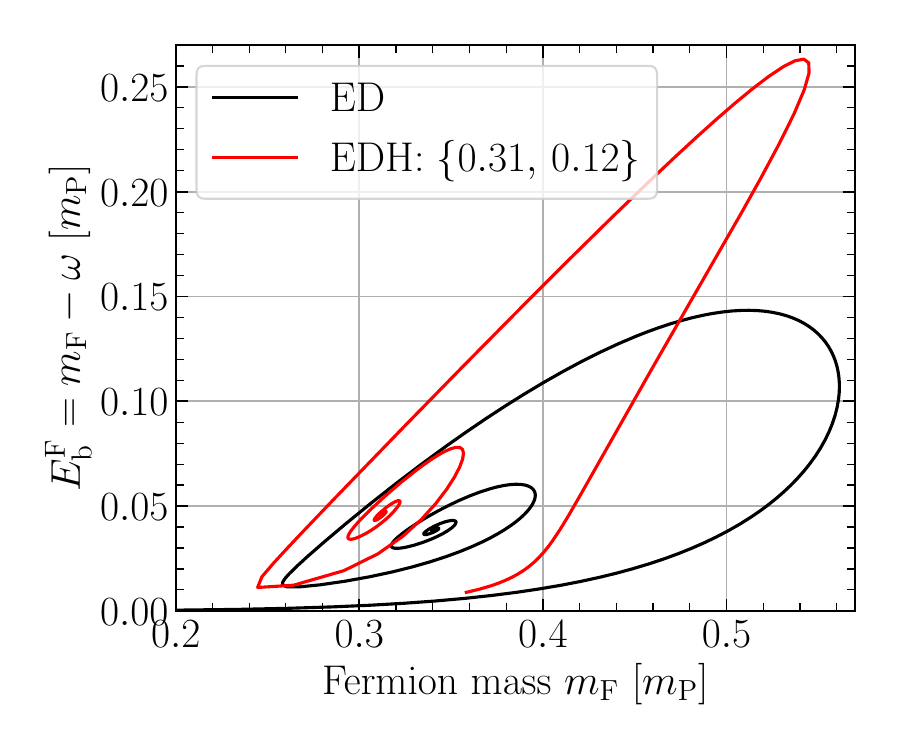}
    \caption{\justifying Fermion binding energy $E_\mathrm{b}^\mathrm{F}$~\eqref{eq:Eb_F} spectrum against asymptotic fermion mass $m_\mathrm{F}$ for $\kappa=6$ families in the ground state~($n=0$). The corresponding FSY ED family~(black) is plotted against a trial EDH family with~$\{\xi=0.31,\,v=0.12\}$~(red).}
    \label{fig:kappa=6spiral}
\end{figure}

\subsection{Families of states}

Turning to families of solutions with many fermions and high angular momenta in Fig.~\ref{fig:logz_manyfermion}~($\kappa=10$), we see the appearance of kinks in the EDH families near $2\leq z\leq3$, which also manifests in the FSY ED families for high $\kappa$~\cite{leith2021}. 
The first kink is aligned with the formation of the first exterior photon sphere to which the fermion fields migrate. 

This first kink also demarcates a significant transition in the values of $m_\mathrm{F}$ and $M$~(middle and right panels of Fig.~\ref{fig:logz_manyfermion}). As can be seen in the middle $m_\mathrm{F}$ panel, the fermion masses all start at $m_\mathrm{F}>0.5$ around $z=1$, with some reaching up to nearly $m_\mathrm{F}\approx 0.75$. Meanwhile, at the same redshift, the ADM mass normalized to the number of fermions does not exceed $M/\kappa\approx0.46$. Therefore a number of the families presented in Fig.~\ref{fig:logz_manyfermion} evidently exhibit some degree of mass-scale separation.

At higher redshifts above this kink, the trend reverses. The fermion masses drop to smaller values, exhibit damped oscillations in $\ln{z}$, and
never venture outside the range $0.2<m_\mathrm{F}<0.4$. 
The ADM mass per fermion similarly oscillates between $0.45<M/\kappa<0.48$. Thus, $M$ consistently exceeds the sum of the constituent fermion masses. 
It then seems that mass-scale separation is suppressed at higher redshifts above the kink, where the photon sphere formed at the kink moves to smaller radii well inside the soliton.

In Fig.~\ref{fig:kappa=6spiral}, we plot a spiral spectrum for fermion binding energy $E_\mathrm{b}^\mathrm{F}$~\eqref{eq:Eb_F} against mass $m_\mathrm{F}$, for $\kappa=6$ families. Compared to the FSY ED family with the same fermion number~(black), the fermion masses of the EDH $\{\xi=0.31,\,v=0.12\}$ family are relatively similar, but $E_\mathrm{b}^\mathrm{F}$ is generally higher than in the ED family, and thus the obtained $\omega$ values are smaller than in the ED family. This indicates a degree of mass-scale separation, though not as strong as for the $\kappa=2$ families in Fig.~\ref{fig:mc_spirals_xi}.
Therefore, as can be seen in both this figure and Fig.~\ref{fig:logz_manyfermion}, many-fermion families display similar characteristics to the two-fermion families.

\section{Mass-scale separation: Cause}\label{sec:MSS1}

\begin{figure*}
    \centering 
    \begin{subfigure}{0.75\textwidth}
        \includegraphics[width=\textwidth]{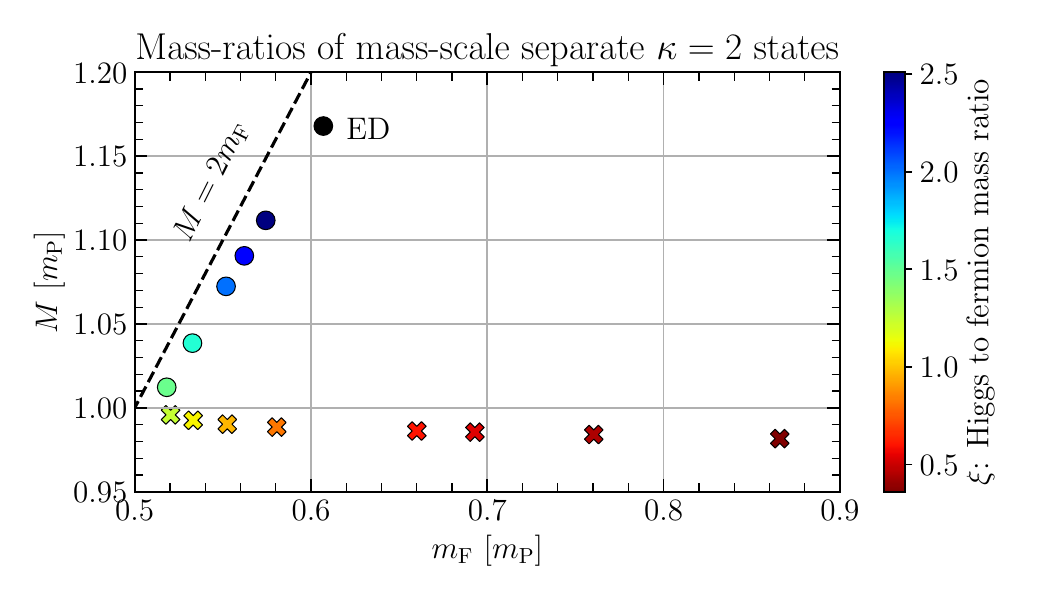}
        \caption[]%
        {$\kappa=2$.}
        \label{fig:mass-scale_phase_k=2}
    \end{subfigure}
    \hfill
    \begin{subfigure}{0.75\textwidth}  
        \includegraphics[width=\textwidth]{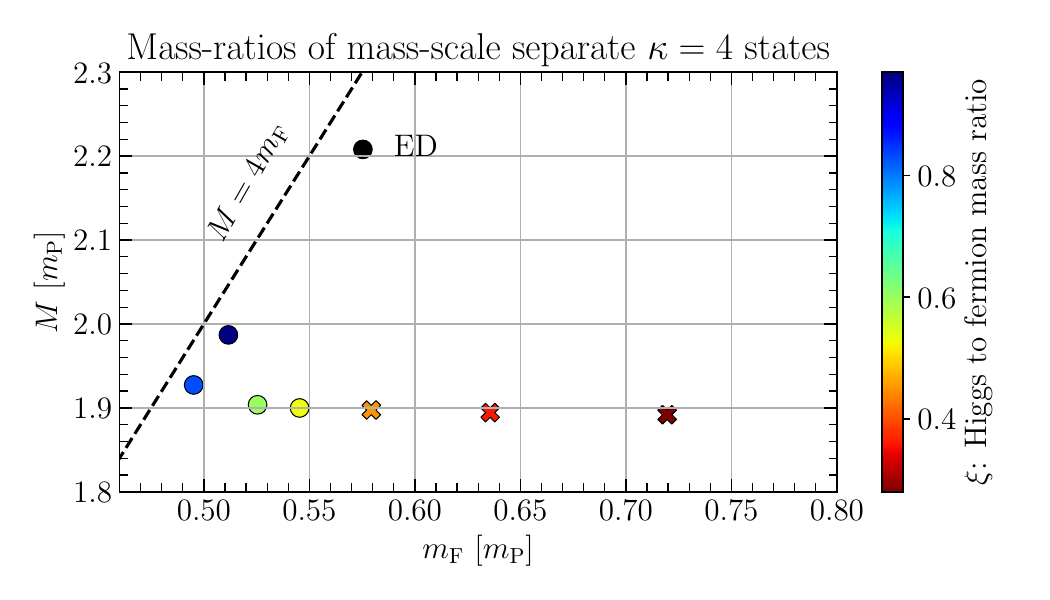}
        \caption[]%
        {$\kappa=4$.} 
        \label{fig:mass-scale_phase_k=4}
    \end{subfigure}
    \caption[]
    {\justifying Mass-scales of maximally bound EDH solutions from ground-state~($n=0$) families with $v=0.11$ and various values of $\xi$. Circles and crosses denote $h_0>0$ and $h_0<0$ respectively.}
    \label{fig:mass-scale_phase} 
\end{figure*}

Having now presented an overview of the various phenomena observable in EDH solitons, we now move onto discussing the main exotic dynamic exhibited by our solutions: the mass-scale separation. As introduced briefly in Sec.~\ref{subsec:2-ferm_fam_sol}, this is a phenomenon whereby the sum of the constituent asymptotic fermion masses not only exceed the ADM mass, but do so quite drastically~\cite{leith2023}. We have also seen in Sec.~\ref{sec:mc_exc} and Sec.~\ref{sec:many_fermion} that the Higgs staircase seems to diminish the effect of this mass-scale separation. 

Of course any solution with a nonzero $E_\mathrm{b}^\mathrm{tot}$ value~\eqref{eq:Eb_tot} must inherently possess some ADM-to-fermion mass disparity in order for the solitonic system to be bound or unbound. However, our terminology refers to the \textit{degree} to which this disparity is observed in the EDH system. 

We now state a qualitative conjecture about the specific cause of the mass-scale separation: 
\begin{quote}
    {The mass-scale separation in EDH solitons is a manifestation of the fermion mass in the core of the soliton being far below the asymptotic evanescent zone fermion mass. This in turn arises from large values of the Yukawa coupling parameter $\mu$ that increase the fermion tilt, which drives the acceleration of $h(r)$ towards the positive VEV.} 
\end{quote} 
To test this, we focus primarily on solution families with a fixed VEV $v=0.11$, and we vary $\xi$. Unless stated otherwise, we probe $\kappa=2$ ground-state~($n=0$) solutions. While a smaller $\xi$ produces larger mass-scale separations, $\xi=0.36$ is small enough to display the effect while avoiding numerical issues such as multivaluedness. We also confirm the validity of our hypothesis with a simplified toy model generated by adapting the FSY ED system to one with a sudden jump in the fermion mass at the inner edge of the evanescent zone, as detailed in Appendix~\ref{app:toy_model}. 

To illustrate the applicability of the above conjecture, we explicitly show in Fig.~\ref{fig:mass-scale_phase} the mass-scales $M$ and $m_\mathrm{F}$ for the maximally bound states of the aforementioned families.
Samples of $\kappa=2$ families are shown in Fig.~\ref{fig:mass-scale_phase_k=2}, whereas those of $\kappa=4$ are shown in Fig.~\ref{fig:mass-scale_phase_k=4}. In Fig.~\ref{fig:mass-scale_phase_k=2}, we see that for $\xi\lesssim1.25$, the mass scales clearly deviate from the $M=2m_\mathrm{F}$ line. In fact, as $\xi$ is decreased, the mass scales turn a corner and follow a trajectory wherein the ADM mass $M$ decreases, while the fermion masses increase. In this light, the mass-scale separation can be interpreted as a significant departure from the expected $M\approx\lvert\kappa\rvert m_\mathrm{F}$ relation. For conciseness, we refer to this deviating branch as the ``Yukawa'' branch, since these states are strongly Yukawa-coupled. 

It is then worth noticing here that, at least for these $\kappa=2$ states, the Higgs field at $r=0$ is negative for all states on the Yukawa branch. While we see in Fig.~\ref{fig:mass-scale_phase_k=4} that this is not necessarily the case for $\kappa>2$ states, it is nevertheless consistent to say that for all of the states on the Yukawa branch $h(r=0)=h_0$ is far below $v$.

\subsection{Mass in the Einstein-Dirac-Higgs system}

\begin{figure}
    \centering
    \includegraphics[width=0.999\linewidth]{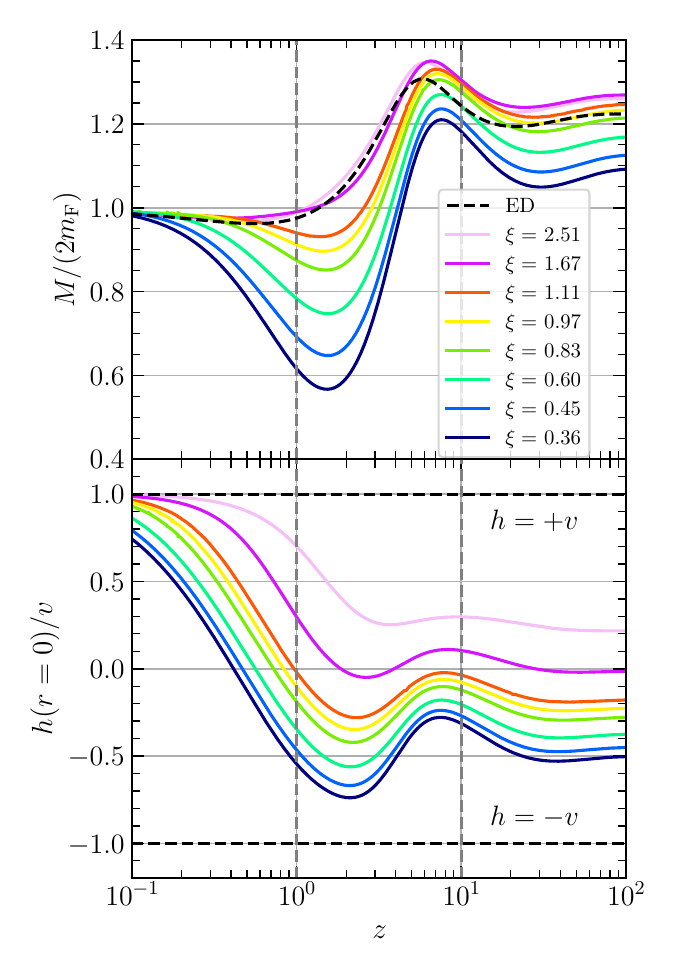}
    \caption{\justifying Depiction of the degree of mass-scale separation ($M/(2m_\mathrm{F})$: upper panel) and the initial value of the Higgs field in the core $h(r=0)=h_0$ (lower panel), both against the central redshift $z$ for different families of EDH solitons. $v=0.11$ for all families in this plot, aside from the FSY ED line~(dashed black on the top panel). $\kappa=2$ for these families.} 
    \label{fig:mssvsh0}
\end{figure}

\begin{figure}
    \centering
    \includegraphics[width=0.999\linewidth]{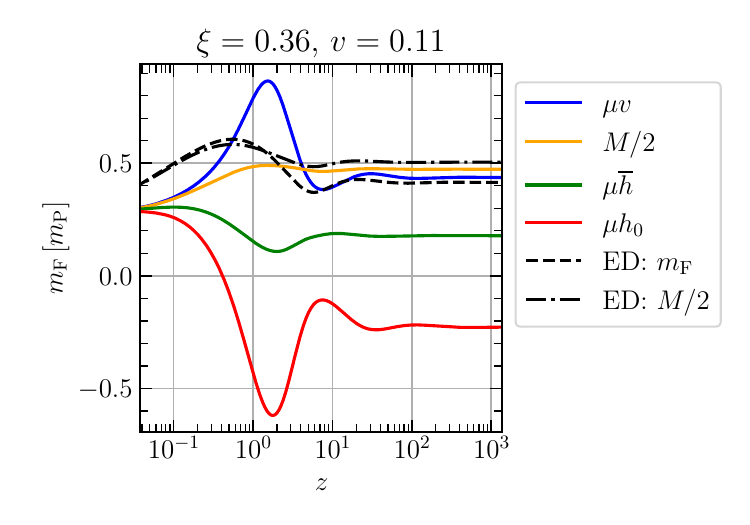}
    \caption{\justifying Fermion masses evaluated at different radii in the EDH system, over redshift $z$. $\mu v$~(blue) is the asymptotic fermion mass, $M/2$ is half the ADM mass~(orange), $\mu \overline{h}$~(green) is the radially averaged mass as defined in~\eqref{eq:mass_avg}, and $\mu h_0$~(red) is the fermion mass at the origin. The FSY ED values for $m_\mathrm{F}$ and $M/2$ are the black dashed and dashdotted lines, respectively.} 
    \label{fig:diff_h}
\end{figure}

As the mass-scale separation pertains to these Yukawa-coupled EDH solitons, it is natural to focus specifically on the mechanism that distinguishes these solutions from those of FSY~\cite{fsy1999}: namely, the interaction of the fermions with a mass-generating Higgs field. 

With solitons of small $\xi$ exhibiting this phenomenon the most, such solutions contain a large Yukawa-coupling constant $\mu$ and a small $\lambda$. The limit $\xi\rightarrow0$ may, through looking at $\mathcal{L}_\mathrm{m}$~\eqref{eq:L_mat}, be interpreted as one wherein $h$ is a free scalar field that does not interact with itself as $V(h)\rightarrow0$, or alternatively as one where the Yukawa-coupling becomes infinite as $\mu\rightarrow\infty$. 

Focusing on the case when $\mu$ is large, it is easy to see by referring to~\eqref{eq:Veff_Higgs} that this steepens the tilt. Therefore, for the same redshift, small-$\xi$ states experience a more drastic change in $\mu h(r)$ than for large-$\xi$ states, from the core to the evanescent zone. It is also clear that the value of the Higgs field at the origin must be further from the VEV, as illustrated in Fig.~\ref{fig:mssvsh0}. The figure also shows that $h(r=0)$ takes a minimum value at the maximally bound states, the latter of which correspond to the minimum of the $M/(2m_\mathrm{F})$ curves. This also links to why families with small $v$ values tend to deviate from the FSY ED family, as seen in Fig.~\ref{fig:mc_spirals_v}: with the VEV being closer to zero, the same change in $h(r)$ for a small-$v$ solution of e.g. $v=0.08$ would be quantitatively more significant than for a solution with e.g. $v=0.14$. 

From the above observation, it is worth discussing whether the asymptotic fermion mass is a sufficiently appropriate measure of the fermion mass, in the sense of whether it should be compared at all to the ADM mass. As discussed previously, there appear to be no states for which $h$ approaches the VEV $v$ from above, which makes sense considering the Higgs field dynamics~(Sec.~\ref{subsec:higgs_mech}). Hence, $\mu v$ is not only \textit{not} equal to the fermion mass anywhere interior to the evanescent zone, but it is in fact the \textit{maximum possible} fermion mass that is reached only after the bulk of the soliton has already localized gravitationally. Juxtaposing the asymptotic fermion masses to a gravitational mass obtained from $A(r)$, which has simply \textit{not} backreacted to fermions of mass $\mu v$ up until the start of the evanescent zone, is perhaps not the most appropriate comparison. 

To test the above, we define a mean value $\overline{h}$ for the Higgs field $h$:
\begin{equation}\label{eq:mass_avg}
    \overline{h} \equiv4\pi \int^\infty_0\dfrac{h(r) T(r)}{\sqrt{A(r)}}(\alpha^2(r)+\beta^2(r))\mathrm{d}r,
\end{equation} 
This integral reduces to a constant when $h$ is constant such as in the ED case, which can be seen by comparing to the normalisation integral~\eqref{eq:norm_rescaled}. 
The motivation for defining $\overline{h}$ is not because it is physically important per se, but rather to test whether the corresponding averaged mass $\overline{m}_\mathrm{F}\equiv\mu \overline{h}$ may serve as a better proxy for fermion mass instead of the asymptotic $m_\mathrm{F}=\mu v$ that applies to non-localized fermions. 

However, we find as seen in Fig.~\ref{fig:diff_h} that a mass $\overline{m}_\mathrm{F}$ defined using~\eqref{eq:mass_avg} deviates significantly from FSY expectations, and cannot be used in the stead of the asymptotic fermion mass. Actually, it turns out that $\overline{m}_\mathrm{F}$ lies far below both $\mu v$ and the FSY ED mass prediction. 

The radially averaged mass $\overline{m}_\mathrm{F}$ is nevertheless not without merit. It is seen that at the redshift of the maximally-bound state where $\mu v$ takes its maximum value, $\overline{m}_\mathrm{F}$ has a \textit{minimum}. This is roughly consistent with Fig.~\ref{fig:mssvsh0}, and therefore it may be seen that the starting value of the Higgs field has a considerable effect on $\overline{m}_\mathrm{F}$, especially for relatively disperse solitons such as those of the maximally-bound state. 

This may tie into why $\kappa>2$ states can exhibit the mass-scale separation despite not having $h_0<0$ on the Yukawa branch. As seen in Fig.~\ref{fig:edh_k>2_a}, the effective Higgs potential $V_\mathrm{eff}$ is effectively untilted in the core and $h(r)$ can even decrease~(e.g. Fig.~\ref{fig:edh_k>2_a}). Through this, even if $h(r=0)$ is not negative, the relatively small fermion masses in the core can pull down the overall $\overline{m}_\mathrm{F}$. 

It is worth discussing here whether the difference between gravitational ADM mass and the sum of the fermion masses is a true measure of the binding energy of the soliton in the first place. In the two-fermion case, it is intuitive to refer to $E_\mathrm{b}^\mathrm{tot}$~\eqref{eq:Eb_tot} as the energy required to gravitationally delocalize an even number of fermions from each other. In contrast, consider the Dirac star constructions of Refs.~\cite{herdeiro2019, herdeiro2022}, which are comprised of a \textit{single} spinning fermion of mass $m_\mathrm{F}$ that has localized gravitationally. What would a nonzero value for $M-\lvert m_\mathrm{F}\rvert$ mean in such a case, and what would it mean to have a bound versus an unbound state for such a system wherein the fermion is not gravitationally bound to any other fermion? For the two-fermion case, it may be argued that it is more appropriate in the present instance to evaluate the binding energy as the difference between the ADM mass of two localized fermions and the sum of the ADM masses of two single spinning fermions. In such a case, the latter would need to be considered in the zero-spin limit but not quite at spin zero, as single-fermion Dirac stars inherently require spin~\cite{herdeiro2019}.

\begin{figure*}
    \centering
    \includegraphics[width=0.99\linewidth]{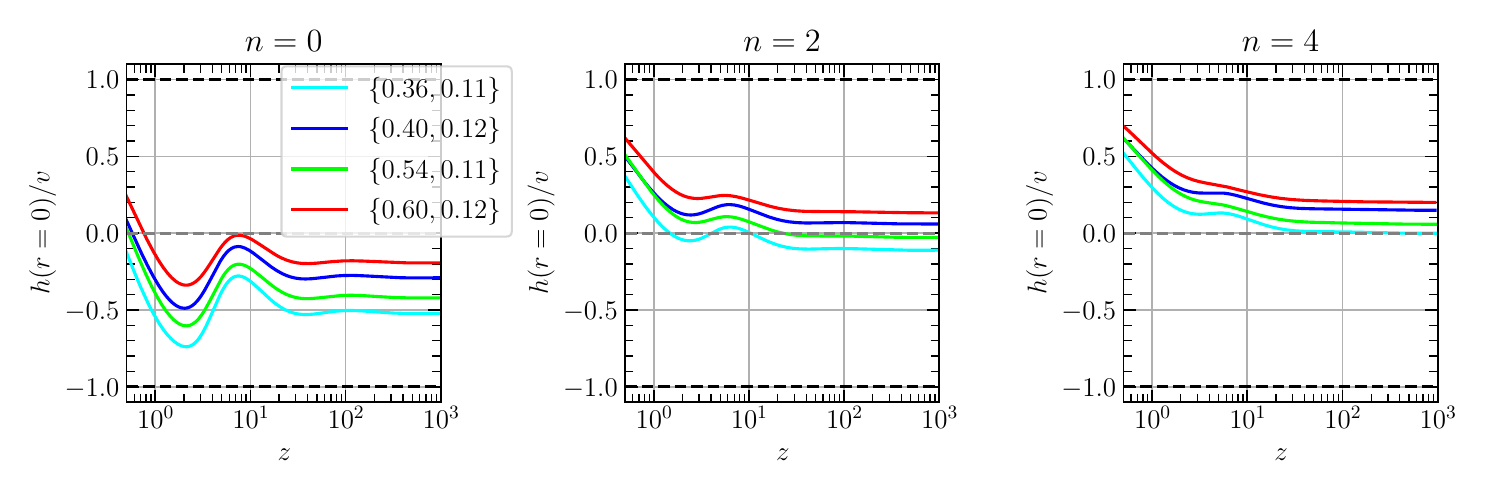}
    \caption{\justifying Starting value of the Higgs field $h(r=0)$ versus redshift $z$ with $\kappa=2$. As in the other figures, each family is parametrized by $\{\xi,\,v\}$. The dashed black lines on each panel correspond to $h(r=0)=\pm v$, and the dashed gray lines are $h(r=0)=0$. Left panel: $n=0$, middle panel: $n=2$, right panel: $n=4$. } 
    \label{fig:tilt_exc}
\end{figure*}

\subsection{Higgs staircase} 

The reason for the $\kappa=2$ excited states~($n\geq2$) exhibiting little or no mass-scale separation in comparison to the ground states is, in our view, a consequence of the Higgs staircase. Our claim here is not that the excited states quell the mass-scale separation entirely, but rather that their zonal structures severely inhibit this effect.

With the Higgs field alternating between $h''>0$ and $h''<0$ as a consequence of the radial intervals with $\beta^2>\alpha^2$ around each $\alpha$ node in the wave zone, $h(r)$ has a punctuated rise up the Higgs staircase before entering the evanescent zone. This can alternatively be reframed as the increase in the Higgs field being hampered by the alternating positive and negative effective accelerations arising from $V_\mathrm{eff}$ tilting back and forth. In ground-state $\kappa=2$ solitons, $h(r)$ rises in one step to arrive at the VEV immediately after entering the evanescent zone.
In contrast, in the excited and higher-redshift many-fermion configurations, the rise of $h(r)$ toward the VEV is repeatedly stalled as it climbs up the staircase. 

It is through a similar mechanism that the many-fermion ($\kappa>2$) cases reduce their degree of mass-scale separation after the formation of the first exterior photon sphere. In such cases, the brief $\beta$-domination provides the effective potential $V_\mathrm{eff}$~\eqref{eq:Veff_Higgs} with a positive tilt, as can be seen in Fig.~\ref{fig:edh_k>2_b}. 

This then implies that, for the same redshift, the Higgs field must start at a value \textit{closer} to the VEV for excited-state solitons or solutions that contain brief instances where $\beta^2>\alpha^2$, to be able to reach $v$ in time for the beginning of the evanescent zone. This is validated in Fig.~\ref{fig:tilt_exc}.

\section{Mass-scale separation: Consequences}\label{sec:MSS2} 

In this section, we highlight some effects of the mass-scale separation discussed in Sec.~\ref{sec:MSS1}, which is not exhibited in the FSY ED system. We still maintain $v$ at $v=0.11$ as in the previous section.

\subsection{Trapping via null geodesics}\label{subsec:phase}

\begin{figure*}
    \centering 
    \begin{subfigure}{0.9\textwidth}
        \includegraphics[width=\textwidth]{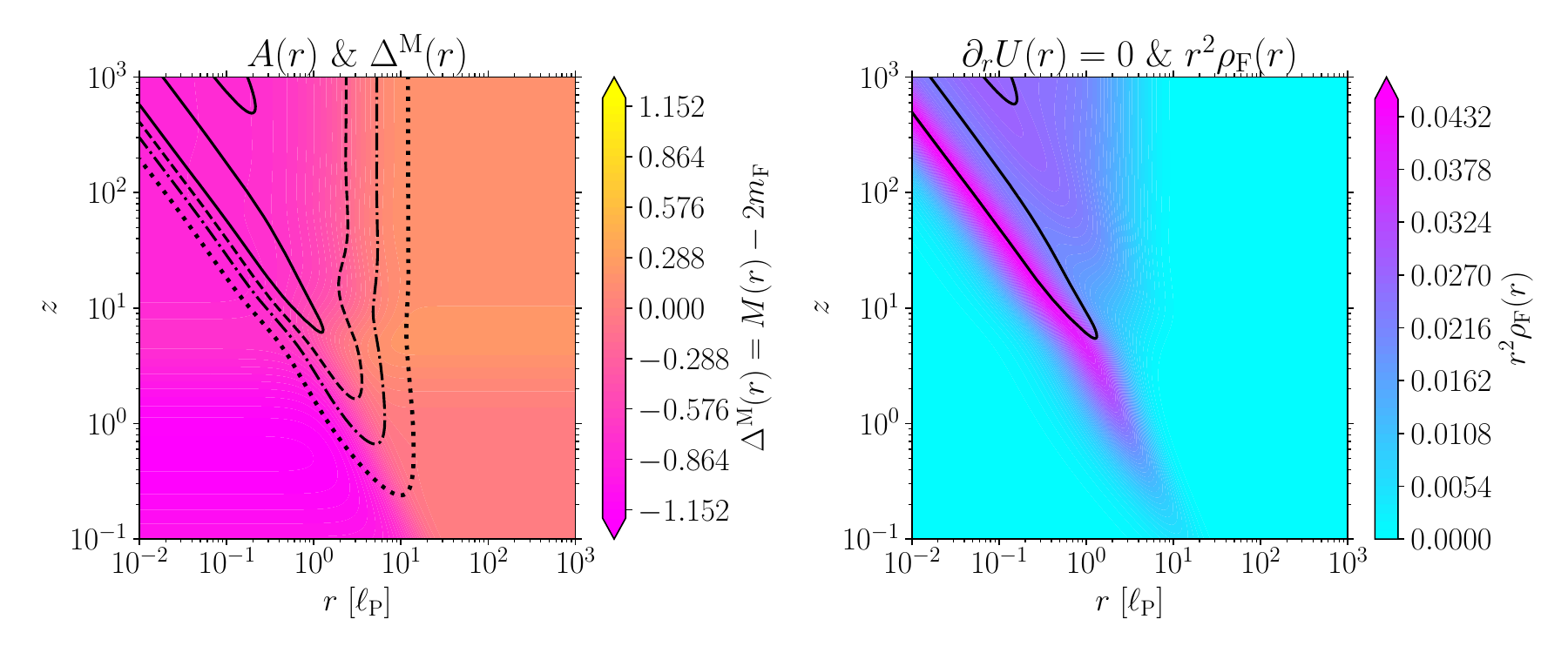}
        \caption[]%
        {FSY ED: equivalent to $\{\xi=\infty\}$.}
        \label{fig:phase_ed}
    \end{subfigure}
    \hfill
    \begin{subfigure}{0.9\textwidth}  
        \includegraphics[width=\textwidth]{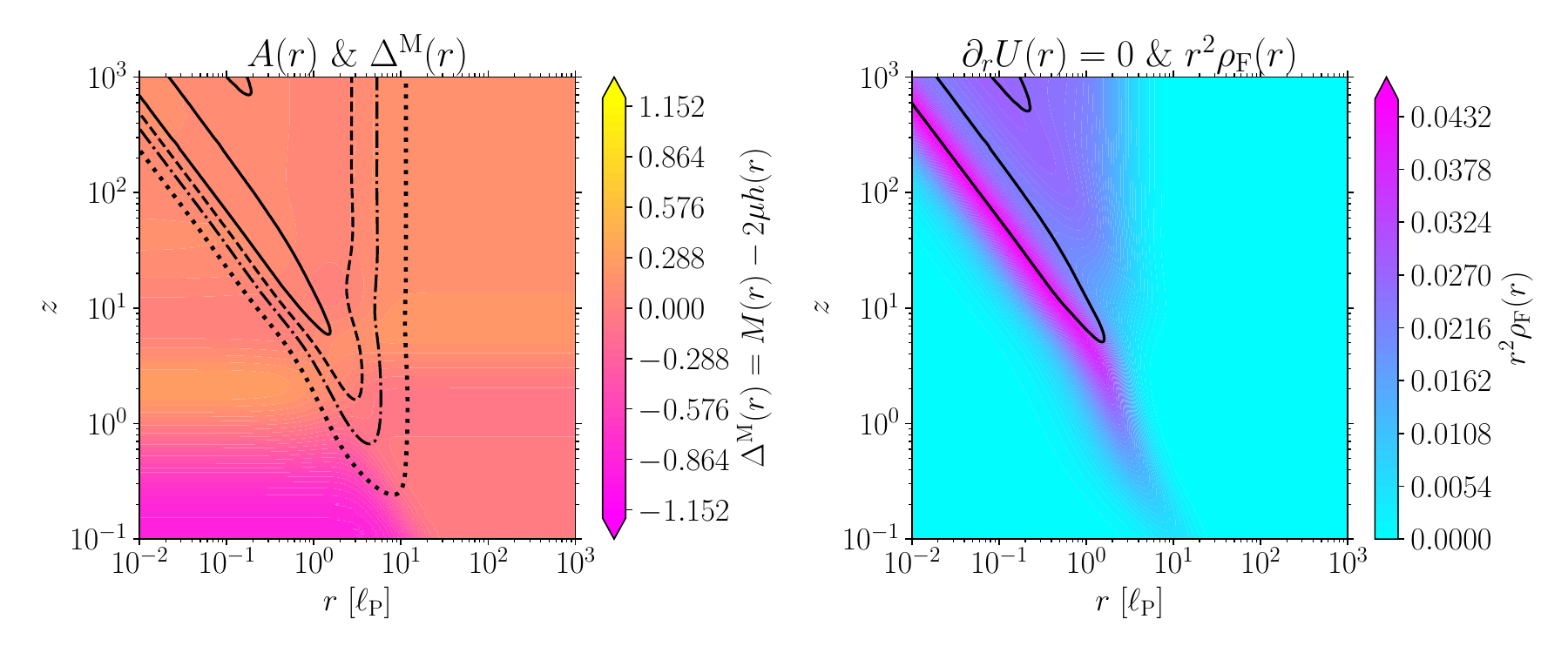}
        \caption[]%
        {EDH: $\{\xi=1.11,\,v=0.11\}$.} 
        \label{fig:phase_edh_111}
    \end{subfigure}
    \hfill
    \begin{subfigure}{0.9\textwidth}  
        \includegraphics[width=\textwidth]{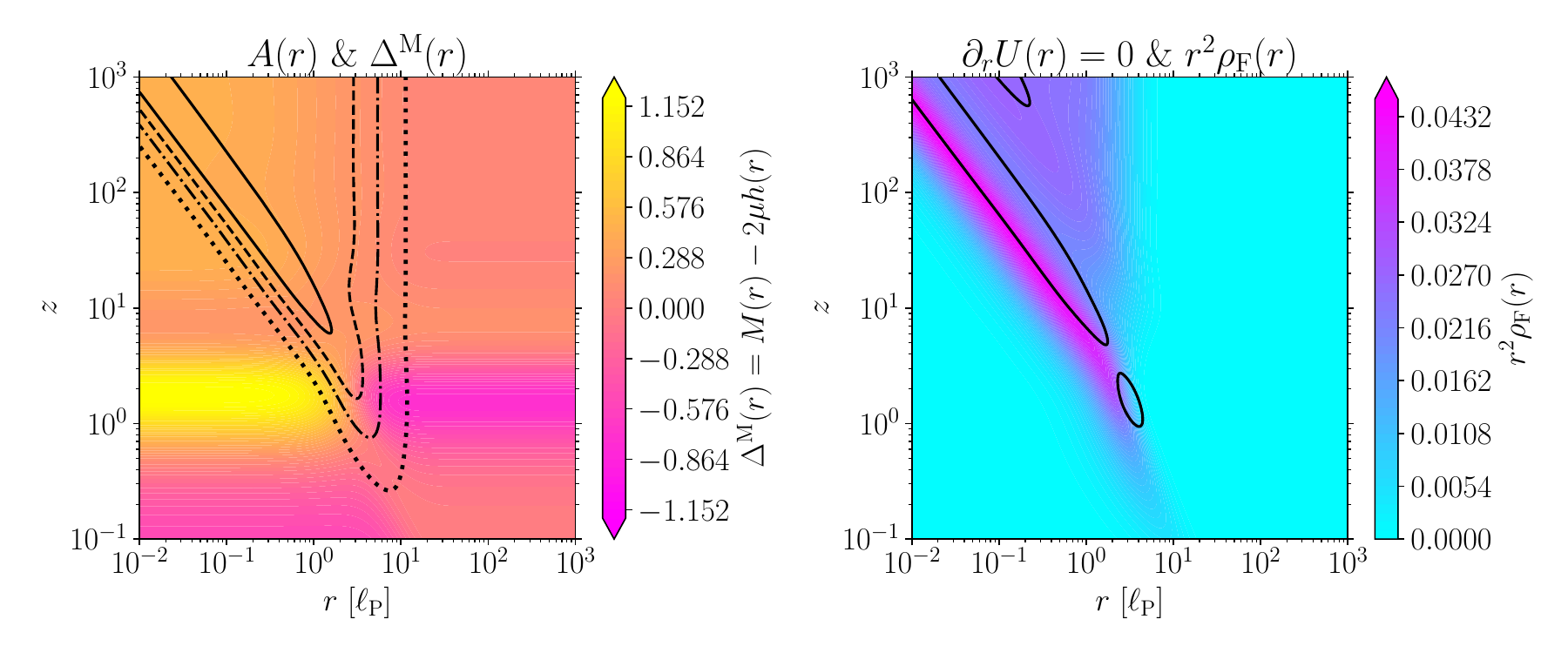}
        \caption[]%
        {EDH: $\{\xi=0.36,\,v=0.11\}$.} 
        \label{fig:phase_edh_036}
    \end{subfigure}
    \caption[]
    {\justifying Phase diagrams for three ground-state~($n=0$) soliton families: FSY ED~(Fig.~\ref{fig:phase_ed}), EDH with $\{\xi=1.11\,,v=0.11\}$~(Fig.~\ref{fig:phase_edh_111}), and EDH with $\{\xi=0.36\,,v=0.11\}$~(Fig.~\ref{fig:phase_edh_036}). The left panels contain black contours of $A(r)=[1/3~(\mathrm{solid}),\,1/2~(\mathrm{dashed}),\,2/3~(\mathrm{dashdotted}),\,5/6~(\mathrm{dotted})]$ overlaid on $\Delta^\mathrm{M}(r)$~\eqref{eq:Eb_local}, while the right panels overlay the contours of $\partial_r U=0$~\eqref{eq:dUeff}~(black) on $r^2 \rho_\mathrm{F}(r)$~\eqref{eq:rhoF(r)}.}
    \label{fig:edh_phases} 
\end{figure*} 

\begin{figure*}
    \centering 
    \begin{subfigure}{0.9\textwidth}
        \includegraphics[width=\textwidth]{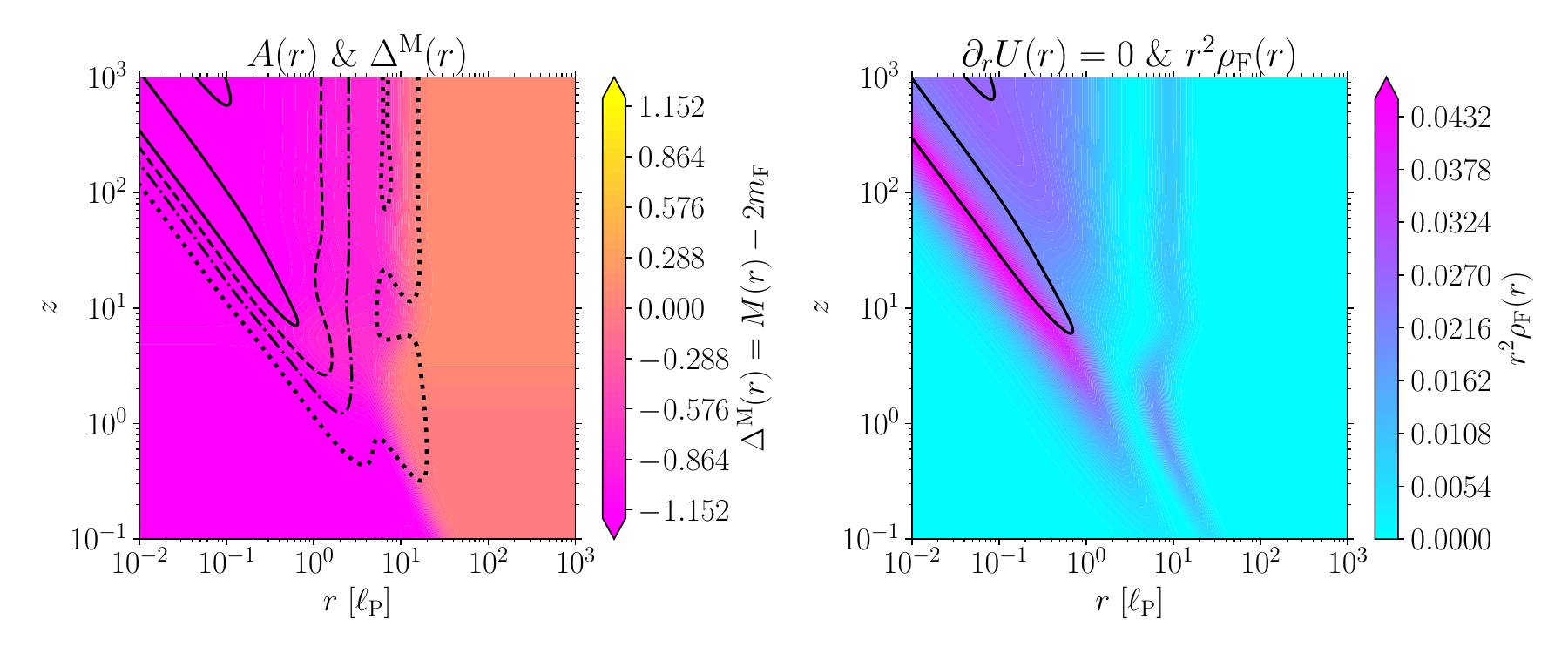}
        \caption[]%
        {FSY ED: equivalent to $\{\xi=\infty\}$.}
        \label{fig:phase_ed_n=2}
    \end{subfigure}
    \hfill
    \begin{subfigure}{0.9\textwidth}  
        \includegraphics[width=\textwidth]{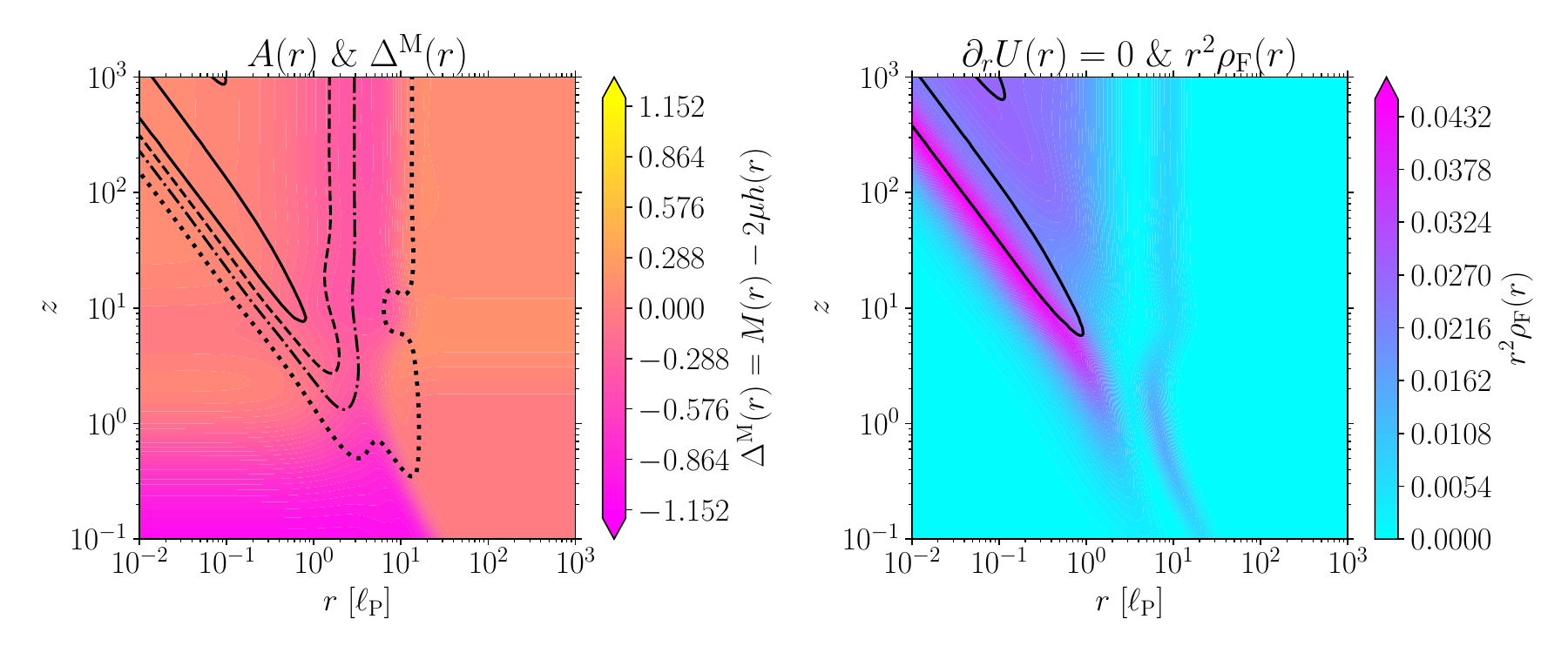}
        \caption[]%
        {EDH: $\{\xi=0.36,\,v=0.11\}$.} 
        \label{fig:phase_edh_036_n=2}
    \end{subfigure}
    \caption[]
    {\justifying Phase diagrams for two second-excited state~($n=2$) soliton families: FSY ED~(Fig.~\ref{fig:phase_ed}) and EDH with $\{\xi=0.36\,,v=0.11\}$~(Fig.~\ref{fig:phase_edh_036}). Like in Fig.~\ref{fig:edh_phases}, the left panels overlay black contours of $A(r)=[1/3~(\mathrm{solid}),\,1/2~(\mathrm{dashed}),\,2/3~(\mathrm{dashdotted}),\,5/6~(\mathrm{dotted})]$ on $\Delta^\mathrm{M}(r)$~\eqref{eq:Eb_local}, while the right panels overlay the contours of $\partial_r U=0$~\eqref{eq:dUeff}~(black) on $r^2 \rho_\mathrm{F}(r)$~\eqref{eq:rhoF(r)}.}
    \label{fig:edh_phases_n=2} 
\end{figure*} 

To uncover the effects of the mass-scale separation, we can consider the radial profiles of not only the Higgs field, but also the fermion and metric fields as well. To do so, we evaluate an entire family of solutions over many decades in redshift $z$, and plot contours obtained from the radial profiles of the fields. 

To measure the radial variation in the mass-scale separation, we adopt a quantity to measure the local disparity between the ADM and fermion masses:
\begin{equation}\label{eq:Eb_local}
    \Delta^\mathrm{M}(r)=M(r)-\lvert\kappa\rvert \mu h(r),
\end{equation}
where $M(r)$ is the Misner-Sharp mass function~\eqref{eq:Misner_Sharp}.

In addition to the above and the null geodesic structure~\eqref{eq:dUeff}, we consider the fermion energy density $\rho_\mathrm{F}$ by isolating the fermion contribution in ${T^t}_t$~\eqref{eq:T_munu_comps}~\cite{leith2022_phd}:
\begin{equation}\label{eq:rhoF(r)}
    \rho_\mathrm{F}(r)=\dfrac{\lvert\kappa\rvert\omega T^2}{r^2}(\alpha^2+\beta^2).
\end{equation}
$r^2\rho_\mathrm{F}(r)$~\eqref{eq:rhoF(r)} peaking where $\partial_r U=0$~\eqref{eq:dUeff} indicates that the solution is so highly relativistic that the fermions can be approximated by null particles that are trapped at the (stable) photon sphere of the metric~\cite{leith2020}.

Using the above, we present the phase diagrams for three different soliton solution families in Fig.~\ref{fig:edh_phases}, where Fig.~\ref{fig:phase_ed} corresponds to the non-Yukawa-coupled ED family, and Figs.~\ref{fig:phase_edh_111} and~\ref{fig:phase_edh_036} show EDH families for $\xi=1.11$ and $\xi=0.36$ respectively, where both have VEVs of $v=0.11$. The color map on the left-hand panels denote the value of the ``local'' mass difference $\Delta^\mathrm{M}(r)$~\eqref{eq:Eb_local}. Overlaid directly on this color map are black contours of $1/3\leq A(r) < 1$ in steps of $1/6$; the lower limit of this interval corresponds to the $A(r)$ value for the power-law infinite-redshift solution~\eqref{eq:infz}, such that the damped oscillations of the power-law zone reaching above and below this value are easily visualized. On the right-hand panel, the color map indicates the value of $\rho_\mathrm{F}(r)$ at each radius, and the black contours denote where $\partial_r U=0$~\eqref{eq:dUeff}. 

In all three cases, no major differences are seen in the general radial profiles of $A(r)$, nor of $\rho_\mathrm{F}$. The first dip below $A=1/3$ occurs at around $z\approx6$, and $\rho_\mathrm{F}(r)$ starts to strongly peak at around $z\approx1$ when the soliton transitions from its nonrelativistic to relativistic regimes. These are both symptomatic of an increased degree of compactness~\cite{leith2020}. 

The more curious departures of the EDH families, particularly those of the $\xi=0.36$ family, are observable in the contours of $\Delta^\mathrm{M}(r)$~\eqref{eq:Eb_local} and $\partial_r U=0$~\eqref{eq:dUeff}. 

The null geodesic structure of the $\xi=0.36$ family~(Fig.~\ref{fig:phase_edh_036}) is unique in comparison to the two others. Specifically, the island of $\partial_r U=0$ seen at $1\lesssim z \lesssim 2$ indicates that, for this small range in redshift, a pair of photon spheres emerges and separates, then converges and annihilates. The photon spheres appear again at higher redshift~($5\lesssim z \lesssim 6$) for all families, but of the three families illustrated only the $\xi=0.36$ family exhibits this behavior at lower redshifts. This coincides exactly with the range of redshift wherein the mass-scale separation is most prominent. 

The presence of photon spheres in Fig.~\ref{fig:phase_edh_036} implies that the most mass-scale separate states would yield a ``soliton shadow''. Astrophysical objects with shadows are often called \textit{ultracompact}~\cite{cardoso2014}, and therefore a soliton with photon spheres can be referred to as such. 

The reasoning behind why these photon spheres emerge in Fig.~\ref{fig:phase_edh_036} can be intuited as the following. We know that for these mass-scale separate states of the $\xi=0.36$ family, $h(r=0)<0$~(Fig.~\ref{fig:mssvsh0}), which means that at a certain point in radius when the potential is tilted, $h=0$ and the fermions actually become null particles. This may also be intuited from Fig.~\ref{fig:diff_h}, wherein $\overline{m}_\mathrm{F}$~\eqref{eq:mass_avg} is much smaller than the asymptotic fermion mass. When $\rho_\mathrm{F}(r)$~\eqref{eq:rhoF(r)} peaks, the fermions have masses of magnitudes far smaller than $\mu v$, and may indeed be null particles if not close to null. However, further inwards in radius they are more massive in magnitude albeit having negative masses. It is then perhaps unsurprising that the fermions, when they are most massive in the core, mimic the signatures of highly relativistic compact objects.

Despite the above discussion, in our view the emergence of the photon spheres seems an unlikely culprit in the emergence of the mass-scale separation, and rather seems to be a consequence. The reasoning for this is in relation to our $\kappa>2$ solutions discussed in Sec.~\ref{sec:many_fermion}. In particular, the multi-shell splitting after the kink in Fig.~\ref{fig:logz_manyfermion} and the resulting region where $\beta^2>\alpha^2$ seems to reduce the mass-scale separation. This is also evidenced by the highly relativistic ($z\gg1$) soliton solutions even with $\kappa=2$, which have multiple photon spheres but little mass-scale separation.

One can also discuss why, for example on Fig.~\ref{fig:edh_phases}, the region with very positive $\Delta^\mathrm{M}(r)$ must transition directly into the region with very negative $\Delta^\mathrm{M}(r)$, for mass-scale separate states. Why is the mass disparity so large in \textit{both} the core and evanescent zones? We assign this once again to the fermion tilt and $\mu$. If $\xi$ is small and $\mu$ is large, this leads to more negative $h(r=0)$ values. A similar relation also holds for the asymptotic fermion masses: if $\mu$ is large, $\mu v$ must be large. The value of $E_\mathrm{b}^\mathrm{tot}$ being very negative thereby arises from a lack of proportional response in the ADM mass $M$ in relation to the asymptotic fermion mass $m_\mathrm{F}$, which in turn is a result of the metric locally backreacting to fermion masses lower than $\mu v$.

Referring to Fig.~\ref{fig:edh_phases_n=2} for phase diagrams of the second excited state~($n=2$), we see some major changes; the $A(r)$ contours exhibit less triangular shapes, owing to the oscillations of the fermion fields in the wave zone. The extra peak in $r^2\rho_\mathrm{F}(r)$ further out in radius is a consequence of the second peak present in second-excited states, and is seen clearly on the right-hand panels. Additionally, the photon sphere pocket formed in the $n=0$ ground states for $\xi=0.36$~(Fig.~\ref{fig:phase_edh_036}) has disappeared completely in Fig.~\ref{fig:phase_edh_036_n=2}. This agrees with the subsidence of the mass-scale separation in excited states, which is also witnessed in the lack of a large variation in $\Delta^\mathrm{M}(r)$~\eqref{eq:Eb_local}.

\subsection{Decay rates}\label{subsec:decayrates}

\begin{figure*} 
    \centering 
    \begin{subfigure}{0.49\textwidth}
        \includegraphics[width=\textwidth]{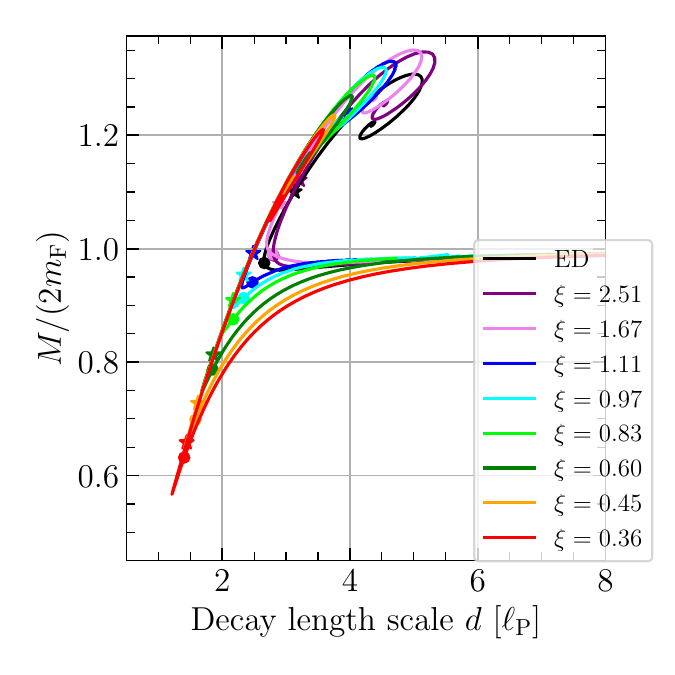}
        \caption[]%
        {Exponential decay scale $d$ for fermion fields $\alpha$ and $\beta$.}
        \label{fig:decay_d}
    \end{subfigure}
    \hfill
    \begin{subfigure}{0.49\textwidth}  
        \includegraphics[width=\textwidth]{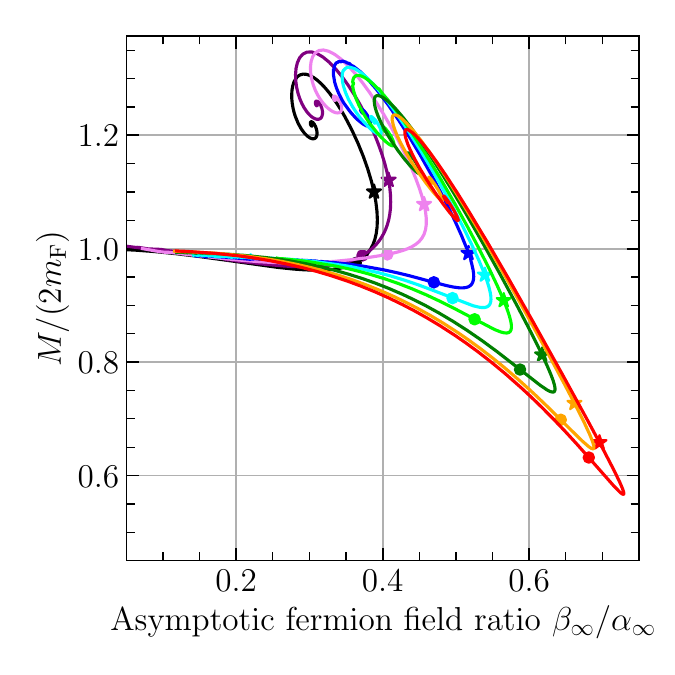}
        \caption[]%
        {Large-$r$ asymptotic ratio $\beta_\infty/\alpha_\infty$.} 
        \label{fig:decay_power}
    \end{subfigure}
    \hfill
    \begin{subfigure}{0.49\textwidth}
        \includegraphics[width=\textwidth]{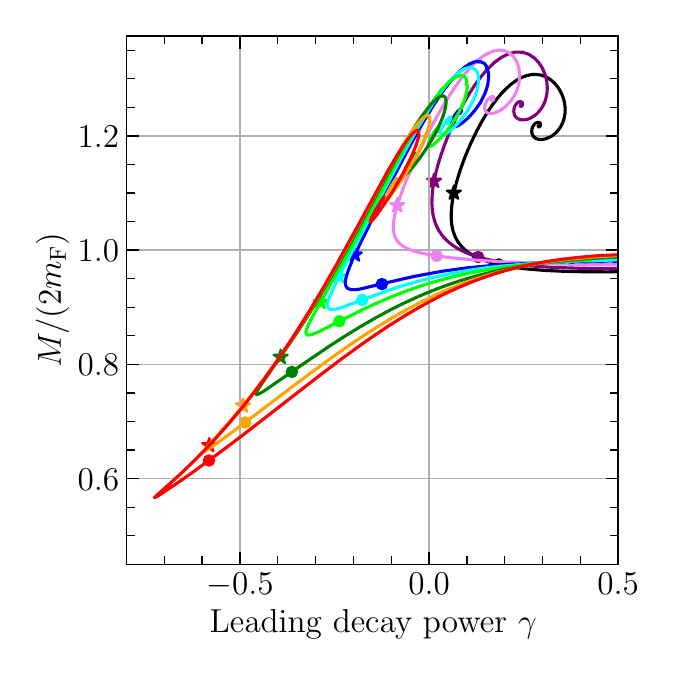}
        \caption[]%
        {Power-law exponent $\gamma$.} 
        \label{fig:decay_gamma}
    \end{subfigure}
    \hfill
    \begin{subfigure}{0.49\textwidth}  
        \includegraphics[width=\textwidth]{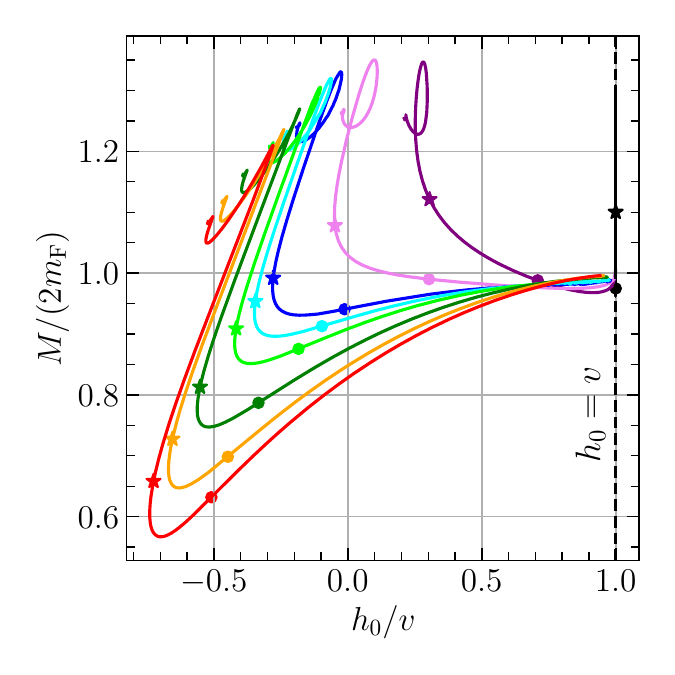}
        \caption[]%
        {Higgs field value at the origin $h_0$.} 
        \label{fig:decay_h0}
    \end{subfigure}
    \caption[]{\justifying 
    Decay parameters of ED(H) solitons in the evanescent zone in relation to the mass-scale separation. Each curve is a family with fixed $v=0.11$ and different values of $\xi$, as indicated in the legend. The circles and stars correspond to states with redshifts $z=1$ and $2.5$ respectively, between which the most bound state occurs. 
    The exponential decay length scale $d$~\eqref{eq:decayexp} is shown in
    Panel\,(a), the power-law exponent $\gamma$~\eqref{eq:gamma} in Panel\,(b), and the asymptotic $\beta/\alpha$ ratio~\eqref{eq:E0D0} in Panel\,(c).
    Panel\,(d) shows the central Higgs field value $h(r=0)=h_0$. 
    \label{fig:edh_decay} }
\end{figure*}

We now discuss the decay rates of EDH solitons in the evanescent zone, for the $\kappa=2$ and $n=0$ case. 

We visualize the evanescent zone parameters $d$~\eqref{eq:decayexp}, $\beta_\infty/\alpha_\infty$~\eqref{eq:E0D0}, and $\gamma$~\eqref{eq:gamma} in Fig.~\ref{fig:edh_decay} against the mass-scale separation quotient $M/(2m_\mathrm{F})$: Fig.~\ref{fig:decay_d} shows $d$, Fig.~\ref{fig:decay_power} shows $\beta_\infty/\alpha_\infty$, Fig.~\ref{fig:decay_gamma} shows $\gamma$, and Fig.~\ref{fig:decay_h0} shows $h_0/v$. In all cases, we find a clear correlation between the extrema of all decay coefficients, and where the mass ratio dips to its minimum value. 

In Fig.~\ref{fig:decay_d}, we find that the decay length scale $d$~\eqref{eq:decayexp} reaches a minimum at the most mass-scale separate state. The $\alpha$ and $\beta$ fields have shorter exponential decay scales as the mass-scale separation becomes more prominent; thus, the fields decay faster in $r$. 

In Fig.~\ref{fig:decay_power}, the asymptotic ratio of the fermion fields $\beta_\infty/\alpha_\infty$ increases at the most mass-scale separate state as $\xi$ is decreased. If this ratio is $0<\beta_\infty/\alpha_\infty<1$, then $\beta$ is subdominant to $\alpha$ in magnitude in the evanescent zone; this is expected for all particle-dominated~($\kappa>0$) states. One may surmise that the ratio $\beta_\infty/\alpha_\infty$ asymptotically approaches one with $\xi\rightarrow0$ for the maximally bound states, as this can be interpreted to be a limit wherein $\mu$ becomes very large. Thus, the mass-scale separation is associated with $\alpha^2\approx\beta^2$ in the evanescent zone. 

Fig.~\ref{fig:decay_gamma} shows that the index $\gamma$ reaches a minimum at the minimum of the ADM-to-fermion mass ratio. This essentially further increases the fermion decay rates. 

Fig.~\ref{fig:decay_h0} is a replot of Fig.~\ref{fig:mssvsh0}, but the two vertical axes are plotted now in correlation to each other. However, the strength of this plot lies in showing that the correlation between the decay coefficients and the mass-scale separation is similar to that between the fermion mass in the core and the mass-scale separation. 

From the above, it is seen that the mass-scale separation is associated with steeply decaying fermion fields. In all cases, the prominent ``beaks'', where the soliton's ADM mass $M$ drops well below the sum of its two fermion masses $m_\mathrm{F}$, occur near redshift $z\approx1.5$ for small $\xi$. 

In our view, the faster decay rates seem more of a consequence, rather than a causation, of the mass-scale separation. As seen from the asymptotic expansions and corresponding power-law relations for the decay coefficients, large $\mu$ values will clearly reduce $d$ and $\gamma$, while pushing $\beta_\infty/\alpha_\infty$ closer to unity.

\section{Conclusion}\label{sec:conc}

In the present work, we investigate the structures and behaviors of nonsingular Dirac stars in general relativity, which are Yukawa-coupled to a Higgs-like self-interacting scalar field with a quartic double-well potential. We summarize our main findings as follows: 
\begin{itemize}
    \item \textit{Equations of motion}: We present the system of equations describing an Einstein-Dirac soliton comprised of an arbitrary even number of fermions Yukawa-coupled to a Higgs field~(Sec.~\ref{sec:eom}). 
    \item \textit{Zonal structure and infinite redshift}: We establish the applicability of the terminology used to describe zonal structure of standard Einstein-Dirac solitons to their Yukawa-coupled counterparts~(Sec.~\ref{subsec:zonal_structure}), specifically in reference to the Higgs field dynamics~(Sec.~\ref{subsec:higgs_mech}). 
    \item \textit{$\kappa\geq2$ and $n\geq0$ solutions}: We present new solutions for ground-state Einstein-Dirac-Higgs solutions~(Sec.~\ref{sec:mc_dynamics}), as well as excited-state~(Sec.~\ref{sec:mc_exc}) and many-fermion~(Sec.~\ref{sec:many_fermion}) solitons. We also explain the resulting deviations of these Einstein-Dirac-Higgs soliton families from the Einstein-Dirac families obtained by Finster, Smoller, and Yau~\cite{fsy1999}. 
    \item \textit{Cause of the mass-scale separation}: We qualitatively diagnose a potential cause of a large ADM-to-fermion mass disparity present in these solutions as originating from a large difference between the interior and exterior fermion masses~(Sec.~\ref{sec:MSS1}). This is validated against the phenomenologies of both our ground-state and excited-state solutions. We also identify that this phenomenon is only seen for when the soliton is mildly relativistic. 
    \item \textit{Mass-scale separation consequences}: We illustrate several unexpected behaviors of mass-scale separate solitons~(Sec.~\ref{sec:MSS2}), such as the formation of stable and unstable photon spheres even for not highly relativistic states~(Sec.~\ref{subsec:phase}), as well as faster decay rates of fermion fields~(Sec.~\ref{subsec:decayrates}). 
\end{itemize}

To conclude, we propose further extensions of the present work.

First, while we have been able to offer a potential qualitative explanation for the origins of the mass-scale separation, a quantitative argument that analytically relates the ADM mass to the Yukawa-coupling $\mu$ is yet to be presented. For example, we have not fully justified the mass-scales diverging off into the Yukawa branch at around $\xi\approx1.2$, as discussed in Fig.~\ref{fig:mass-scale_phase}. Such an extended analysis of this phenomenon may involve a simplified toy model that patches the analytically known relations in the core zone and evanescent zone of the soliton. 

Additionally, this Einstein-Dirac-Higgs system is yet to be explored in great detail, and has many avenues for further development. For example, this may involve the inclusion of conformal coupling of the Higgs field to the geometry via a $Rh/12$ term in the matter action instead of a handpicked quartic potential, or alternatively a spinning generalization to our solutions. We detail some problems with conformally coupled solutions in Appendix~\ref{app:conformal_coupling}. 

Furthermore, a thorough dynamical time-dependent exploration of Dirac stars is warranted. These Einstein-Dirac objects have not enjoyed extensive studies in numerical relativistic contexts, unlike boson stars~\cite{bosonstar_gw, marks2026}. Works which consider the stability and temporal evolutions of Dirac stars~\cite{fsy1999, Daka_2019} are so far limited to spherically symmetric cases. Of course, this symmetry naturally does not permit the emission of gravitational waves. Therefore, a study of the inspiral, merger, and ringdown of an Einstein-Dirac(-Higgs) soliton binary may be a fruitful endeavor. Particularly, the collisions of massive boson stars can lead to the formation of black holes~\cite{bosonstar-bh}; however, as stated earlier, static black hole metrics are forbidden for Einstein-Dirac systems in spherical symmetry~\cite{fsy1999_noBHEDM, fsy2000_noRNBH, fsy2000_noKerrBH, finster2002, bernard2006} and seem to be incompatible with spinning metrics too~\cite{herdeiro2019}. In this case, what are the possibilities for the end states of a Dirac star binary merger? We hope to address this in a future work, for which we may adapt a numerical relativity routine like \textsc{ExoZvezda}~\cite{exozveda1,exozveda2,exozveda3} from the GRTL collaboration~\cite{Andrade2021}. 

It is also of interest to expand the current formulation of the gravitational localization of fundamental particles to other gravitational theories as well. One such theory is conformal Weyl gravity. Proposed by Weyl~\cite{weyl1918} and Bach~\cite{bach1921}, the theory extends conformal invariance to the gravitational sector. Because of the additionally imposed symmetry, the theory enjoys an enhanced degree of compatibility with the forces of the Standard Model of particle physics. In particular, since the theory is scale-invariant, conformal gravity inherently requires particle masses to be generated dynamically through interacting with a mass-generating Higgs-like scalar field~\cite{horne2016}. However, due to the higher-order nature of the theory and its correspondingly complex field equations, only a limited number of vacuum~\cite{mannheim1989,VarieschiFlyby,Varieschishadow,yulo2025} and nonvacuum solutions are known~\cite{mannheim1991_1, brihaye2009, verbin2011, brihaye2013, kusano2026_1, kusano2026_2}. Therefore, a full exploration of Dirac stars in conformal gravity, expanding on the work in Ref.~\cite{leggat2017_phd}, would undoubtedly yield interesting results, even for a classical field theory description of quantum gravity such as the one explored in the present work.



\appendix

\section{Data for individual solutions}\label{app:indivsol}

\begin{table*}[t!]
\begin{tabular}{|ccccccccccccc|}
\hline
Figure                             & $\kappa$   & $n$ & $\xi$  & $v$    & $z$          & $\mu$      & $\omega$  & $\lambda$ & $M$       & $\overline{R}$ & $\widetilde{\alpha}_1$ & $h_0$      \\ \hline
\ref{fig:edh_lowz}                 & $2$        & $0$ & $0.36$ & $0.11$ & $0.10047$    & $3.17420$  & $0.31178$ & $0.16322$ & $0.68488$ & $11.5692$      & $0.02831$              & $+0.08160$ \\ 
\ref{fig:edh_maxb}                 & $2$        & $0$ & $0.36$ & $0.11$ & $1.54859$    & $7.87171$  & $0.26521$ & $1.00381$ & $0.98168$ & $2.67553$      & $0.09283$              & $-0.07705$ \\ 
\ref{fig:edh_highz}                & $2$        & $0$ & $0.36$ & $0.11$ & $1183.15$    & $3.97321$  & $0.26928$ & $0.25574$ & $0.94766$ & $3.07997$      & $4.18129$              & $-0.05736$ \\ 
\ref{fig:indiv_kappa=2_n>2}        & $2$        & $2$ & $0.42$ & $0.10$ & $1.49035$    & $6.64488$  & $0.47698$ & $0.97361$ & $1.31060$ & $5.94442$      & $0.13250$              & $-0.01078$ \\
\ref{fig:indiv_kappa=2_n>2}        & $2$        & $4$ & $0.42$ & $0.10$ & $1.49018$    & $8.32209$  & $0.67657$ & $1.52712$ & $1.67928$ & $8.87166$      & $0.13250$              & $+0.01127$ \\
\ref{fig:indiv_kappa=2_n>2}        & $2$        & $8$ & $0.42$ & $0.10$ & $1.49516$    & $11.9319$  & $1.01854$ & $3.13926$ & $2.37303$ & $13.5984$      & $0.12625$              & $+0.02712$ \\
\ref{fig:edh_k>2_a}                & $10$       & $0$ & $0.65$ & $0.12$ & $1.06017$    & $3.88693$  & $0.32145$ & $0.79790$ & $4.76585$ & $12.3138$      & $0.00010$              & $+0.11075$ \\
\ref{fig:edh_k>2_b}                & $10$       & $0$ & $0.65$ & $0.12$ & $14.9453$    & $2.77813$  & $0.29701$ & $0.40761$ & $4.79906$ & $14.8768$      & $10.0000$              & $+0.09632$ \\
\ref{fig:edh_k>2_c}                & $10$       & $4$ & $0.65$ & $0.12$ & $5.64730$    & $4.04832$  & $0.42364$ & $0.86554$ & $5.68022$ & $27.3568$      & $0.17500$              & $+0.09884$ \\
\hline
\end{tabular}
\caption{\justifying Data for each of the individual Einstein-Dirac-Higgs soliton solutions shown in the present work. }
\label{tab:data}
\end{table*}

In Table~\ref{tab:data}, we summarize the initial conditions required to generate each individual localized solution presented in the work, as well as their corresponding characteristic constants.

\section{Numerical method}\label{app:num}

In this Appendix, we discuss the numerical method used to generate our Einstein-Dirac-Higgs soliton solutions. 

The large-$r$ boundary conditions enforce that $A$ asymptotes to $1$ and $T$ asymptotes to some constant $\tau$; these constraints guarantee not a \textit{normalized} solution but a \textit{normalizable} one. Once such a normalizable solution is obtained through integrating the equations of motion for the unscaled fields, we apply gauge freedom to rescale the fields such that the fermion fields are correctly normalized and $T(r)$ asymptotes to $1$. 

Unscaled fields and constants are marked with a tilde. Initializing our fields at a small but nonzero radius $\widetilde{r}_0$, for a nonsingular soliton with $\kappa=2$, we have the following small-$r$ Taylor expansions: 
\begin{equation}\label{eq:smallr_2}
    \begin{aligned}
        \widetilde{\alpha}(\widetilde{r})&=\widetilde{\alpha}_1 \widetilde{r}+...,\\[2.ex]
        \widetilde{\beta}(\widetilde{r})&=\dfrac{\widetilde{\omega}\,\widetilde{T}_0 - \widetilde{\mu}h_0}{3}\widetilde{\alpha}_1 \widetilde{r}^2+...,\\[2.ex]
        \widetilde{T}(\widetilde{r})&=\widetilde{T}_0 +\dfrac{4\pi G}{3}\widetilde{T}_0\left(\widetilde{V}(h_0)+\widetilde{\alpha}_1^2\widetilde{T}_0(\widetilde{\mu}h_0 - \widetilde{\omega}\,\widetilde{T}_0)\right)\widetilde{r}^2+...,\\[2.ex]
        \widetilde{A}(\widetilde{r})&=1-\dfrac{8\pi G}{3}\left(\widetilde{V}(h_0)+2\widetilde{\omega}\widetilde{\alpha}_1^2\widetilde{T}_0^2\right)\widetilde{r}^2+...,\\[2.ex]
        \widetilde{h}(\widetilde{r})&=h_0+\dfrac{2\widetilde{\lambda} h_0(h_0^2-v^2)+\widetilde{\mu}\widetilde{\alpha}_1^2\widetilde{T}_0}{3}\widetilde{r}^2 + ...,
    \end{aligned}
\end{equation}
where $\widetilde{T}(\widetilde{r}=0)=\widetilde{T}_0$. $h(r)$ is only sensitive to a radial rescaling, so $\widetilde{h}(\widetilde{r}_0)=\widetilde{h}_0=h_0$ and $\widetilde{v}=v$.

For $\kappa>2$, the fermion contributions at small $r$ are negligible compared to those from the Higgs field, such that the next-to leading order for $\widetilde{T}$, $\widetilde{A}$, and $\widetilde{h}$ is $\widetilde{r}^2$: 
\begin{equation}\label{eq:smallr>2}
    \begin{aligned}
        \widetilde{\alpha}(\widetilde{r})&=\widetilde{\alpha}_1 \widetilde{r}^{\kappa/2}+...,\\
        \widetilde{\beta}(\widetilde{r})&=\dfrac{\widetilde{\omega}\widetilde{T}_0-\widetilde{\mu}h_0}{\kappa+1}\widetilde{\alpha}_1 \widetilde{r}^{\kappa/2+1}+...,\\
        \widetilde{T}(\widetilde{r})&=\widetilde{T}_0+\dfrac{4\pi G}{3}\widetilde{T}_0\widetilde{V}(h_0)\widetilde{r}^2+...,\\
        \widetilde{A}(\widetilde{r})&=1-\dfrac{8\pi G}{3}\widetilde{V}(h_0)\widetilde{r}^2+...,\\
        \widetilde{h}(\widetilde{r})&=h_0+\dfrac{2}{3}\widetilde{\lambda}h_0(h_0^2-v^2)\widetilde{r}^2+...
    \end{aligned}
\end{equation}
We have eight free parameters: $\widetilde{\alpha}_1$, $\widetilde{\mu}$, $\widetilde{\lambda}$, ${v}$, $\widetilde{\omega}$, and ${h}_0$, on top of the relativistic quantum number $\kappa$ and excited-state number $n$. 

We initialize our unscaled fields $\widetilde{\alpha}$, $\widetilde{\beta}$, $\widetilde{T}$, $\widetilde{A}$, $\widetilde{h}$, and $\widetilde{h}'$ at a small radius $\widetilde{r}_0=10^{-4}$ using the small-$r$ expansions. Having further defined $\widetilde{T}_0=1$, $\widetilde{\mu}=1/v$ and $\widetilde{\lambda}$ through~\eqref{eq:xi}, we integrate the six field equations over $\ln{\widetilde{r}}$ up to a maximum radius of integration $\widetilde{r}_\mathrm{max}$ using the equations of motion~\eqref{eq:Dirac_EDH},~\eqref{eq:Einstein_EDH}, and~\eqref{eq:Higgs_EDH}. 

We remark here that in the FSY ED system~\cite{fsy1999}, only $\widetilde{\omega}$ needs to be found as an eigenvalue. However, in the EDH system, we also need to finely tune $h(r=0)=h_0$ so that the Higgs field $h$ comes to rest exactly at the VEV $v$ at $r\rightarrow\infty$. 

As we have two separate eigenvalues to look for~($\widetilde{\omega}$ and $h_0$), we apply a two-parameter sequential shooting method to obtain solutions. Specifically, for a trial value of $h_0$, we effectively binary chop in $\widetilde{\omega}$ for normalizable fermion fields, just as in the standard Einstein-Dirac system~\cite{leith2021}. We do this by checking the number of nodes/zeros encountered in $\widetilde{\alpha}$ and $\widetilde{\beta}$. Increasing $\widetilde{\omega}$ increases the anticlockwise circulation in the $\widetilde{\alpha}$-$\widetilde{\beta}$ plane, starting in the first quadrant. Hence, if too many nodes are found in $\widetilde{\beta}$ we increase $\widetilde{\omega}$ by a stepsize $\Delta \widetilde{\omega}$, while we decrease if too many are found in $\widetilde{\alpha}$. We also increase $\widetilde{\omega}$ if the correct number of nodes is achieved until too many nodes are found in $\widetilde{\alpha}$. If the direction of shooting changes, we halve $\Delta \widetilde{\omega}$. This routine is iterated until $\Delta\widetilde{\omega}$ is small enough to be negligible to double precision ($\Delta\widetilde{\omega}/\widetilde{\omega}<10^{-16}$), at which point we terminate the shooting routine. 

Then, we check the large-$r$ value of $\widetilde{h}(\widetilde{r})$. If $\widetilde{h}(\widetilde{r}_\mathrm{max})>v$, we decrease $h_0$ by $\Delta h_0$, while we do the opposite if $\widetilde{h}(\widetilde{r}_\mathrm{max})<v$, with the halving logic also adopted from the shooting procedure in $\widetilde{\omega}$ from the inner loop. Once $\widetilde{h}(\widetilde{r}_\mathrm{max})\approx v$, we stop the integration as we have obtained a normalizable solution. We do this for \textit{all} values of $\xi$; Ref.~\cite{leith2023} reverses the shooting order for $\xi>2$ such that an $h_0$ value is first found for each trial $\widetilde{\omega}$, but we refrain from this since we find this approach to be numerically unstable. This may have resulted in our difficulty to generate low-redshift (excited state) solutions consistently. 

The present system can suffer from multivaluedness for small $\xi$, wherein multiple $h_0$ values become viable for a single $\widetilde{\alpha}_1$. While we refrain from generating such solutions due to their increased numerical complications, the desired phenomena discussed in the present work are nevertheless observable even for states without this multivaluedness.

However, numerically evaluating these fields does not by itself generate a physical solution.
As stated, the Dirac fields must be normalized, $T$ and $A$ must asymptote to Minkowski, and the Higgs field must asymptote to its VEV $v$. 

Because of this, we find the two constants $\tau$ and $\chi$:
\begin{equation}\label{eq:norm_consts}
\begin{aligned}
    \tau &= \lim_{\widetilde{r}\rightarrow\infty} \widetilde{T}(\widetilde{r})<\infty, \\
    \chi^2&=4\pi\int^\infty_0\dfrac{\widetilde{T}}{\sqrt{\widetilde{A}}}({\widetilde{\alpha}}^2+{\widetilde{\beta}}^2)\mathrm{d}\widetilde{r}<\infty,
\end{aligned}
\end{equation}
with $\tau$ and $\chi$ respectively corresponding to the temporal and radial rescaling constants for our system. Numerically, it is easier to find $\tau$ by noting that ${T}$ and ${A}$ match onto the Schwarzschild metric at large $\widetilde{r}$ for a localized particle, whereupon $-g_{tt}=1/g_{rr}$. At a maximum radius of integration $\widetilde{r}=\widetilde{r}_\mathrm{max}$, $\tau$ is~\cite{bakuczcanario2020}
\begin{equation}\label{eq:tau_alt}
    \left.\tau=\widetilde{T}(\widetilde{r})\sqrt{\widetilde{A}(\widetilde{r})}\,\right\rvert_{\widetilde{r}=\widetilde{r}_\mathrm{max}},
\end{equation}
where we also ensure that $\widetilde{r}_\mathrm{max}$ is large enough to make an extrapolation to infinity not necessary. 

Having obtained $\tau$ and $\chi$ through~\eqref{eq:norm_consts} and~\eqref{eq:tau_alt}, we scale our solution anisotropically in time and radius:
\begin{equation}\label{eq:rescalings_fields}
    \begin{gathered}
        \alpha(r) \equiv \sqrt{\dfrac{\tau}{\chi}}\widetilde{\alpha}(\chi r),\hspace{0.5cm}
        \beta(r) \equiv  \sqrt{\dfrac{\tau}{\chi}}\widetilde{\beta}(\chi r),\\[2.5ex]
        A(r) \equiv \widetilde{A}(\chi r) ,\hspace{0.5cm}
        T(r) \equiv \dfrac{1}{\tau}\widetilde{T}(\chi r) ,\\[2.5ex]
        h(r) \equiv \widetilde{h}(\chi r),\hspace{0.5cm}
        h'(r) \equiv \chi h'(\chi r).
    \end{gathered}
\end{equation} 
and therefore naturally $r=\widetilde{r}/\chi$. While $\widetilde{T}(r\rightarrow0)=1$, using the temporal rescaling by $\tau$ stretches $T$ such that $T(r\rightarrow0)=1+z$~\eqref{eq:redshift_def}. 

The characteristic constants of our system must also be rescaled as follows: 
\begin{equation}\label{eq:rescalings_consts}
    \omega \equiv \chi\tau\widetilde{\omega}, \quad
    \mu \equiv \chi\widetilde{\mu}, \quad
    \lambda \equiv \chi^2\widetilde{\lambda}.
\end{equation}

\section{Einstein-Dirac toy model}\label{app:toy_model}

\begin{figure*}
    \centering
    \includegraphics[width=0.99\linewidth]{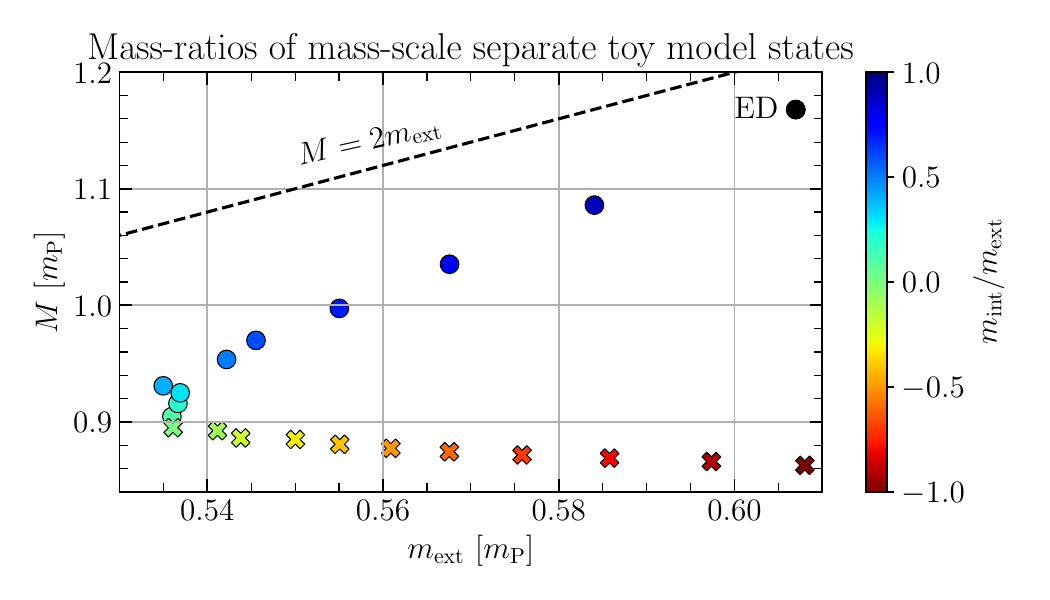}
    \caption{\justifying Mass-scales of maximally-bound states from the toy Einstein-Dirac model. Just as in Fig.~\ref{fig:mass-scale_phase}, circles and crosses denote where $m_\mathrm{int}$ is positive and negative, respectively. $m_\mathrm{int}$ and $m_\mathrm{ext}$ denote interior and exterior masses before and after the jump, respectively.}
    \label{fig:mass-scale_phase_toy}
\end{figure*}

In this appendix, we detail the toy model which is used to recreate the mass-scale separation in the absence of a mass-generating scalar Higgs field, which was briefly mentioned in Sec.~\ref{sec:MSS1}.

We first reduce the Einstein-Dirac-Higgs model of the present work to the standard Einstein-Dirac system. This requires $h=v$, $h',\,h''\rightarrow0$, and $\lambda\rightarrow\infty$. 

As a result, we obtain the equations~\cite{leith2020, leith2021}:
\begin{equation}
    \begin{aligned}
    \sqrt{A}\alpha'&=+\dfrac{\kappa\alpha}{2r}-(\omega T+m_\mathrm{F})\beta;\\
        \sqrt{A}\beta'&=-\dfrac{\kappa\beta}{2r}+(\omega T-m_\mathrm{F})\alpha;\\
        -1+A+rA'&=-8\pi G \lvert\kappa\rvert\omega T^2(\alpha^2+\beta^2);\\
        -1+A-\dfrac{2rAT'}{T}&=8\pi G \lvert\kappa\rvert T\sqrt{A}(\alpha\beta'-\alpha'\beta).
    \end{aligned}
\end{equation}
We further modify the mass terms in the Dirac equations, such that 
\begin{equation}
    \begin{aligned}
        m_\mathrm{F}&=m_\mathrm{int},\quad \omega\,T(r)>m_\mathrm{ext},\\
        m_\mathrm{F}&=m_\mathrm{ext},\quad \omega\,T(r)<m_\mathrm{ext}.
    \end{aligned}
\end{equation}
Through this manoeuvre, we manually force the emergence of the evanescent zone by forcing the mass to jump from an internal mass $m_\mathrm{int}$ to an external/evanescent $m_\mathrm{ext}$ at where the evanescent zone would start for the usual FSY Einstein-Dirac solution. In some capacity, this is a relatively crude approximation of the radial profile of the Higgs field. 

As seen in Fig.~\ref{fig:mass-scale_phase_toy}, despite the approximative approach taken in the present toy model, it manages to recreate the mass-scale separation exhibited in Einstein-Dirac-Higgs solitons to a reasonable degree. From this toy model, it is clear that the fermion mass disparity between the interior and exterior of the extended soliton distribution is contributing to the mass-scale separation.

\section{The trouble with conformal coupling}\label{app:conformal_coupling}

We highlight the difficulties in generating particlelike solutions wherein the Higgs field is conformally coupled to the curvature through a $R h /12$ term in $\mathcal{L}_\mathrm{m}$~\eqref{eq:L_mat}. 

Because intrinsic scales are forbidden in conformal systems, $v=0$. Therefore, in asymptotically flat spacetimes, the Ricci scalar $R=0$. 

This means that the Higgs potential becomes a quartic without the double well shape, leading to the asymptotic masses of the constituent fermions being zero. In fact, at no point in the radial profile of the soliton will these fermions have positive masses, which is required for asymptotic exponential decay of fermion fields as can be seen from the Dirac equations. It is also debatable whether one can even refer to such a scalar field as ``Higgs'' in asymptotically flat spacetimes since the scalar field potential $V(h)$ is no longer a double-well.

In asymptotically flat spacetimes, our numerical calculations have suggested that the spinor fields do not decay due to the lack of a positive fermion mass. As Ref.~\cite{bakuczcanario2020} claims, a positive fermion mass is required to drive down the fermion fields in the evanescent zone.

\nocite{*}

\bibliography{EDH}

@article{bakuczcanario2020,
   title={{Infinite-redshift localized states of Dirac fermions under Einsteinian gravity}},
   volume={102},
   DOI={10.1103/physrevd.102.084049},
   number={8},
   pages={084049},
   journal={Phys. Rev. D},
   publisher={American Physical Society (APS)},
   author={Bakucz Canário, Daniel and Lloyd, Sam and Horne, Keith and Hooley, Chris A.},
   year={2020},
   month=Oct }

@article{leith2020,
   title={{Fermion self-trapping in the optical geometry of Einstein-Dirac solitons}},
   volume={101},
   doi={10.1103/physrevd.101.106012},
   number={10},
   pages={106012},
   journal={Phys. Rev. D},
   author={Leith, Peter E. D. and Hooley, Chris A. and Horne, Keith and Dritschel, David G.},
   year={2020},
   month=May }

@article{leith2021,
   title={{Nonlinear effects in the excited states of many-fermion Einstein-Dirac solitons}},
   volume={104},
   doi={10.1103/physrevd.104.046024},
   number={4},
    pages={046024},
   journal={Phys. Rev. D},
   publisher={American Physical Society (APS)},
   author={Leith, Peter E. D. and Hooley, Chris A. and Horne, Keith and Dritschel, David G.},
   year={2021},
   month=Aug }

@article{leith2023,
   title={{Gravitationally localized states of two neutral fermions interacting with a Higgs field}},
   volume={107},
   doi={10.1103/physrevd.107.106020},
   number={10},
   pages={106020},
   journal={Phys. Rev. D},
   publisher={American Physical Society (APS)},
   author={Leith, Peter E. D. and Leggat, Alasdair Dorkenoo and Hooley, Chris A. and Horne, Keith and Dritschel, David G.},
   year={2023},
   month=May }

@article{leith2026,
  title = {Stealth singularities from self-gravitating fermions},
  author = {Leith, Peter E. D. and Hooley, Chris A. and Horne, Keith and Dritschel, David G.},
  journal = {Phys. Rev. D},
  volume = {113},
  issue = {12},
  pages = {L121507},
  numpages = {6},
  year = {2026},
  month = {Jun},
  publisher = {American Physical Society},
  doi = {10.1103/81mh-mdt5},
  url = {https://link.aps.org/doi/10.1103/81mh-mdt5}
}

@phdthesis{leith2022_phd,
    author = "Leith, Peter Edward Duncan",
    title = "{{Dirac solitons in general relativity: many-fermion states, singular solutions {\&} the inclusion of a Higgs mechanism}}",
    doi = "10.17630/sta/234",
    school = "University of St. Andrews",
    year = "2022"
}

@phdthesis{leggat2017_phd,
    author = "Dorkenoo Leggat, Alasdair",
    title = "{Dirac solitons in general relativity and conformal gravity}",
    school = "University of St. Andrews",
    year = "2017"
}

@article{bartnikmckinnon1988,
  title = "{Particlelike solutions of the Einstein-Yang-Mills equations}",
  author = {Bartnik, Robert and McKinnon, John},
  journal = {Phys. Rev. Lett.},
  volume = {61},
  issue = {2},
  pages = {141--144},
  numpages = {0},
  year = {1988},
  month = {Jul},
  publisher = {American Physical Society},
  doi = {10.1103/PhysRevLett.61.141},
  url = {https://link.aps.org/doi/10.1103/PhysRevLett.61.141}
}

@misc{finster2002,
      title="{Non-existence of black hole solutions to static, spherically symmetric Einstein-Dirac systems - A critical discussion}", 
      author={Felix Finster and Joel Smoller and Shing-Tung Yau},
      year={2002},
      eprint={gr-qc/0211043},
      archivePrefix={arXiv},
      primaryClass={gr-qc},
}

@article{bernard2006,
doi = {10.1088/0264-9381/23/13/009},
year = {2006},
month = {jun},
volume = {23},
number = {13},
pages = {4433},
author = {Bernard, Yann},
title = {{Non-existence of black-hole solutions for the electroweak Einstein–Dirac–Yang/Mills equations}},
journal = {Class. Quantum Grav.}
}

@article{berti2024,
   title={{Tidal Love numbers and approximate universal relations for fermion soliton stars}},
   volume={109},
   DOI={10.1103/physrevd.109.124008},
   number={12},
   journal={Phys. Rev. D},
   publisher={American Physical Society (APS)},
   author={Berti, Emanuele and De Luca, Valerio and Del Grosso, Loris and Pani, Paolo},
   year={2024},
   pages={124008},}

@article{blazquezsalcedo2020,
   title="{Constructing spherically symmetric {Einstein–Dirac} systems with multiple spinors: {Ansatz}, wormholes and other analytical solutions}",
   volume={80},
   ISSN={1434-6052},
   url={http://dx.doi.org/10.1140/epjc/s10052-020-7706-3},
   DOI={10.1140/epjc/s10052-020-7706-3},
   number={2},
   pages={174},
   journal={Eur. Phys. J. C},
   publisher={Springer Science and Business Media LLC},
   author={Blázquez-Salcedo, Jose Luis and Knoll, Christian},
   year={2020},}

@article{cunha2025,
doi = {10.1088/1475-7516/2025/05/083},
url = {https://doi.org/10.1088/1475-7516/2025/05/083},
year = {2025},
volume = {2025},
number = {05},
pages = {083},
author = {Cunha, Pedro V.P.},
title = "{Backreaction of perturbations around a stable light ring}",
journal = {J. Cosmol. Astropart. Phys.}
}

@article{difilippo2025,
  title = "{Can light-rings self-gravitate?}",
  author = {Di Filippo, Francesco and Rezzolla, Luciano},
  journal = {Phys. Rev. D},
  volume = {111},
  issue = {2},
  pages = {L021504},
  numpages = {6},
  year = {2025},
  month = {Jan},
  publisher = {American Physical Society},
  doi = {10.1103/PhysRevD.111.L021504},
  url = {https://link.aps.org/doi/10.1103/PhysRevD.111.L021504}
}

@article{fsy1999,
  title = {{Particlelike solutions of the Einstein-Dirac equations}},
  author = {Finster, Felix and Smoller, Joel and Yau, Shing-Tung},
  journal = {Phys. Rev. D},
  volume = {59},
  number = {10},
  pages = {104020},
  numpages = {19},
  year = {1999},
  month = {Apr},
  publisher = {American Physical Society},
  doi = {10.1103/PhysRevD.59.104020},
}

@article{fsy1999_EDM,
title = {{Particle-like solutions of the Einstein–Dirac–Maxwell equations}},
journal = {Phys. Lett. A},
volume = {259},
number = {6},
pages = {431-436},
year = {1999},
doi = {https://doi.org/10.1016/S0375-9601(99)00457-0},
author = {Felix Finster and Joel Smoller and Shing-Tung Yau}
}

@article{fsy2000_noKerrBH,
author = {Finster, Felix and Kamran, Niky and Smoller, Joel and Yau, Shing-Tung},
title = {{Nonexistence of time-periodic solutions of the Dirac equation in an axisymmetric black hole geometry}},
journal = {Commun. Pure Appl. Math.},
volume = {53},
number = {7},
pages = {902-929},
doi = {https://doi.org/10.1002/(SICI)1097-0312(200007)53:7<902::AID-CPA4>3.0.CO;2-4},
year = {2000}
}

@article{fsy1999_noBHEDM,
   title={{Non-existence of black hole solutions for a spherically symmetric, static Einstein-Dirac-Maxwell system}},
   volume={205},
   DOI={10.1007/s002200050675},
   number={2},
   journal={Comm. Math. Phys.},
   publisher={Springer Science and Business Media LLC},
   author={Finster, Felix and Smoller, Joel and Yau, Shing-Tung},
   year={1999},
   month=Aug, pages={249–262} }

@article{fsy2000_noRNBH,
   title={{Non-existence of time-periodic solutions of the Dirac equation in a Reissner-Nordström black hole background}},
   volume={41},
   DOI={10.1063/1.533234},
   number={4},
   journal={J. Math. Phys.},
   publisher={AIP Publishing},
   author={Finster, Felix and Smoller, Joel and Yau, Shing-Tung},
   year={2000},
   month=Apr, pages={2173–2194} }

@article{herdeiro2017,
title = "{Asymptotically flat scalar, Dirac and Proca stars: Discrete vs. continuous families of solutions}",
journal = {Phys. Lett. B},
volume = {773},
pages = {654-662},
year = {2017},
issn = {0370-2693},
doi = {https://doi.org/10.1016/j.physletb.2017.09.036},
url = {https://www.sciencedirect.com/science/article/pii/S0370269317307451},
author = {Carlos A.R. Herdeiro and Alexandre M. Pombo and Eugen Radu}
}

@article{herdeiro2019,
title = {{Asymptotically flat spinning scalar, Dirac and Proca stars}},
journal = {Phys. Lett. B},
volume = {797},
pages = {134845},
year = {2019},
doi = {https://doi.org/10.1016/j.physletb.2019.134845},
author = {C. Herdeiro and I. Perapechka and E. Radu and Ya. Shnir},
}

@article{herdeiro2020,
    author = "Herdeiro, Carlos A. R. and Radu, Eugen",
    title = "{{Asymptotically flat, spherical, self-interacting scalar, Dirac and Proca stars}}",
    doi = "10.3390/sym12122032",
    journal = "Symmetry",
    volume = "12",
    number = "12",
    pages = "2032",
    year = "2020"
}

@article{herdeiro2022,
    author = "Herdeiro, Carlos and Perapechka, Ilya and Radu, Eugen and Shnir, Ya.",
    title = "{{Spinning gauged boson and Dirac stars: A comparative study}}",
    doi = "10.1016/j.physletb.2021.136811",
    journal = "Phys. Lett. B",
    volume = "824",
    pages = "136811",
    year = "2022"
}

@article{iwasawa2025,
  title = {{Multishell Dirac fermions in the Einstein-Dirac system}},
  author = {Iwasawa, Shunsuke and Sawado, Nobuyuki},
  journal = {Phys. Rev. D},
  volume = {112},
  number = {6},
  pages = {064046},
  numpages = {17},
  year = {2025},
  month = {Sep},
  publisher = {American Physical Society},
  doi = {10.1103/3btt-5621},
}

@article{kain2023,
   title={{Einstein-Dirac system in semiclassical gravity}},
   volume={107},
   DOI={10.1103/physrevd.107.124001},
   number={12},
   journal={Phys. Rev. D},
   publisher={American Physical Society (APS)},
   author={Kain, Ben},
   year={2023},
   pages={124001},}

@article{karlovini2001,
doi = {10.1088/0264-9381/18/5/305},
url = {https://doi.org/10.1088/0264-9381/18/5/305},
year = {2001},
month = {mar},
volume = {18},
number = {5},
pages = {817},
author = {Max Karlovini and Kjell Rosquist and Lars Samuelsson},
title = {Constructing stellar objects with multiple necks},
journal = {Class. Quantum Grav.}
}

@article{karlovini2002,
   title={"Ultracompact stars with multiple necks"},
   volume={17},
   ISSN={1793-6632},
   url={http://dx.doi.org/10.1142/S0217732302006400},
   DOI={10.1142/s0217732302006400},
   number={04},
   journal={Mod. Phys. Lett. A},
   publisher={World Scientific Pub Co Pte Lt},
   author={Karlovini, Max and Rosquist, Kjell and Samuelsson, Lars},
   year={2002},
   month=Feb, pages={197–203} }

@article{liang2023,
   title={Dirac-boson stars},
   volume={2023},
   DOI={10.1007/jhep02(2023)249},
   pages={249},
   number={2},
   journal={J. High Energy Phys.},
   publisher={Springer Science and Business Media LLC},
   author={Liang, Chen and Ren, Ji-Rong and Sun, Shi-Xian and Wang, Yong-Qiang},
   year={2023},
   month=Feb }

@article{wheeler1955,
  title = {Geons},
  author = {Wheeler, John Archibald},
  journal = {Phys. Rev.},
  volume = {97},
  issue = {2},
  pages = {511--536},
  numpages = {0},
  year = {1955},
  month = {Jan},
  publisher = {American Physical Society},
  doi = {10.1103/PhysRev.97.511},
  url = {https://link.aps.org/doi/10.1103/PhysRev.97.511}
}

@article{kaup1968,
  title = {Klein-{G}ordon Geon},
  author = {Kaup, David J.},
  journal = {Phys. Rev.},
  volume = {172},
  issue = {5},
  pages = {1331--1342},
  numpages = {0},
  year = {1968},
  month = {Aug},
  publisher = {American Physical Society},
  doi = {10.1103/PhysRev.172.1331},
  url = {https://link.aps.org/doi/10.1103/PhysRev.172.1331}
}

@article{bosonstar_gw,
  title = "{Gravitational-wave data analysis with high-precision numerical relativity simulations of boson star mergers}",
  author = {Evstafyeva, Tamara and Sperhake, Ulrich and Romero-Shaw, Isobel M. and Agathos, Michalis},
  journal = {Phys. Rev. Lett.},
  volume = {133},
  issue = {13},
  pages = {131401},
  numpages = {7},
  year = {2024},
  month = {Sep},
  publisher = {American Physical Society},
  doi = {10.1103/PhysRevLett.133.131401},
  url = {https://link.aps.org/doi/10.1103/PhysRevLett.133.131401}
}

@misc{marks2026,
    author = "Marks, Gareth Arturo and Staelens, Seppe J. and Sperhake, Ulrich",
    title = "{Black hole-boson star binaries: Gravitational wave signals and tidal disruption}",
    eprint = "2604.06312",
    archivePrefix = "arXiv",
    primaryClass = "gr-qc",
    month = "4",
    year = "2026"
}

@article{sun2024,
    author = "Sun, Shi-Xian and Cui, Si-Yuan and Huang, Long-Xing and Fang, Tie-Feng and Wang, Yong-Qiang",
    title = "{$\kappa$-Dirac stars}",
    journal = "Eur. Phys. J. C",
    year = "2024",
    volume="84",
    pages="699",
    issue="7",
    DOI="10.1140/epjc/s10052-024-12942-z"
}

@ARTICLE{bach1921,
   author       = "Bach, R.", 
   title        = "Zur {W}eylschen {R}elativitätstheorie und der {W}eylschen {E}rweiterung des {K}rümmungstensorbegriffs. ", 
   journal      = "Math. Z.", 
   volume       = "9", 
   pages        = "110-135", 
   year         = "1921", 
   doi          = "10.1007/BF01378338",
   doilink      = "https://doi.org/10.1007/BF01378338"
}

@ARTICLE{weyl1918,
   author       = "H. Weyl", 
   title        = "Reine {I}nfinitesimalgeometrie. ", 
   journal      = "Math. Z.", 
   volume       = "2", 
   pages        = "384--411", 
   year         = "1918", 
   doi          = "10.1007/BF01199420",
   doilink      = "https://doi.org/10.1007/BF01199420"
}

@ARTICLE{mannheim1989,
   author       = "Mannheim, P. D. and Kazanas, D.", 
   title        = "Exact vacuum solution to conformal {W}eyl gravity and galactic rotation curves. ", 
   journal      = "Astrophys. J.", 
   volume       = "342", 
   pages        = "635-638", 
   year         = "1989", 
   doi          = "10.1086/167623",
   doilink      = "https://ui.adsabs.harvard.edu/link_gateway/1989ApJ...342..635M/doi:10.1086/167623"
}

@ARTICLE{mannheim1991_1,
   author       = "Mannheim, P. D. and Kazanas, D.", 
   title        = "Solutions to the {R}eissner-{N}ordstrom, {K}err, and {K}err-{N}ewman problems in fourth-order conformal {W}eyl gravity. ", 
   journal      = "Phys. Rev. D", 
   volume       = "44", 
   issue        = "2",
   pages        = "417-423", 
   year         = "1991", 
   doi          = "10.1103/PhysRevD.44.417",
   doilink      = "https://doi.org/10.1103/PhysRevD.44.417?_gl=1*agefc6*_gcl_au*MjI0NjgyMjM5LjE3MzEwOTE1MTE.*_ga*MTc2MTkzNTIzMy4xNzIzMzEyNDc1*_ga_ZS5V2B2DR1*MTczNDk2NDE1My4xOS4xLjE3MzQ5NjYxMTYuNjAuMC4xNzcxMTg3OTA."
}

@ARTICLE{brihaye2009,
   author       = "Brihaye, Y. and Verbin, Y.", 
   title        = "Spherical structures in conformal gravity and its scalar-tensor extension. ", 
   journal      = "Phys. Rev. D", 
   volume       = "80", 
   issue        = "12",
   pages        = "124048", 
   year         = "2009", 
   doi          = "10.1103/PhysRevD.80.124048",
   doilink      = "https://doi.org/10.1103/PhysRevD.80.124048?_gl=1*a4myd5*_gcl_au*MjI0NjgyMjM5LjE3MzEwOTE1MTE.*_ga*MTc2MTkzNTIzMy4xNzIzMzEyNDc1*_ga_ZS5V2B2DR1*MTczNDk5NDU5Mi4yMy4xLjE3MzQ5OTYwNDYuNjAuMC4xODA2MzY5OTc1"
}

@article{verbin2011,
   title={Exact string-like solutions in conformal gravity},
   volume={43},
   doilink={http://dx.doi.org/10.1007/s10714-011-1209-3},
   doi={10.1007/s10714-011-1209-3},
   issue={10},
   journal={Gen. Rel. Grav.},
   author={Verbin, Y. and Brihaye, Y.},
   year={2011},
   month=jun, pages={2847–2863}}

@article{brihaye2013,
   title={{AdS} solitons with conformal scalar hair},
   volume={88},
   pages={104006},
   url={http://dx.doi.org/10.1103/PhysRevD.88.104006},
   DOI={10.1103/physrevd.88.104006},
   number={10},
   journal={Phys. Rev. D},
   publisher={American Physical Society (APS)},
   author={Brihaye, Yves and Hartmann, Betti and Tojiev, Sardor},
   year={2013},
   month=Nov }

@article{kusano2026_1,
doi = {10.1088/1361-6382/ae30c8},
url = {https://doi.org/10.1088/1361-6382/ae30c8},
year = {2026},
month = {feb},
publisher = {IOP Publishing},
volume = {43},
number = {3},
pages = {035008},
author = {Kusano, Reinosuke and Yulo Asuncion, Miguel and Horne, Keith},
title = "{Charged black holes in Weyl conformal gravity}",
journal = {Class. Quantum Grav.}
}

@article{kusano2026_2,
doi = {10.1088/1361-6382/ae415f},
url = {https://doi.org/10.1088/1361-6382/ae415f},
year = {2026},
month = {mar},
publisher = {IOP Publishing},
volume = {43},
number = {5},
pages = {055002},
author = {Kusano, Reinosuke and Horne, Keith and Koenig, Friedrich and Yulo Asuncion, Miguel},
title = "{Backreactions from loading the stable photon sphere in Weyl conformal gravity}",
journal = {Class. Quantum Grav.},
}

@article{Daka_2019,
   title="{Perturbing the ground state of Dirac stars}",
   volume={100},
   DOI={10.1103/physrevd.100.084042},
   number={8},
   journal={Phys. Rev. D},
   publisher={American Physical Society (APS)},
   author={Daka, Emanuel and Phan, Nhon N. and Kain, Ben},
   year={2019},
   month=Oct }

@article{exozveda1,
    author = "Helfer, Thomas and Sperhake, Ulrich and Croft, Robin and Radia, Miren and Ge, Bo-Xuan and Lim, Eugene A.",
    title = "{Malaise and remedy of binary boson-star initial data}",
    doi = "10.1088/1361-6382/ac53b7",
    journal = "Class. Quant. Grav.",
    volume = "39",
    number = "7",
    pages = "074001",
    year = "2022"
}

@article{exozveda2,
    author = "Evstafyeva, Tamara and Sperhake, Ulrich and Helfer, Thomas and Croft, Robin and Radia, Miren and Ge, Bo-Xuan and Lim, Eugene A.",
    title = "{Unequal-mass boson-star binaries: Initial data and merger dynamics}",
    doi = "10.1088/1361-6382/acc2a8",
    journal = "Class. Quant. Grav.",
    volume = "40",
    number = "8",
    pages = "085009",
    year = "2023"
}

@article{exozveda3,
   title={Local continuity of angular momentum and {N}oether charge for matter in general relativity},
   volume={40},
   DOI={10.1088/1361-6382/accc6a},
   number={10},
   journal={Class. Quantum Grav.},
   author={Croft, Robin},
   year={2023},
   month=Apr, pages={105007} }

@article{Andrade2021,
  doi = {10.21105/joss.03703},
  url = {https://doi.org/10.21105/joss.03703},
  year = {2021},
  publisher = {The Open Journal},
  volume = {6},
  number = {68},
  pages = {3703},
  author = {Tomas Andrade and Llibert Areste Salo and Josu C. Aurrekoetxea and Jamie Bamber and Katy Clough and Robin Croft and Eloy de Jong and Amelia Drew and Alejandro Duran and Pedro G. Ferreira and Pau Figueras and Hal Finkel and Tiago Fran\c{c}a and Bo-Xuan Ge and Chenxia Gu and Thomas Helfer and Juha Jäykkä and Cristian Joana and Markus Kunesch and Kacper Kornet and Eugene A. Lim and Francesco Muia and Zainab Nazari and Miren Radia and Justin Ripley and Paul Shellard and Ulrich Sperhake and Dina Traykova and Saran Tunyasuvunakool and Zipeng Wang and James Y. Widdicombe and Kaze Wong},
  title = "{$\textsc{GRChombo}$: An adaptable numerical relativity code for fundamental physics}",
  journal = {J. Open Source Softw.}
}

@article{brito2016,
   title="{Proca stars: Gravitating Bose–Einstein condensates of massive spin 1 particles}",
   volume={752},
   ISSN={0370-2693},
   url={http://dx.doi.org/10.1016/j.physletb.2015.11.051},
   DOI={10.1016/j.physletb.2015.11.051},
   journal={Phys. Lett. B},
   publisher={Elsevier BV},
   author={Brito, Richard and Cardoso, Vitor and Herdeiro, Carlos A.R. and Radu, Eugen},
   year={2016},
   month=Jan, pages={291–295} }

@article{cardoso2014,
  title = "{Light rings as observational evidence for event horizons: Long-lived modes, ergoregions and nonlinear instabilities of ultracompact objects}",
  author = {Cardoso, Vitor and Crispino, Lu\'{\i}s C. B. and Macedo, Caio F. B. and Okawa, Hirotada and Pani, Paolo},
  journal = {Phys. Rev. D},
  volume = {90},
  issue = {4},
  pages = {044069},
  numpages = {10},
  year = {2014},
  month = {Aug},
  publisher = {American Physical Society},
  doi = {10.1103/PhysRevD.90.044069},
  url = {https://link.aps.org/doi/10.1103/PhysRevD.90.044069}
}

@article{yulo2025,
doi = {10.1088/1361-6382/ae1109},
year = {2025},
volume = {42},
issue = {21},
pages = {215005},
author = {Yulo Asuncion, Miguel and Horne, Keith and Kusano, Reinosuke and Dominik, Martin},
title = "{Structure of {Kerr} black hole spacetimes in {Weyl} conformal gravity}",
journal = {Class. Quantum Grav.},
}

@article{bhusal2026,
doi = {10.1088/1475-7516/2026/06/064},
year = {2026},
volume = {2026},
number = {06},
pages = {064},
author = {Bhusal, Nabeen and M. Chávez, Ernesto and G. Garcia, Marcos A. and G. Menkara, Adriana and Pierre, Mathias},
title = "{Fermion (non)reheating with a quartic inflaton potential}",
journal = {	J. Cosmol. Astropart. Phys.},}

@article{bosonstar-bh,
  title = "{Head-on collisions of $\ensuremath{\ell}$-boson stars}",
  author = {Jaramillo, V\'{\i}ctor and Sanchis-Gual, Nicolas and Barranco, Juan and Bernal, Argelia and Degollado, Juan Carlos and Herdeiro, Carlos and Megevand, Miguel and N\'u\~nez, Dar\'{\i}o},
  journal = {Phys. Rev. D},
  volume = {105},
  issue = {10},
  pages = {104057},
  numpages = {23},
  year = {2022},
  month = {May},
  publisher = {American Physical Society},
  doi = {10.1103/PhysRevD.105.104057},
  url = {https://link.aps.org/doi/10.1103/PhysRevD.105.104057}
}

@article{dafermos2003,
   title="{On ``time-periodic'’ black-hole solutions to certain spherically symmetric Einstein-matter systems}",
   volume={238},
   DOI={10.1007/s00220-003-0870-0},
   number={3},
   journal={Comm. Math. Phys.},
   author={Dafermos, Mihalis},
   year={2003},
   pages={411–427}}

@ARTICLE{horne2016,
       author = {{Horne}, K.},
        title = "{Conformal Gravity rotation curves with a conformal {Higgs} halo}",
      journal = {Mon. Not. R. Astron. Soc.},
         year = 2016,
        month = jun,
       volume = {458},
       issue = {4},
        pages = {4122-4128},
          doi = {10.1093/mnras/stw506},
archivePrefix = {arXiv}
}

@article{freese2018,
   title="{The Higgs boson can delay reheating after inflation}",
   volume={2018},
   DOI={10.1088/1475-7516/2018/05/067},
   issue={05},
   journal={J. Cosmol. Astropart. Phys.},
   author={Freese, Katherine and Sfakianakis, Evangelos I. and Stengel, Patrick and Visinelli, Luca},
   year={2018},
   pages={067}}

@article{datta2024,
  title = "{Effects of reheating on charged lepton Yukawa equilibration and leptogenesis}",
  author = {Datta, Arghyajit and Roshan, Rishav and Sil, Arunansu},
  journal = {Phys. Rev. Lett.},
  volume = {132},
  issue = {6},
  pages = {061802},
  numpages = {7},
  year = {2024},
  month = {Feb},
  doi = {10.1103/PhysRevLett.132.061802}
}

@article{litsa2021,
  title = "{Large density perturbations from reheating to standard model particles due to the dynamics of the Higgs boson during inflation}",
  author = {Litsa, Aliki and Freese, Katherine and Sfakianakis, Evangelos I. and Stengel, Patrick and Visinelli, Luca},
  journal = {Phys. Rev. D},
  volume = {104},
  issue = {12},
  pages = {123546},
  numpages = {25},
  year = {2021},
  month = {Dec},
  publisher = {American Physical Society},
  doi = {10.1103/PhysRevD.104.123546},
  url = {https://link.aps.org/doi/10.1103/PhysRevD.104.123546}
}

@article{VarieschiFlyby,
   title="{Kerr metric, geodesic motion, and flyby anomaly in fourth-order conformal gravity}",
   volume={46},
   DOI={10.1007/s10714-014-1741-z},
   issue={6},
   journal={Gen. Rel. Grav.}, 
   publisher={Springer Science and Business Media LLC},
   author={Varieschi, G. U.},
   year={2014},
   pages={1741},
   month=may }

@article{Varieschishadow,
   title="{Black hole shadows in fourth-order conformal Weyl gravity}",
   volume={95},
   DOI={10.1139/cjp-2017-0241},
   issue={12},
   journal={Can. J. Phys.},
   publisher={Canadian Science Publishing},
   author={Mureika, J. R. and Varieschi, G. U.},
   year={2017},
   month=dec, pages={1299–1306}}

\end{document}